\documentclass[12pt]{article}
\pdfoutput=1
\usepackage{epsf,amsfonts,amssymb,epsfig,amsmath,mathtools,graphics,slashed}
\usepackage{hyperref,hep,graphicx,subfig}
\usepackage{rotating}
\usepackage{xcolor}
\usepackage{verbatim}
\usepackage[nottoc]{tocbibind}
\allowdisplaybreaks

\definecolor{greenish}{RGB}{0,190,0}

\definecolor{yellowish}{RGB}{190,190,0}

\definecolor{bluish}{RGB}{0,0,190}

\newcommand{\nn}{\notag \\}

\makeatletter

\@addtoreset{equation}{section}
\makeatother

\begin{document}

\begin{titlepage}

\vfill

\begin{flushright}
DCPT-21/01, CPHT-RR001.012021
\end{flushright}

\vfill

\begin{center}
   \baselineskip=16pt
   {\Large\bf Incoherent hydrodynamics of density waves in magnetic fields}
  \vskip 1.5cm
  \vskip 1.5cm
      Aristomenis Donos$^1$, Christiana Pantelidou$^2$ and Vaios Ziogas$^3$\\
   \vskip .6cm
      \begin{small}
      \textit{$^1$ Centre for Particle Theory and Department of Mathematical Sciences,\\ Durham University,
       Durham, DH1 3LE, U.K.}\\
        \textit{$^2$ School of Mathematics, Trinity College Dublin, Dublin 2, Ireland}\\
            \textit{ $^3$ CPHT, CNRS, \'{E}cole Polytechnique, IP Paris, F-91128 Palaiseau, France}
        \end{small}\\   
         
\end{center}

\vfill

\begin{center}
\textbf{Abstract}
\end{center}
\begin{quote}
We use holography to derive effective theories of fluctuations in spontaneously broken phases of systems with finite temperature, chemical potential, magnetic field and momentum relaxation in which the order parameters break translations. We analytically construct the hydrodynamic modes corresponding to the coupled thermoelectric and density wave fluctuations and all of them turn out to be purely diffusive for our system. Upon introducing pinning for the density waves, some of these modes acquire not only a gap, but also a finite resonance due to the magnetic field. Finally, we study the optical properties and perform numerical checks of our analytical results. A crucial byproduct of our analysis is the identification of the correct current which describes the transport of heat in our system.
\end{quote}

\vfill

\end{titlepage}

\tableofcontents
\setcounter{equation}{0}
%\newpage

\section{Introduction}
The holographic conjecture predicts that in a certain large-$N$ limit, large classes of conformal field theories possess a dual classical gravitational description. Apart from its fundamental implications about quantum gravity, it provides a powerful tool to study strongly interacting regimes of quantum field theories which are inaccessible by standard perturbative techniques.

Over the last decade, the duality has been used to study aspects of strongly coupled systems. One of its exciting applications concerns condensed matter systems at finite temperature, chemical potential and magnetic field \cite{Hartnoll:2007ih,Hartnoll:2007ip,Herzog:2007ij,Hartnoll:2007ai,Buchbinder:2008dc,Hansen:2008tq,Buchbinder:2008nf,Blake:2015hxa}. In that context, the discussion was sparkled by the discovery of electrically charged black hole instabilities which lead to superfluids/superconductors \cite{Hartnoll:2008vx,Hartnoll:2008kx,Herzog:2009xv} from the field theory point of view. In this case, the order parameter is given by the expectation value of a complex operator which breaks an internal $U(1)$ symmetry.

Soon after the discovery of holographic superfluid phases, black hole instabilities which spontaneously break translations were found in \cite{Nakamura:2009tf}.  These phases are expected to play a crucial role in understanding particular physical aspects of various condensed matter systems which exhibit instabilities such as charge and spin density waves, including the cuprate superconductors. In this paper we wish to construct the effective theory of long wavelength excitations in holographic phases with spontaneously broken translations.

In order to make contact with realistic condensed matter systems, one needs to tackle the extra complication of the ionic lattice which relaxes the momentum of charge and energy carriers in the system. Momentum relaxation is an essential ingredient in discussing the low frequency transport properties of real materials. In order to accomplish this holographically, we need to deform our UV conformal field theory by relevant operators with source parameters which break translations. In other words, apart from the spontaneous, we also need to implement explicit breaking of translations.

The construction of these inhomogeneous black hole backgrounds and the study of the corresponding thermodynamics is technically challenging mainly due to the fact that unstable modes naturally lead to inhomogeneous backgrounds where the only expected symmetry left is time translations. However, for certain classes of holographic theories with a bulk action which is invariant under global $U(1)$ symmetries one can follow a Q-lattice construction \cite{Donos:2013eha} in order to implement both the holographic lattice as well as the order parameter that spontaneously breaks translations by simply solving ODEs. This system was introduced in \cite{Donos:2014uba,Donos:2014yya} where the focus was on the transport properties and the derivation of analytic formulae for the DC transport coefficients.

In \cite{Donos:2019txg} it was shown that the spontaneous breaking of the global $U(1)$ in the bulk introduced additional diffusive hydrodynamic degrees of freedom to the system, which are separate from the universal ones associated to the conservation of heat and electric charge in the system. From that point of view, the system we are studying is different from the modulated phases of holography where apart from translations, no additional symmetry breaking occurs. In this work, our aim is to generalise the results of \cite{Donos:2019hpp} in order to include an arbitrary number of internal broken symmetries as well as a finite magnetic field.\footnote{For numerical computations of quasinormal modes in $3+1$ boundary dimensions, in the presence of magnetic fields (for systems preserving translations and without spontaneous symmetry breaking), as well as the effects of chiral anomaly, see \cite{Janiszewski:2015ura,Ammon:2017ded,Ammon:2020rvg}.}

Similar systems with spontaneous breaking of translations via an internal symmetry have been studied before in \cite{Delacretaz:2017zxd}. In the absence of an explicit lattice and at zero magnetic field, the longitudinal hydrodynamic modes included one pair of sound and two diffusive modes. One of the diffusive modes can be accounted to the incoherent thermoelectric mode while the second one was associated to the diffusive mode of the internal symmetry breaking described in \cite{Donos:2019txg}. Similarly, the transverse sector of the system contains a single pair of sound modes. It is known that a finite magnetic field has the effect of combining the transverse and longitudinal sound modes to produce a gapped mode and a mode whose frequency was growing quadratically with the wavenumber \cite{PhysRevB.46.3920,PhysRevB.18.6245}. Based on the hydrodynamic models of \cite{Delacretaz:2019wzh,Armas:2020bmo}, reference \cite{Baggioli:2020edn} argued that the corresponding constant of proportionality is complex, using numerical techniques in a holographic massive gravity model; see also \cite{Amoretti:2021fch} for related work in effective theories for weak explicit background lattices. 

In contrast, in our work, we consider phases with strong explicit background lattices, and we analytically show that all hydrodynamic modes remain diffusive even in the presence of a magnetic field. Another aspect of our effective theory is the inclusion of explicit deformation parameters which perturbatively pin the density waves in the system. As one might expect, such deformations introduce a collection of gaps for some of the diffusive modes in our theory. Interestingly, we find that at finite magnetic field pinning also introduces resonance frequencies. As we show, from the retarded Green's functions point of view these show up as poles in the lower half plane.

In section \ref{sec:setup} we discuss the class of holographic models we are considering along with some important aspects of their thermodynamics. In section \ref{sec:lin_hydro} we start by introducing the model of hydrodynamics that provides an effective description of the long wavelength excitations, meanwhile identifying the correct current that describes the transport of heat in our system. We then move on to include the effects of pinning in order to compute the resulting gap and resonance of the density waves, as well as compute the retarded Green's functions and extract their optical properties. We conclude the section by discussing how to decouple the Goldstone modes from the $U(1)$ and heat currents, and by deriving the dispersion relations of our hydrodynamic modes. Section \ref{sec:numerics} contains a number of non-trivial numerical validity checks of the effective theory of section \ref{sec:lin_hydro}. We summarise our most important observations and conclude in section \ref{sec:discussion}. Finally, the appendix contains technical details of the analytical calculations of section \ref{sec:lin_hydro}.

\section{Setup}\label{sec:setup}
In this section we introduce the holographic model that captures all the necessary ingredients that we would like to include in our theory. For this reason, we consider holographic theories which in addition to the metric, they contain a gauge field $A_{\mu}$ and $N_{Y}+N_{Z}$ complex scalar fields $Y^{J}$ and $Z^{I}$ with a global $U(1)^{N_{Y}+N_{Z}}$ symmetry. Essentially, this is a generalisation of the model that we considered in \cite{Donos:2019hpp} to include an arbitrary number of complex scalars in the bulk.

The gauge field will be used to introduce the chemical potential $\mu$ and the magnetic field $B$ in the dual field theory. The first $N_{Y}$ complex scalars are going to implement the explicit lattice and should therefore be relevant operators with respect to the UV theory. The remaining $N_{Z}$ will provide the density wave order parameters in our system. For simplicity, we consider only four bulk spacetime dimensions, corresponding to a $2+1$ dimensional conformal field theory on the boundary, but all our results can easily be generalized to higher dimensional theories as well.

The class of theories we are considering is described by the two-derivative bulk action\footnote{Note that, throughout this paper, the scalar indices $I,J$ will not be summed over unless explicitly stated.}
\begin{align}\label{eq:bulk_action}
S_{bulk}&=\int d^4 x \sqrt{-g}\,\Bigl(R-V-\frac{1}{2}\left(\sum_{I=1}^{N_{Z}}G_{I}\,\partial Z^{I}\partial \bar{Z}^{I} +\sum_{J=1}^{N_{Y}}W_{J}\,\partial Y^{J}\partial \bar{Y}^{J} \right) -\frac{\tau}{4}\,F^{2} \Bigr)  \,,
\end{align}
with $G_{I}$, $W_{J}$, $\tau$ and $V$ being only functions of the moduli $b_{I}=|Z_{I}|^{2}$ and $n_{J}=|Y_{J}|^{2}$.  In this case, the global internal symmetries are represented by the invariance under the global transformations $Z^{I}\to e^{i\theta_{I}}Z^{I}$ and $Y^{J}\to e^{i\omega_{J}}Y^{J}$.

The equations of motion are
\begin{align}\label{eq:eom1}
L_{\mu\nu}&\equiv R_{\mu\nu}-\frac{\tau}{2} (F_{\mu\rho}F_{\nu}{}^{\rho}-\frac{1}{4}g_{\mu\nu}F^2)-\frac{1}{2}g_{\mu\nu} V\nn
&-\frac{1}{2}\left(\sum_{I} G_{I}\, \partial_{(\mu}Z^{I}\partial_{\nu)}\bar{Z}^{I}+\sum_{J} W_{J}\, \partial_{(\mu}Y^{J}\partial_{\nu)}\bar{Y}^{J}\right)=0\,,\notag\\
&\nabla^{\mu}\left(G_{L} \nabla_{\mu}Z^{L}\right)-\partial_{b_{L}}V\,Z^{L}-\frac{\partial_{b_{L}}\tau}{4}Z^{L}\,F^{2}\nn
&-\frac{1}{2}\left(\sum_{I}\partial_{b_{L}}G_{I}\,\partial Z^{I}\partial \bar{Z}^{I} +\sum_{J}\partial_{b_{L}}W_{J}\,\partial Y^{J}\partial \bar{Y}^{J} \right)\,Z^{L}=0\,\notag\,,\\
&\nabla^{\mu}\left(W_{K} \nabla_{\mu}Y^{K}\right)-\partial_{n_{K}}V\,Y^{K}-\frac{\partial_{n_{K}}\tau}{4}Y^{K}\,F^{2}\nn
&-\frac{1}{2}\left(\sum_{I}\partial_{n_{K}}G_{I}\,\partial Z^{I}\partial \bar{Z}^{I} +\sum_{J}\partial_{n_{K}}W_{J}\,\partial Y^{J}\partial \bar{Y}^{J} \right)\,Y^{K}=0\,\notag\,,\\
&C^{\nu}\equiv \nabla_{\mu}\left( \tau\,F^{\mu\nu}\right) =0\,.
\end{align}

In order for our bulk theory to admit an $AdS_{4}$ solution of unit radius, we demand the small $Z^{I}$ and $Y^{J}$ expansions
\begin{align}
V&=-6-\frac{1}{2}\sum_{I=1}^{N_{Z}}m_{Z_{I}}^{2}|Z^{I}|^{2}-\frac{1}{2}\sum_{J=1}^{N_{Y}}m_{Y_{J}}^{2}|Y^{J}|^{2}+\cdots\nn
G_{I}&=1+\cdots,\qquad W_{J}=1+\cdots,\qquad \tau=1+\cdots\,.
\end{align}
In this case, the conformal dimensions $\Delta_{I}$ and $\tilde{\Delta}_{J}$ of the dual complex operators satisfy $\Delta_{I}(\Delta_{I}-3)=m_{Z_{I}}^{2}$ and $\tilde{\Delta}_{J}(\tilde{\Delta}_{J}-3)=m_{Y_{J}}^{2}$. For the $AdS_{4}$ vacuum we use a coordinate system in which the metric reads
\begin{align}
ds^{2}=r^{2}(-dt^{2}+dx_{1}^{2}+dx_{2}^{2})+\frac{dr^{2}}{r^{2}}\,.
\end{align}
In these coordinates, the near conformal boundary expansion for the scalars takes the form
\begin{align}\label{eq:ZY_bulk_exp}
Z^{I}(t,x^{i},r)&=z^{I}_{s}(t,x^{i})\,\frac{1}{r^{3-\Delta_{I}}}+\cdots+z^{I}_{v}(t,x^{i})\,\frac{1}{r^{\Delta_{I}}}+\cdots\,,\nn
Y^{J}(t,x^{i},r)&=y^{J}_{s}(t,x^{i})\,\frac{1}{r^{3-\tilde{\Delta}_{J}}}+\cdots+y^{J}_{v}(t,x^{i})\,\frac{1}{r^{\tilde{\Delta}_{J}}}+\cdots\,,
\end{align}
where $\bar{z}^{I}_{s}$ and $\bar{y}^{J}_{s}$ are the source parameters for the dual operators $\mathcal{O}_{Z^{I}}$ and $\mathcal{O}_{Y^{J}}$.\footnote{In this paper we are using the canonical definition of the one-point function of any real or complex scalar operator $\mathcal{O}$.}
% by differentiating the partition function $\mathcal{Z}$ with respect to the corresponding source $s_\mathcal{O}$, i.e. $\langle\mathcal{O}\rangle=\frac{\partial  \mathcal{Z}}{\partial s_\mathcal{O}}$.}}
The constants of integration $z^{I}_{v}$ and $y^{J}_{v}$ are related to the VEVs $\langle\mathcal{O}_{Z^{I}}\rangle$ and $\langle\mathcal{O}_{Y^{J}}\rangle$ as explained below.

Provided that the operators $\mathcal{O}_{Y^{J}}$ are relevant with $\tilde{\Delta}_{J}<3$, the Q-lattice construction \cite{Donos:2013eha} picks the deformation parameters $y^{J}_{s}(t,x^{i})=\psi_{s}^{J}\,e^{ik^{J}_{si}x^{i}}$. At this point we see that we have a set of $N_{Y}$ wavevectors $k_{si}^{J}$ which are determined externally as part of the sources related to the explicit breaking of the lattice. In our theory, the operators $\mathcal{O}_{Z^{I}}$ take a VEV spontaneously suggesting that for the bulk fields $Z^{I}$ we should have $z^{I}_{s}=0$. In this case, the bulk field $Z^{I}$ are going to be zero above a certain critical temperature. As we lower the temperature of the system, they start developing instabilities which will yield new branches of black holes breaking the global bulk $N_{Z}$ $U(1)$'s spontaneously. However, in section \ref{sec:pinning} we will turn on $z^{I}_{s}$ perturbatively in order to study the pinning of the density waves.

Specifically, within the Q-lattice ansatz for the backgrounds, we have that
\begin{align}
\langle\mathcal{O}_{Z^{I}}\rangle=\left(\Delta_{I}-\frac{3}{2}\right)z_{v}^{I}(t,x^{i})=\left(\Delta_{I}-\frac{3}{2}\right)\phi_{v}^{I}\,e^{i(k_{i}^{I}x^{i}+c^{I})}\,,
\end{align}  
up to potential contact terms. Thus, we have a set $N_{Z}$ wavevectors $k_{i}^{I}$ associated to the order parameters $\langle\mathcal{O}_{Z^{I}}\rangle$. These are dynamically chosen by the system in a way that the free energy is minimised, in contrast to $k_{si}^{J}$ which are fixed by us as part of the boundary conditions. For the particular holographic systems we are studying, the free energy is minimised when $k_{i}^{I}=0$. However, following the logic of \cite{Amoretti:2017frz,Donos:2019hpp}, we will still consider background solutions with $k_{i}^{I}\neq 0$. In this way the order parameters of the spontaneous breaking also break translations apart from the internal $U(1)$'s. 

It is useful to define the real operators
\begin{align}
\Omega^{I}=\frac{1}{2}\,\left(e^{-i(k_{i}^{I}x^{i}+c^{I})} \mathcal{O}_{Z^{I}}+e^{i(k_{i}^{I}x^{i}+c^{I})}\bar{\mathcal{O}}_{Z^{I}}\right)\,,
\end{align}
which have a constant expectation value $\langle\Omega^{I}\rangle=\left(\Delta_{I}-\frac{3}{2}\right)\,\phi_{v}^{I}=|\langle\mathcal{O}_{Z^{I}}\rangle|$ in the broken phase. We now perform an internal infinitesimal rotation $\delta\varepsilon^{I}$ to the bulk scalar $Z^{I}$. The VEVs of the operators $\mathcal{O}_{Z^{I}}$ and $\bar{\mathcal{O}}_{Z^{I}}$ transform according to $\delta\langle\mathcal{O}_{Z^{I}}\rangle=-i\,\langle\mathcal{O}_{Z^{I}}\rangle\,\delta\varepsilon^{I}$ and $\delta\langle\bar{\mathcal{O}}_{Z^{I}}\rangle=i\,\langle\bar{\mathcal{O}}_{Z^{I}}\rangle\,\delta\varepsilon^{I}$ and therefore
\begin{align}
\delta\langle\Omega^{I}\rangle=\langle S^{I}\rangle\delta\varepsilon^{I}\equiv \frac{1}{2i}\,\left(e^{-i(k_{i}^{I}x^{i}+c^{I})} \langle\mathcal{O}_{Z^{I}}\rangle-e^{i(k_{i}^{I}x^{i}+c^{I})}\langle\bar{\mathcal{O}}_{Z^{I}}\rangle\right)\delta\varepsilon^{I}\,.
\end{align}
The above suggests that from a microscopic point of view, the operator
\begin{align}
S^{I}=\frac{1}{2i}\,\left(e^{-i(k_{i}^{I}x^{i}+c^{I})} \mathcal{O}_{Z^{I}} -e^{i(k_{i}^{I}x^{i}+c^{I})}\bar{\mathcal{O}}_{Z^{I}}\right)\,,
\end{align}
is the right object to focus on in order to study the gapless fluctuations of the system. In order to make this point clearer, we parametrise the spacetime fluctuations of the VEVs $\langle\mathcal{O}_{Z^{I}}\rangle$ according to
\begin{align}
\delta\langle\mathcal{O}_{Z_{I}}\rangle(t,x^{i})=\left(\Delta_{I}-\frac{3}{2}\right)e^{i(k_{i}^{I}x^{i}+c^{I})}\,\left(\delta\phi^{I}_{v}(t,x^{i})+i \phi^{I}_{v}\,\delta c^{I}(t,x^{i})\right)\,,
\end{align}
where $\delta c^{I}(t,x^{i})$ parametrises fluctuations of the phase around its value in the thermal state. Correspondingly, we see that the fluctuations of the VEV of $S^{I}$ are 
\begin{align}\label{eq:gapless_op}
\delta\langle S^{I}\rangle(t,x^{i})= \langle\Omega^{I}\rangle\,\delta c^{I}(t,x^{i})\,.
\end{align}
Therefore, the operators $S^{I}$ capture the gapless mode we wish to study.

Note that the bulk expansions \eqref{eq:ZY_bulk_exp} imply that the source for $\Omega^{I}$ is $2\,\textrm{Re}[e^{i(k_{i}^{I}x^{i}+c^{I})} \bar{z}^{I}_{s}]$ and the source for $S^{I}$ is $-2\,\textrm{Im}[e^{i(k_{i}^{I}x^{i}+c^{I})} \bar{z}^{I}_{s}]$.

In order to solve the bulk equations of motion more efficiently, we find convenient to make the field transformations
\begin{align}\label{eq:polar_decomposition}
Y^{J}=\psi^{J}\,e^{i \sigma^{J}},\quad Z^{I}=\phi^{I}\,e^{i \chi^{I}}\,.
\end{align}
bringing the bulk action to the form
\begin{align}\label{eq:bulk_action_alt}
S_{bulk}&=\int d^4 x \sqrt{-g}\,\Bigl(R-V-\frac{1}{2}\left(\sum_{J}W_{J}\,(\partial \psi^{J})^{2} +\sum_{I}G_{I}\,(\partial \phi^{I})^{2} \right)\notag\\
&\qquad\qquad -\frac{1}{2}\left(\sum_{J}\Psi_{J}\,(\partial \sigma^{J})^{2} +\sum_{I}\Phi_{I}\,(\partial \chi^{I})^{2} \right)-\frac{\tau}{4}\,F^{2} \Bigr)\,,\notag\\
\Psi_{J}&\equiv W_{J}\,(\psi^{J})^{2},\quad \Phi_{I}\equiv G_{I}\,(\phi^{I})^{2}\,.
\end{align}
A consistent ansatz that captures all the necessary ingredients we discussed so far for the thermal state is
\begin{align}
\label{eq:ansatz}
&ds^2=-U(r) dt^2+\frac{1}{U(r)} dr^2+ g_{ij}(r)dx^{i}dx^{j}\,,\nonumber\\
&A=a_{t}(r) dt-B x^2 dx^1\,,\nonumber\\
&\phi^{I}=\phi^{I}(r)\,,\qquad \chi^{I}= k^{I}_i x^i+c^{I}\,,\nonumber\\
&\psi^{J}=\psi^{J}(r)\,,\qquad \sigma^{J}= k^{J}_{si} x^i\,.
\end{align}
According to \eqref{eq:polar_decomposition}, the constants $c^{I}$ in our ansatz \eqref{eq:ansatz} for the background black holes translate to an overall phase for the complex scalars $Z^{I}$. The absence of explicit sources in our asymptotic expansion for the corresponding field does not fix it and we have to leave it arbitrary. These are essentially the gapless modes associated with the symmetry breaking in the bulk that we wish to promote to hydrodynamic ones in section \ref{sec:lin_hydro}.

We choose our coordinate system so that the conformal boundary of $AdS_{4}$ is approached as we take $r\to\infty$. In this case, the asymptotic expansions of the functions in our ansatz \eqref{eq:ansatz} take the form
\begin{align}\label{asymptsol}
U&\to (r+R)^2+\cdots+W\,(r+R)^{-1}+\cdots,\nn g_{ij}&\to \delta_{ij}\,(r+R)^{2}+\cdots+g_{ij}^{(3)}\,(r+R)^{-1}+\cdots,\qquad a\to\mu+Q\,(r+R)^{-1}+\cdots,\nn
\psi^{J}&\to  \psi^{J}_{s}\,(r+R)^{-3+\tilde{\Delta}_{J}}+\cdots+\psi^{J}_{v}\,(r+R)^{-\tilde{\Delta}_{J}}+\cdots, \nn
\phi^{I}& \to   \phi^{I}_{s}\,(r+R)^{-3+\Delta_{I}}+\cdots+\phi^{I}_{v}\,(r+R)^{-\Delta_{I}}+\cdots\,,
\end{align}
where we chose to only show the terms where constants of integration of the relevant ODEs appear. The constants of integration $g_{ij}^{(3)}$ that appear in the expansion of the metric have to satisfy $\delta^{ij}g_{ij}^{(3)}=-\frac{1}{2}\sum_{J}\left(\tilde{\Delta}_{J}-3\right)\left(\frac{2}{3}\tilde{\Delta}_{J}-1\right)\psi^{J}_{s}\psi^{J}_{v}$ which is the gravitational constraint and yields the conformal anomaly for the stress tensor.

The constant of integration $R$ in \eqref{asymptsol} represents a global shift in the radial coordinate $r$. This is fixed by demanding that the finite temperature horizon is at $r=0$. Demanding our background solutions to be regular imposes the near horizon expansion
\begin{align}\label{nhexpbh}
U\left(r\right)&=4\pi\,T\,r+\cdots\,,\qquad g_{ij}=g_{ij}^{(0)}+\cdots\,,\qquad a=a^{(0)}\,r+\cdots\,,\nn
\phi^{I}&=\phi^{I(0)}+\cdots\,,\qquad \psi^{J}=\psi^{J(0)}+\cdots\,.
\end{align}

The equations of motion \eqref{eq:eom1} lead to the following equations for the phases of the complex scalars associated to spontaneous breaking
\begin{align}\label{eq:chi_eom}
\nabla_{\mu}\left(\Phi_{I}\nabla^{\mu}\chi^{I} \right)=0\,.
\end{align}
At this point, it is useful to note that the fields $\sigma^{J}$ and $\chi^{I}$ are not well defined when either $\psi^{J}$ or $\phi^{I}$ are equal to zero. This certainly happens close to the conformal boundary and in order to avoid misinterpretations with the holographic dictionary, we discuss asymptotic expansions in terms of the complex fields $Y^{J}$ and $Z^{I}$ through \eqref{eq:polar_decomposition}. This is well defined in the regime of perturbation theory that we are interested in.

The asymptotic expansions for the perturbations of $\phi^{I}$ and $\chi^{I}$ are
\begin{align}
\delta\phi^{I}(t,x^{i}, r)&=\delta\phi_{s}^{I}(t,x^{i})\frac{1}{(r+R)^{3-\Delta_{I}}}+\cdots+\delta\phi_{v}^{I}(t,x^{i})\frac{1}{(r+R)^{\Delta_{I}}}+\cdots\,,\nn
\delta\chi^{I}(t,x^{i}, r)&=\frac{\zeta_{S^{I}}}{\phi^{I}_{v}}(t,x^{i})\frac{1}{(r+R)^{3-2\Delta_{I}}}+\cdots+\delta c^{I}(t,x^{i})+\cdots\,.
\end{align}
From these expansions and using \eqref{eq:polar_decomposition} we obtain
\begin{align}\label{eq:pert_sources}
	\delta z^{I}_{s}&=e^{i(k_{i}^{I}x^{i}+c^{I})}(i \zeta_{S^{I}}(t,x^{i})+\delta\phi^{I}_{s}(t,x^{i}))\,,\nn
	\delta z^{I}_{v}&=e^{i(k_{i}^{I}x^{i}+c^{I})}(i \phi^{I}_{v}\,\delta c^{I}(t,x^{i})+\delta\phi^{I}_{v}(t,x^{i}))\,, 
\end{align}
where we have used that $\bar{z}^{I}_{s}=0$, or equivalently $\phi^{I}_{s}=0$, in the phases we are interested in. This shows that, up to contact terms, $\delta \langle S^{I}\rangle=\left(\Delta_{I}-\frac{3}{2}\right)\,\phi_{v}^{I}\,\delta c^{I}$, and that $2\,\delta\phi^{I}_{s}$ is a source for $\Omega^{I}$ while $2\,\zeta_{S^{I}}$ is a source for $S^{I}$, consistent with equation \eqref{eq:gapless_op} and the discussion below it.

In the next subsection we discuss aspects of the thermodynamics of our broken phase black holes. This will give us the opportunity to define certain quantities that will appear later in the context of hydrodynamics.

\subsection{Thermodynamics}\label{sec:thermo}
In this subsection we would like to consider the thermodynamics of the background black holes we are interested in. In order to do this we need to regularise the bulk action \eqref{eq:bulk_action} by adding suitable boundary terms which act as counter-terms \cite{Balasubramanian:1999re,Skenderis:2002wp}. The purpose of these terms is dual, the first is to render the total on-shell action finite. The second is to make the variational problem well defined, provided we have a unique way to fix the boundary conditions on our bulk fields.

It is often the case that such terms are not unique and for the purposes of our paper it is enough to list the following terms \cite{Skenderis:2002wp}
\begin{align}\label{eq:bdy_action}
S_{bdr}=&\int_{\partial M}d^{3}x\,\sqrt{-\gamma}\,\left(-2K + 4 +R_{bdr}\right)\notag\\
&\quad-\frac{1}{2}\int_{\partial M}d^{3}x\,\sqrt{-\gamma}\left[\sum_{I}(3-\Delta_{I})\bar{Z}^{I}Z^{I}+\sum_{J}(3-\tilde{\Delta}_{J})\bar{Y}^{J}Y^{J}\right]\notag\\
&\quad+\frac{1}{2}\int_{\partial M}d^{3}x\,\sqrt{-\gamma}\left[\sum_{I}\frac{1}{2\Delta_{I}-5}\,\partial_{a}\bar{Z}^{I}\partial^{a}Z^{I}+\sum_{J}\frac{1}{2\tilde{\Delta}_{J}-5}\,\partial_{a}\bar{Y}^{J}\partial^{a}Y^{J}\right]+\cdots\,.
\end{align}
Further counter-terms can be added \cite{Papadimitriou:2011qb} but these will introduce extra contact terms in the retarded Green's functions that we wish to compute from the bulk theory.

In order to compute the free energy of the system we need to consider the Euclidean version of the total action $I_{E}=-iS_{tot}$. We then need to evaluate the value of $I_{E}$ on the solution with the analytically continued time $t=-i t_E$ and the periodic identification $t_E \sim t_E+ T^{-1}$. Since our system extends infinitely in the spatial field theory directions, the total free energy $W_{FE}=T\,I_{E}$ is not meaningful and we instead consider the free energy density $w_{FE}$
\begin{align}
w_{FE}=\epsilon- T\,s-\mu\,\rho\,,
\end{align}
where $\epsilon$, $s$ and $\rho$ denote the energy density, entropy density and electric charge density respectively. Apart from the thermodynamic data $T$, $\mu$, $B$ and the explicit lattice data $\psi^{J}_{s}$, $k^{J}_{si}$ our solutions also depend on the wavenumbers $k^{I}_{i}$ which are related to the spontaneous breaking. Even though different values of $c^{I}$ in \eqref{eq:ansatz} yield different solutions, the free energy is independent of those in the spontaneous case, when $\phi_{s}^{I}=0$.\footnote{Since the bulk $U(1)$ symmetries which shift $c^I$ are global, the dual boundary theory does not possess the corresponding local Noether charges and currents \cite{Amado:2013xya,Donos:2019txg}. Based on this fact,  \cite{Donos:2019txg,Baggioli:2020nay,Baggioli:2020haa} argued that the corresponding gapless modes behave like phasons.} 

The first variation of the free energy yields the first law
\begin{align}\label{eq:first_law}
\delta w_{FE}=-\rho\,\delta\mu-s\,\delta T+ \sum_{I}w_{I}^{i}\,\delta k^{I}_{i}-M\,\delta B\,.
\end{align}
where the electric charge and entropy densities can be computed from the black hole horizon data
\begin{align}\label{eq:rhos_hor}
\rho=\sqrt{g_{(0)}}\,\tau^{(0)}\,a^{(0)},\quad s=4\pi\,\sqrt{g_{(0)}}\,,
\end{align}
and $M$ is the magnetisation and by $\tau^{(0)}$ we denote the value of $\tau$ when $\phi^I$ and $\psi^J$ are evaluated on the horizon. In order to compute the variation $w_{I}^{i}$ of the free energy with respect to the wavenumber $k^{I}_{i}$, we simply vary the total action $S_{tot}$ and use the background equations of motion to find\footnote{Here and below, $\sqrt{g}$ denotes the square root of the determinant of the spatial metric $g_{ij}$.}
\begin{align}\label{eq:wkvariation}
w^{i}_{I}=\partial_{k^{I}_{i}}w=\int_{0}^{\infty}dr\,\sqrt{g}\,\Phi_{I}\,g^{ij} k_{j}^{I}\,.
\end{align}
Notice that in the spontaneous case this is coming entirely from the variation of the bulk action \eqref{eq:bulk_action} and it is a finite number. Potential contributions from finite counter-terms other than those listed in \eqref{eq:bdy_action} would be possible but these would constitute contact terms which we can ignore for our purposes. 

In order to obtain the electric magnetisation $M^{ij}$ of the dual field theory, we need to perform a straightforward variation of the bulk action \eqref{eq:bulk_action} with respect to the magnetic field $B$. Apart from the electric magnetisation, our backgrounds are also going to have a non-trivial thermal magnetisation $M_{T}^{ij}$. Similarly to the electric magnetisation, this is not immediately obvious from the backgrounds in \eqref{eq:ansatz} since the homogeneity of our solutions prevents the appearance of explicit heat magnetisation currents. In order to define it, we would need to introduce a larger background ansatz than \eqref{eq:ansatz} in order to include NUT charges in our metric \cite{Donos:2015bxe}. Instead of doing that, we simply give the expression for its value in terms of the background
\begin{align}
M^{ij}&=-\frac{B\varepsilon^{ij}}{2}\,\int_{0}^{\infty}dr\, \frac{\tau}{\sqrt{g}}=M\,\varepsilon^{ij}\,,\label{eq:magnetisation}\\
M^{ij}_{T}&=B\varepsilon^{ij}\,\int_{0}^{\infty}dr\,\frac{\tau}{\sqrt{g}}\,a_{t}=M_{T}\,\varepsilon^{ij}\,.\label{eq:magnetisation_th}
\end{align}

Apart from the quantities that appear in the first variation of the free energy, we also find it useful to introduce a set of susceptibilities through the second variation of the free energy
\begin{align}
\delta s&=T^{-1}c_{\mu}\,\delta T+\xi\,\delta\mu+\sum_{I}\nu^{i}_{I}\,\delta k^{I}_{i}\,,\notag\\
\delta\rho&=\xi\,\delta T+\chi_{q}\,\delta \mu+\sum_{I}\beta_{I}^{i}\,\delta k^{I}_{i}\,,\notag\\
\delta w^{i}_{I}&=-\nu_{I}^{i}\,\delta T-\beta_{I}^{i}\,\delta\mu+\sum_{L}w_{IL}^{ij}\,\delta k^{L}_{j}\,.
\end{align}
An expression for the susceptibilities $\nu_{I}^{i}$, $\beta_{I}^{i}$ and $w_{IL}^{ij}$ in terms of the background can be obtained by simply varying e.g. equation \eqref{eq:wkvariation}. Moreover, the susceptibilities $\nu_{I}^{i}$ and $\beta_{I}^{i}$ can also be found by varying the densities in equation \eqref{eq:rhos_hor} with respect to the spontaneous wavenumbers $k_{i}^{I}$. Even though it is not obvious from the bulk expressions that these two approaches lead to the same result, this is guaranteed by the thermodynamic Maxwell relations. This observation will become important in section \ref{sec:lin_hydro}, when we derive the constitutive relations for the currents and the Josephson equation in a derivative expansion.

\subsection{Perturbations}\label{sec:perts}

The first step in order to understand the subset of hydrodynamic fluctuations is to consider general gravitational perturbations around the background black holes \eqref{eq:ansatz}. As we wish to study fluctuations which involve both spatial directions on the boundary, we need to write down the consistent ansatz
\begin{align}\label{eq:sep_var}
\delta (ds^2) & =e^{-i\omega\, v_{EF} +i q_{j} x^{j}}\,\left[\delta g_{\mu\nu}(r)dx^{\mu}dx^{\nu} + 2 (i\omega)^{-1}\, U \zeta_{i}\, dtdx^i\right]\,,\nn
\delta A & =e^{-i\omega\, v_{EF} +i q_{j} x^{j}}\,\left[\delta a_{\mu}(r) dx^{\mu} + (i\omega)^{-1} (E_{i}- a_{t} \zeta_{i})\, dx^i\right]\,,\nn
\delta \phi^{I} & = e^{-i\omega\, v_{EF} +i q_{j} x^{j}} \delta\phi^{I}(r)\,,\qquad \delta \chi^{I} = e^{-i\omega\, v_{EF} +i q_{j} x^{j}} \delta\chi^{I}(r)\,,\nn
\delta \psi^{J} & = e^{-i\omega\, v_{EF} +i q_{j} x^{j}} \delta\psi^{J}(r)\,,\qquad \delta \sigma^{J} = e^{-i\omega\, v_{EF} +i q_{j} x^{j}} \delta\sigma^{J}(r)\,,
\end{align}
where we have also performed a separation of variables. We have introduced the combination
 \begin{align}\label{eq:vef}
 v_{EF}=t+S(r)\,,
 \end{align}
 with $S(r)$ such that close to the horizon at $r=0$ it approaches $S(r)\sim \frac{1}{4\pi T} \ln r +\cdots$. In this case, the function $v_{EF}$ approaches the infalling Eddington Finkelstein coordinate and the perturbation is regular infalling by demanding that
% $\delta X\equiv\{\delta g_{tt},\delta g_{ti},\delta g_{rr},\delta g_{ri},\delta g_{ij},\delta a_{t},\delta a_{r},\delta a_{i},\delta \psi^{J}, \delta\sigma^{J},\delta \phi^{I}, \delta\chi^{I}\}$
% \begin{align}\label{eq:sep_var}
% \delta X(t,r,x_{1})=e^{-i\omega\, v_{EF} +i q_{i} x^{i}}\,\delta X(r)
% \end{align}
% where we have introduced 
\begin{align}\label{eq:gen_exp}
\delta g_{tt}(r)&= 4\pi T\,r\, \delta g_{tt}^{(0)}+\cdots\,,\quad
\delta g_{rr}(r)=\frac{\delta g_{rr}^{(0)}}{4\pi T\,r}+\cdots\, \,,\nn
\delta g_{ti}(r)&=\delta g_{ti}^{(0)}+r\,\delta g_{ti}^{(1)}+\cdots\,,\quad 
\delta g_{ri}(r)=\frac{\delta g_{ri}^{(0)}}{4\pi T\,r}+\delta g_{ri}^{(1)}+\cdots\,,\nn
\delta g_{ij}(r)&=\delta g_{ij}^{(0)}+\cdots\,,\quad\quad
\delta g_{tr}(r)=\delta g_{tr}^{(0)}+\cdots,\quad \delta a_{i}(r)=\delta a_{i}^{(0)}+\cdots\,,\nn
\delta a_{t}(r)&=\delta a_{t}^{(0)}+\delta a_{t}^{(1)}\,r+\cdots\,,\quad
\delta a_{r}(r)=\frac{1}{4\pi T\,r} \delta a_{r}^{(0)}+\delta a_{r}^{(1)}+\cdots\,,\nn
\delta \psi^{J}(r)&=\delta \psi^{J(0)}+\cdots,\quad \delta \phi^{I}(r)=\delta \phi^{I(0)}+\cdots,\nn \delta \chi^{I}(r)&=\delta \chi^{I(0)}+\cdots,\quad \delta \sigma^{J}(r)=\delta \sigma^{J(0)}+\cdots\,.
\end{align}
which are compatible with the equations of motion. In order to achieve regularity, the above need to be supplemented by
\begin{align}\label{eq:nh_reg}
-2\pi T(\delta g_{tt}^{(0)}+\delta g_{rr}^{(0)})=-4\pi T\,\delta g_{rt}^{(0)}&\equiv p\,,\notag\\
\delta g_{ti}^{(0)}=\delta g_{ri}^{(0)}&\equiv-v_{i},\notag\\
\delta a_{r}^{(0)}=\delta a_{t}^{(0)}&\equiv \varpi\,.
\end{align}

It is useful to note that at the current stage of the discussion, the $13+2\, N_{Z}+2\,N_{Y}$ constants $\delta g_{tt}^{(0)}$, $\delta g_{ti}^{(1)}$, $\delta g_{ij}^{(0)}$, $\delta a_{i}^{(0)}$, $\delta a_{t}^{(1)}$, $\delta \psi^{J(0)}$, $\delta \phi^{I(0)}$, $\delta \chi^{I(0)}$, $\delta \sigma^{J(0)}$, $\varpi$, $p$ and $v_{i}$ are constants of integration and therefore free. Moreover, we haven't fixed a gauge choice and coordinate system for our fluctuations. For example, the choice of $\delta g_{r\mu}$ and $\delta a_{r}$ can be completely arbitrary, as long as we satisfy the regular boundary conditions prescribed in \eqref{eq:gen_exp} and \eqref{eq:nh_reg}. After doing so, all the remaining functions will satisfy $9+2N_{Z}+2N_{Y}$ second order ODEs in the radial direction as well as 4 constraints which come from diffeomorphism invariance and the Gauss constraint coming from gauge invariance.

In order to discuss the constraints which we need to impose, we choose to work in a radial foliation and we define the normal one form $n=dr$ normal to constant $r$ hypersurfaces. We now use our equations of motion in \eqref{eq:eom1} to define $L^{\mu}=E^{\mu}{}_{\rho}n^{\rho}$ and $C=C^{\rho}n_{\rho}$ with $E_{\mu\nu}=L_{\mu\nu}-\frac{1}{2}g_{\mu\nu}\,L^{\rho}{}_{\rho}$. The four gravitational constraints are simply $L^{\mu}=0$ and the Gauss constraint is $C=0$.

The constraints that we need to satisfy can be imposed on any hypersurface with e.g. constant radial coordinate $r$. Close to the conformal boundary, they are equivalent to the Ward identities of charge conservation, diffemorphism and Weyl invariance of the dual conformal field theory. Similar to \cite{Donos:2019hpp}, we will derive an effective hydrodynamic theory in section \ref{sec:lin_hydro} by utilizing a subset of these constraints on a surface close to the background black hole horizon at $r=0$, in terms of the constants that appear in \eqref{eq:gen_exp} and \eqref{eq:nh_reg}. 

We now define the horizon currents
\begin{align}\label{eq:J_hor2}
\delta Q_{(0)}^{i} &=4\pi T \sqrt{g_{(0)}} v^i\,,\nn
\delta J_{(0)}^{i}&=\sqrt{g_{(0)}}\tau^{(0)} \left( i q^{i} \varpi +i\omega g_{(0)}^{ij}\delta a^{(0)}_{j} +E^i+v^{i}a_{t}^{(0)}+F^{ij}_{(0)}v_j\right)\,.
\end{align}
Building on \cite{Donos:2017ihe,Donos:2015bxe}, we derive the horizon constraints that the above currents should satisfy\footnote{It is relatively straightforward to check that the $r$ and $t$ components of the gravitational constraints are equivalent in the $r\to 0$ limit.} %in terms of which the horizon constraints can be written as \cite{Donos:2017ihe,Donos:2015bxe}, 
\begin{subequations}\label{hconstraints_all}
\begin{align}
&iq_{i}\, \delta  Q_{(0)}^{i}=2\pi T\, i\omega \sqrt{g_{(0)}} g_{(0)}^{ij}\delta g^{(0)}_{ij}\,, \label{hamconstraint}\\
&i q_{i} \delta  J^{i}_{(0)} = i\omega \sqrt{g_{(0)}}\Big[
\tau^{(0)}\left(a^{(0)}\,\left(\delta g^{(0)}_{tt}+\frac{p}{4\pi T}\right) +\delta a^{(1)}_{t}-\frac{i \omega}{4\pi T}\left( \delta a^{(1)}_{t}-\delta a^{(1)}_{r} \right)\right) \nn
&\qquad \qquad +\frac{1}{2}\tau^{(0)} a^{(0)}g_{(0)}^{ij}\delta g^{(0)}_{ij} 
+\partial_{\phi^{I}}\tau^{(0)}a^{(0)}\,\delta\phi^{I(0)}+\partial_{\psi^{J}}\tau^{(0)}a^{(0)}\,\delta\psi^{J(0)}\Big]\,, \label{gaugeconstraint}\\
&i\omega\left(-\delta g_{ti}^{(1)}-g^{(1)}_{il}v^{l} +iq_{i}(\delta g_{tr}^{(0)}-\delta g_{rr}^{(0)}) +\frac{i\,\omega}{4\pi T}(\delta g_{ti}^{(1)}-\delta g_{ri}^{(1)})+\zeta_{i} +i q^{k}\delta g^{(0)}_{ki} \right) \nn
&+q^2 v_{i} +q_i q_j v^{j} +i q_{i} \left(1+\frac{i\omega}{2\pi T}\right)p -4\pi T\,\zeta_{i}-\tau^{(0)}a^{(0)}\,\left( i q_{i}\varpi+i\omega\delta a_{i}^{(0)}+E_i \right) \nn
& +\Psi_{J}^{(0)}k_{s\,i}^{J}\,\left(k_{s\,J}^{I}v^{j}-i\omega \delta\sigma^{J(0)} \right) +\Phi_{I}^{(0)}k_{i}^{I}\,\left(k_{j}^{I}v^{j}-i\omega \delta\chi^{I(0)} \right) - F_{ij}^{(0)}\,\left(\sqrt{g_{(0)}}\right)^{-1} \delta  J_{(0)}^{j}=0\,. \label{imomconstraint}
\end{align}
\end{subequations}

Close to the conformal boundary at $r=\infty$, the expansion of our functions is
\begin{align}\label{eq:gen_UVexp}
\delta g_{tt}(r)&=\cdots+ \frac{\delta g_{tt}^{(v)}}{r+R}+\cdots\,,\quad
\delta g_{rr}(r)=\mathcal{O}(r^{-4})\,,\quad\delta g_{ti}(r)=\cdots+ \frac{\delta g_{ti}^{(v)}}{r+R}+\cdots\,,\nn 
\delta g_{ri}(r)&=\mathcal{O}(r^{-3})\,,\quad
\delta g_{ij}(r)=\cdots+ \frac{\delta g_{ij}^{(v)}}{r+R}+\cdots\,,\quad
\delta g_{tr}(r)=\mathcal{O}(r^{-2})\,,\nn
\delta a_{i}(r)&=\frac{\delta a_{i}^{(v)}}{r+R}+\cdots\,,\quad \delta a_{t}(r)=\frac{\delta a_{t}^{(v)}}{r+R}+\cdots\,,\quad \delta a_{r}(r)=\mathcal{O}(r^{-2})\,,\nn
\delta \psi^J(r)&=\frac{\delta\psi^{J(v)}}{(r+R)^{\tilde{\Delta}_{J}}}+\cdots\,,\quad \delta \phi^I(r)=\frac{\delta\phi^{I(v)}}{(r+R)^{\Delta_{I}}}+\cdots\,,\nn
\delta \chi^I(r)&=\frac{\zeta_{S^{I}}}{\phi_{v}^{I}}\,(r+R)^{2\Delta_{I}-3}+\cdots+ \delta c^{I}+\cdots \,,\quad \delta \sigma^J(r)=\delta\sigma^{J(v)}\,(r+R)^{2\tilde{\Delta}_{J}-3}+\cdots\,,
\end{align}
where we have chosen to only show the undetermined terms involving the constants of integration of the second order ODE's we need to solve. We have also included the constants $\zeta_{S^{I}}$ which represent the sources for the operators $S^{I}$ we discussed in section \ref{sec:setup} and which we need to fix. For the components which we are free to choose by using diffeomorphism and gauge invariance, we only show their desired behaviour close to the boundary as they offer no additional information as far as the constants of integration are concerned.

The remaining $9+2N_{Z}+2N_{Y}$ constants of integration $\delta g_{tt}^{(v)}$, $\delta g_{ti}^{(v)}$, $\delta g_{ij}^{(v)}$, $\delta a_{t}^{(v)}$, $\delta a_{i}^{(v)}$, $\delta \psi^{J(v)}$, $\delta \phi^{I(v)}$, $\delta c^{I}$ and $\delta\sigma^{J(v)}$ together with the $13+2N_{Z}+2N_{Y}$ coming from the horizon expansion, will fix a unique solution of the $9+2N_{Z}+2N_{Y}$ second order ODE's and the 4 constraints. In the next section we construct solutions which correspond to the late time, long wavelength hydrodynamic modes of the boundary theory.

\section{Linearised hydrodynamics}\label{sec:lin_hydro}
In this section we study the hydrodynamic limit of the fluctuations that we introduced in subsection \ref{sec:perts}. In subsection \ref{sec:hydro_perts} we discuss the construction of these modes up to second order in the derivative expansion. This allows us to write an effective hydrodynamic theory for the conserved currents of the system and its gapless modes related to spontaneous breaking in the bulk in subsection \ref{sec:eft}. 

Having a complete effective theory for our fluctuations, in subsection \ref{sec:pinning} we examine the gap induced for the spontaneous density waves sliding modes, by perturbatively small deformations for the operators $\Omega_{I}$. To illustrate, we specialise to the isotropic case where we find that apart from a gap, the theory also develops resonance frequencies. In subsection \ref{sec:transport} we study the retarded Green's functions of the operators in our theory at finite frequency and we give the precise way that the poles of subsection \ref{sec:pinning} influence the transport properties. 

We then move on to give more general, model-independent, Kubo formulas for some of the transport coefficients in subsection \ref{subsec:decoupling}. There, we also define heat and electric current operators which decouple from the Goldstone modes, and we discuss some of their properties. Finally, in subsection \ref{sec:diffusion} we give an algebraic equation whose solutions yield the dispersion relations of our hydrodynamic modes. Even though we are not able to find the solutions in closed form, we can show that all our modes are purely diffusive at the order we are working. An interesting outcome of our results is that, after correctly identifying the heat current, the thermodynamic coefficients $w_{I}^{i}$ all drop out from physically interesting quantities.

\subsection{Hydrodynamic perturbations}\label{sec:hydro_perts}
In their infinite wavelength $q_{i}\to 0$ limit, the hydrodynamic modes we wish to study will reduce to a uniform distribution of energy, electric charge and phase shift of the complex scalar $Z_{I}$ which break the global symmetries in the bulk. In order to keep track of our expansion we scale $q^{i}\to\varepsilon q^{i}$ with $\varepsilon$ a small number and expand the frequencies and radial functions in the bulk according to
\begin{align}\label{eq:X_epsilon_exp}
\omega&=\varepsilon\,\omega_{[1]}+\varepsilon^{2}\,\omega_{[2]}+\cdots\notag\\
\delta X(r)&=\delta X_{[0]}(r)+\varepsilon\delta X_{[1]}(r)+\varepsilon^{2}\delta X_{[2]}(r)+\cdots\,,
\end{align}
where $\delta X(r)$ can be any of the functions that appear in the ansatz for the perturbation \eqref{eq:sep_var}.

A key point of our construction is the leading piece of the $\varepsilon$ expansion which according to our earlier discussion has to reduce to
\begin{align}\label{eq:deltaX0}
\delta X_{[0]}=\frac{D X_{b}}{D T}\,\delta T_{[0]}+\frac{D X_{b}}{D \mu}\,\delta \mu_{[0]} +\sum_{I}\frac{\partial X_{b}}{\partial c^{I}}\,\delta c^{I}_{[0]}\,.
\end{align}
The functions $X_{b}$ represent the background fields of the black hole in equation \eqref{eq:ansatz}. In order to generate the perturbations which satisfy the correct boundary conditions \eqref{eq:gen_exp}, \eqref{eq:nh_reg} and \eqref{eq:gen_UVexp}, at the same time with a simple partial derivative with respect to $T$, $\mu$ and $c^{I}$ we also need to perform the perturbative coordinate and gauge transformation \cite{Donos:2017ihe}
\begin{align}
t\to t+\delta T_{[0]}\,T^{-1}\,g(r),\qquad A\to A-\delta\mu_{[0]}\,d(t+g(r))\,.
\end{align}
As expected, the boundary condition requirements for our perturbations do not uniquely fix $g(r)$ in the bulk. It is enough to choose it such that close to the conformal boundary it vanishes sufficiently fast while close to the horizon at $r=0$ it approaches $g(r)\to \ln r/(4\pi T)+g^{(1)}\,r+\cdots$.

Choosing the perturbations as in \eqref{eq:sep_var} leads to an inhomogeneous system of differential equations coming from the bulk equations of motion \eqref{eq:eom1} at the perturbative level. It is clear from the above construction that the seed solution \eqref{eq:deltaX0} satisfies the corresponding homogeneous system of equations \cite{Donos:2019hpp}. This suggests that we can add them at each order in the $\varepsilon$ expansion \eqref{eq:X_epsilon_exp} and therefore consider the split,
\begin{align}
\delta X_{[n]}=\delta \tilde{X}_{[n]}+\frac{D X_{b}}{D T}\,\delta T_{[n]}+\frac{D X_{b}}{D \mu}\,\delta \mu_{[n]}+\sum_{I}\frac{\partial X_{b}}{\partial c^{I}}\,\delta c^{I}_{[n]}\,,
\end{align}
with $\delta \tilde{X}_{[n]}$ a solution to the inhomogeneous problem which is sourced by lower order terms of the solution. Following closely the analysis of \cite{Donos:2019hpp}, we can show that the eigenmodes of the system necessarily have $\omega_{[1]}=\delta T_{[0]}=\delta \mu_{[0]}=0$. Therefore the variation of the temperature and chemical potential starts at order $\mathcal{O}(\varepsilon)$.

The next to leading part of the bulk solution $\delta X_{[1]}$ will only be driven by a shift in the background phases of the complex scalars according to $\delta\chi^{I}=e^{i\varepsilon\,q_{i}x^{i}}\,\delta c^{I}_{[0]}$. Moreover, since we are examining the equations of motion at order $\mathcal{O}(\varepsilon)$, it is only the first derivatives of the varying exponential that enter the source of the inhomogeneous part $\delta\tilde{X}_{[1]}$. Effectively we can say that up to order $\varepsilon$ we have
\begin{align}
\chi^{I}\approx k_{i}^{I}x^{i}+c^{I}+e^{i\varepsilon\,q_{i} x^{i}}\,\delta c^{I}_{[0]}\approx (k_{i}^{I}+i\varepsilon\,q_{i}\,\delta c^{I})\,x^{i}+c^{I}+\delta c^{I}+\cdots\,.
\end{align}
The above implies that $\delta\tilde{X}_{[1]}$ will simply be the change of the background solution under $\delta k_{i}^{I}=i\,\varepsilon q_{i}\,\delta c^{I}$. The same pattern will appear at all orders in the $\varepsilon$ expansion, leading us to the further split of the solution to the inhomogeneous problem according to \cite{Donos:2019hpp}
\begin{align}
\delta \tilde{X}_{[n]}=\delta\mathbb{X}_{[n]}+i\sum_{I}\frac{\partial X_{b}}{\partial k_{i}^{I}}\, q_{i}\,\delta c^{I}_{[n-1]}\,.
\end{align}

In the end, the whole solution is determined by the variations $\delta T$, $\delta\mu$ and $\delta c^{I}$ and so does the horizon fluid velocity $v^{i}$, the local chemical potential $\varpi$, and the vector potential $\delta a_{j}^{(0)}$. More specifically, we can identify
\begin{align}\label{eq:hor_fluid_exp}
&p=4\pi \left( \varepsilon\, \delta T_{[1]}+\varepsilon^2\, \delta T_{[2]}  +\cdots\right),\,\,
v^{i}=\varepsilon^2\,v_{[2]}^{i}+\cdots,\,\,
\varpi=\,- \left( \varepsilon\,\delta\mu_{[1]} +\varepsilon^2\,\delta\mu_{[2]}+\cdots\,\right),\nn
&\delta g_{ij}^{(0)}=\varepsilon\,\left(\frac{\partial g^{(0)}_{ij}}{\partial T}\,\delta T_{[1]}+\frac{\partial g^{(0)}_{ij}}{\partial \mu}\,\delta \mu_{[1]}+i \sum_{I}\frac{\partial g^{(0)}_{ij}}{\partial k_{i}^{I}}\,q_{i}\delta c^{I}_{[0]}+\varepsilon\, \delta g_{[2]ij}^{(0)}+\cdots\right)\,,\nn
&\delta\phi^{I(0)}=\varepsilon\,\left(\frac{\partial \phi^{I(0)}}{\partial T}\,\delta T_{[1]}+\frac{\partial \phi^{I(0)}}{\partial\mu}\,\delta\mu_{[1]} +i \sum_{L}\frac{\partial\phi^{I(0)}}{\partial k_{i}^{L}}\,q_{i}\delta c^{L}_{[0]}+\varepsilon\,\delta\phi_{[2]}^{I(0)}+\cdots \right)\,,\nn
&\delta\psi^{J(0)}=\varepsilon\,\left(\frac{\partial \psi^{J(0)}}{\partial T}\,\delta T_{[1]}+\frac{\partial \psi^{J(0)}}{\partial\mu}\,\delta\mu_{[1]} +i \sum_{I}\frac{\partial \psi^{J(0)}}{\partial k_{i}^{I}}\,q_{i}\delta c^{I}_{[0]}+\varepsilon\,\delta\psi_{[2]}^{J(0)}+\cdots \right)\,,\nn
&\delta\chi^{I(0)}=\, \delta c_{[0]}^{I} +\varepsilon\,\delta\chi_{[1]}^{I(0)}+\cdots \,,\qquad \delta\sigma^{J(0)}=\,\varepsilon\,\delta\sigma_{[1]}^{J(0)}+\cdots \,.
\end{align}
The above identification will prove useful in the next subsection where we give the effective theory of hydrodynamics that governs the fluctuations up to and including $\delta X_{[1]}$ in our expansion.

\subsection{The effective theory}\label{sec:eft}
An important point of our construction is the way that we choose to impose the gravitational and Gauss constraints. As we discussed in section \ref{sec:setup}, these constraints should be imposed at once on any hypersurface at e.g. constant radial coordinate $r$. From the dual field theory point of view, the natural choice for this hypersurface would be near the conformal boundary as they become equivalent to the Ward identities of diffeomorphism, Weyl and global $U(1)$ invariance. More specifically, the constraints $L^{b}=0$ and $C=0$ give
\begin{align}
\nabla_{a}J^{a}&=0\notag\\
\nabla_{a}T^{a}{}_{b}&=F_{b a}J^{a}+\frac{1}{2}\left(\sum_{J}\mathcal{O}_{Y^{J}}\nabla_{b}\bar y_{s}^{J}+ \sum_{I}\mathcal{O}_{Z^{I}}\nabla_{b}\bar z_{s}^{I} +\mathrm{c.c.}\right)\,,
\end{align}
with $F=dA$ the field strength of the external source one-form $A_{a}$ and $\bar y_{s}^{J}$, $\bar z_{s}^{I}$ are the sources for the complex scalar operators.

Contracting the stress tensor Ward identity with a vector $\varLambda^{b}$ gives
\begin{align}
\nabla_{a} \left[\left(T^a{}_b + A_b J^a \right)\varLambda^b\right] &=\frac{1}{2}T^{ab}\,\mathcal{L}_{\varLambda} g_{ab} +J^{a}\mathcal{L}_{\varLambda} A_{a}\nn &+\frac{1}{2}\left(\sum_{J}\mathcal{O}_{Y^{J}}\mathcal{L}_{{\varLambda}} \bar y_{s}^{J}+ \sum_{I}\mathcal{O}_{Z^{I}}\mathcal{L}_{{\varLambda}}\bar z_{s}^{I}+\mathrm{c.c.}\right)\,.
\end{align}
The thermal gradient $\zeta$ and electric field $E$ perturbations enter the boundary metric $g_{ab}$ and external field $A_{a}$ according to
\begin{align}\label{eq:sources}
\delta \left(ds^{2}\right)&=2\,(i\omega)^{-1}\,\zeta_{i}\,e^{-i\omega\,t+i q_{j} x^{j}}\,dt\,dx^{i},\quad \delta A=(i\omega)^{-1}\,(E_{i}-\mu\,\zeta_{i})\,e^{-i\omega\,t+i q_{j} x^{j}}\,dx^{i}\,,
\end{align}
along with the source $\delta z_{s}^{I}$ for the scalar field
\begin{align}\label{eq:sources2}
\delta z^{I}_{s}&=e^{i(k_{i}^{I}x^{i}+c^{I})}(i \zeta_{S^{I}}+\delta\phi^{I}_{s})\,\,e^{-i\omega\,t+i q_{i} x^{i}}\,.
\end{align}
We are now going to make the choice $\varLambda=\partial_{t}$ and perturbatively expand the contracted Ward identity to give the electric charge and heat conservation
\begin{align}\label{eq:conservation_laws}
\partial_{a} \delta J^{a}=&0\notag\\
\partial_{a}\delta Q^{a}=&0
\end{align}
with $\delta Q^{a}=-\delta T^{a}{}_{t}-\mu\,\delta J^{a}$. Equations \eqref{eq:conservation_laws} define two conserved currents at the level of first order perturbation theory. From the point of view of the effective theory, this is a good starting point in order to give a closed system of equations, provided that we can express these currents in terms of the hydrodynamic variables $\delta\hat{\mu}$, $\delta\hat{T}$ and $\delta\hat{c}^{I}$ \cite{Donos:2019hpp}. However, we will see soon that, in the phases we are interested in, the current $\delta J_{H}^{a}$ which describes the transport of heat is different from $\delta Q^{a}$. As we have argued in the previous subsection, the time derivatives scale according to $\partial_{t}\propto \mathcal{O}(\varepsilon^{2})$ while for the spatial derivatives we have $\partial_{i}\propto \mathcal{O}(\varepsilon)$. This suggests that we need to consider the charge densities up to order $\mathcal{O}(\varepsilon)$ and the transport currents up to order $\mathcal{O}(\varepsilon^{2})$ in \eqref{eq:conservation_laws}.

We will write our theory in position space where all our functions will be denoted by hats. Moreover, from now on, we find it useful to define hatted thermodynamic quantities which are local functions of the dynamical temperature, chemical potential and phasons. For example, we define $\hat{w}\equiv w(T+\delta\hat{T},\mu+\delta\hat{\mu},k_{i}^{I}+\partial_{i}\delta\hat{c}^{I})$. 

In appendix \ref{app:js} we relate the boundary currents $\delta J^{a}$, $\delta Q^{a}$ to the horizon currents we have defined in equation \eqref{eq:J_hor2} in the $\varepsilon$-expansion. More specifically, we can write
\begin{align}\label{eq:currents_consti}
\delta\langle \hat{S}^{I}\rangle&=\langle \Omega^{I}\rangle\,\delta \hat{c}^{I}\,,\nn
\delta \hat{J}^{i}&=\delta \hat{J}_{(0)}^{i}+\delta \hat{j}_{m}^{i}\,,\nn
\delta \hat{Q}^{i}&=\delta \hat{Q}_{(0)}^{i}-\sum_{I}w^{i}_{I}\,\partial_{t}\delta\hat{c}^{I}+\delta \hat{q}_{m}^{i}\,,
\end{align}
where we have defined the divergence free magnetisation currents
\begin{align}\label{eq:currents_mag_consti}
\delta \hat{j}_{m}^{i}&=-M^{ij}\hat{\zeta}_{j}+\partial_{j}\hat{M}^{ij}\,,\nn
\delta \hat{q}_{m}^{i}&=-M^{ij}\hat{E}_{j}-2\,M_{T}^{ij}\hat{\zeta}_{j}+\partial_{j}\hat{M}_{T}^{ij}\,.
\end{align}

At this point it is important to identify the correct current $\delta J_{H}^{i}$ that describes the transport of heat. For this reason, using the first law \eqref{eq:first_law}, we write the conservation equations as\footnote{It would be interesting to go beyond linear hydrodynamics and examine whether demanding positivity of entropy production (according to an appropriately defined entropy current) leads to constraints on the transport coefficients defined below, as happens generally \cite{Bhattacharya:2011tra,Kovtun:2012rj}, as well as in related contexts \cite{Delacretaz:2017zxd,Amoretti:2018tzw,Amoretti:2019cef,Baggioli:2020edn,Grozdanov:2016tdf}.}
\begin{align}\label{eq:current_cons}
\partial_{t}\hat{\rho}+\partial_{i}\delta\hat{J}^{i}=&0\,,\nn
\partial_{t}\left(T\,\hat{s}+\sum_{I}w_{I}^{i}\,\partial_{i}\delta \hat{c}^{I}\right)+\partial_{i}\delta\hat{Q}^{i}=&0\Rightarrow\nn
T\,\partial_{t}\hat{s}+\partial_{i}\left(\delta\hat{Q}^{i}+\sum_{I}w^{i}_{I}\,\partial_{t}\delta\hat{c}^{I}\right)=&0\,.
\end{align}
The conservation equation \eqref{eq:current_cons} implies that up to magnetisation current contributions, the currents that correctly describe the transport of heat and electric charge are $ \delta \hat{J}_{H}^{i}= \delta \hat{Q}^{i}+\sum_{I}w_{I}^{i} \partial_{t} \delta \hat{c}^{I}=\delta \hat{Q}^{i}_{(0)}$ and $\delta \hat{J}^{i}=\delta \hat{J}^{i}_{(0)}$. In terms of operators, we can write
\begin{align}\label{eq:pheat_current}
\hat J_{H}^{i}=\hat{Q}^{i}+\sum_{I}\frac{w_{I}^{i}}{\langle \Omega^{I}\rangle} \partial_{t} \hat S^{I}\,.
\end{align}
At first sight, it might seem surprising that the horizon heat current correctly describes the transport of heat. However, one might have expected this to happen since at the level of thermodynamics the entropy density of the system is determined by the horizon. This also ties well with the common lore in holography that dissipation is captured by horizon physics.
%\footnote{\vzb{It is intruguing to compare the above with the case of holographic superfluids. There, naively the horizon contribution to the entropy is different from the thermodynamic entropy. However, they turn out to be the same after imposing that the entropy current (carried by the normal component of the fluid) is divergenceless [1004.2707]. Here, the first law \eqref{eq:first_law} implies the identification of the entropies, and this forces the choice of the entropy current as above.}}

The above discussion suggests that the physically relevant current to discuss is $\hat{J}_{H}^{i}$ rather than $\hat{Q}^{i}$. In the end, we would like to build our theory of hydrodynamics around the conserved currents $\hat{J}_{H}^{i}$ and $\hat{J}^{i}$ and the light modes associated to the operators $S^{I}$. The corresponding triplet of sources is $\left\{\zeta_{i},E_{i},\xi_{S^{I}} \right\}$ for our choice of operators and by examining the source terms in the deformed action we can easily show that
\begin{align}
\xi_{S^{I}}=\zeta_{S^{I}}-\frac{w_{I}^{i}}{\langle \Omega^{I}\rangle}\zeta_{i}\,.
\end{align}
Note that, as discussed in section \ref{sec:thermo}, thermodynamically preferred phases satisfy $w_{I}^{i}=0$, in which case $\hat J_{H}^{i}=\hat{Q}^{i}$.

Given these results, we are able to express the boundary transport currents in terms of $\delta T_{[1]}$, $\delta \mu_{[1]}$, $\delta c_{[0]}^{I}$ and $v_{[2]}^{i}$ at order $\mathcal{O}(\varepsilon^{2})$. However, in \cite{Donos:2019hpp} we have shown that within perturbation theory we can choose to impose the constraint $L^{i}=0$ on a hypersurface close to the horizon. This allows us to integrate out the horizon fluid velocity $v_{[2]}^{i}$ and therefore obtain local expressions for the currents in terms of our hydrodynamic variables
\begin{align}\label{eq:currents_dis_consti}
\delta J^{i}&=\sigma_{H}^{ij}\,\left( \hat{E}_{j}-\partial_{j}\delta\hat{\mu}\right)+T\,\alpha_{H}^{ij}\left(\hat{\zeta}_{j}-T^{-1}\partial_{j}\delta\hat{T} \right)-\sum_{I}\gamma_{I}^{i}\,\partial_{t}\delta\hat{c}^{I}\nn
\delta J_{H}^{i}&=T\,\bar{\alpha}_{H}^{ij}\,\left( \hat{E}_{j}-\partial_{j}\delta\hat{\mu}\right)+T\,\bar{\kappa}_{H}^{ij}\left(\hat{\zeta}_{j}-T^{-1}\partial_{j}\delta\hat{T} \right)-\sum_{I}\lambda_{I}^{i}\,\partial_{t}\delta\hat{c}^{I}\,.
\end{align}
For the specific holographic model we are considering, the transport coefficients can be expressed in terms of horizon data and thermodynamic susceptibilities according to
\begin{align}\label{eq:transc_defs}
\sigma_{H}^{ik}&=\sigma_{0}^{ik}+\frac{4\pi\rho^{2}}{s}\,\mathcal{N}^{i}{}_{j}\left(\mathcal{B}^{-1}\right)^{jl}\mathcal{N}_{l}{}^{k}\,,\qquad &&\alpha_{H}^{ik}=4\pi\rho\, \mathcal{N}^{i}{}_{j}\left(\mathcal{B}^{-1}\right)^{jk}\,, \nn
\bar{\alpha}_{H}^{ik}&=4\pi\rho\,\left(\mathcal{B}^{-1}\right)^{ij}\mathcal{N}_{j}{}^{k}\,, &&\bar{\kappa}_{H}^{ik}=4\pi T s\,\left( \mathcal{B}^{-1}\right)^{ik}\,,\nn
\gamma_{I}^{i}&=4\pi T \rho\,\mathcal{N}^{i}{}_{j}\left(\mathcal{B}^{-1}\right)^{jk} \eta^{I}_{k}\,,\qquad &&\lambda_{I}^{i}=4\pi T^2 s\,\left(\mathcal{B}^{-1}\right)^{ij} \eta^{I}_{j}\,,
\end{align}
where we have used the definitions
\begin{align}\label{eq:matrix_defs}
&\sigma_{0}^{ij}=\frac{\tau^{(0)}s}{4\pi}g^{ij}_{(0)}\,,\qquad \mathcal{N}_{i}{}^{k}=\delta_{i}{}^{k}+\frac{B}{\rho}\varepsilon_{ij}\sigma_{0}^{jk}\,,\qquad \eta^{I}_{i}=\frac{1}{4\pi T}\Phi_{I}^{(0)}k^{I}_{i}\,,\nn
&\mathcal{B}_{ij}=\sum_{J}\Psi_{J}^{(0)}k^{J}_{si}k^{J}_{sj}+\sum_{I}\Phi_{I}^{(0)}k_{i}^{I}k_{j}^{I}+\tau^{(0)}B^{2}\,\varepsilon_{ik}\varepsilon_{jl}g^{kl}_{(0)}-\frac{4\pi\rho}{s}B\varepsilon_{ij}\,.
\end{align}
from appendix \ref{subsec:horizon_constraint}, and indices in $\mathcal{N}$ are raised and lowered with the horizon metric $g_{(0)ij}$.
%For later convenience, we also denote 
%\begin{align}\label{eq:sigma0_def}
%\sigma_{0}^{ij}&=\frac{\tau^{(0)}s}{4\pi}g^{ij}_{(0)}\,,
%\end{align}

We now turn our attention to the pseudo gapless degrees of freedom $\delta c^{I}$ related to the density waves in our system. In order to introduce pinning to our system we turn on a perturbative deformation $\delta\phi_{s}^{I}\propto \mathcal{O}(\varepsilon^{2})$ in the background asymptotics \eqref{asymptsol}. The constitutive relations \eqref{eq:currents_dis_consti} remain unchanged \cite{Donos:2019hpp}, while in appendix \ref{app:chi} we integrate the equation of motion \eqref{eq:chi_eom} to obtain the effective Josephson relation
%\begin{align}\label{eq:josephson}
%\sum_{L}\theta^{I}{}_{L}\,\left( \partial_{t}\delta \hat{c}^{L}+\sum_{M}\Omega^{L}{}_{M}\delta\hat{c}^{M}\right)+\varrho^{I}_{i}\delta \hat{J}_{H}^{i}-\partial_{i}\hat{w}^{i}_{I}=\langle \Omega^{I}\rangle\,\hat{\xi}_{S^{I}}\,,
%\end{align}
\begin{align}\label{eq:josephson}
\theta^{I}\, \partial_{t}\delta \hat{c}^{I} +\langle \Omega^{I}\rangle\,\delta\phi^{I}_{s}\,\delta\hat{c}^{I} +\eta^{I}_{i}\delta \hat{J}_{H}^{i}-\partial_{i}\hat{w}^{i}_{I}=\langle \Omega^{I}\rangle\,\hat{\xi}_{S^{I}}\,,
\end{align}
with the transport coefficient
%\begin{align}
%\theta^{I}{}_{L}=\frac{s\,\Phi_{I}^{(0)}}{4\pi}\delta^{I}_{L},\quad \varrho^{I}_{i}=\frac{1}{4\pi T}\Phi_{I}^{(0)}k^{I}_{i},\quad \Omega^{L}{}_{M}=\langle \Omega^{M}\rangle\,\delta\phi^{M}_{s}\,\left( \theta^{-1}\right)^{L}{}_{M}\,.
%\end{align}
\begin{align}
\theta^{I}=\frac{s\,\Phi_{I}^{(0)}}{4\pi}\,.
\end{align}
Equation \eqref{eq:josephson} holds for each capital index $I$ separately, so we have a set of $N_Z$ Josephson relations. Along with the current conservation equations \eqref{eq:current_cons}, it defines a closed system of equations for the dynamical fields $\delta\hat{T}$, $\delta\hat{\mu}$ and $\delta\hat{c}^{I}$. 

Let us now make some comments on the above holographic results. From \eqref{eq:currents_dis_consti}--\eqref{eq:matrix_defs} we observe that, as in the case without spontaneous symmetry breaking \cite{Donos:2015bxe}, the horizon DC conductivities $\sigma_{H}^{ik},\, \alpha_{H}^{ik},\, \bar{\alpha}_{H}^{ik},\, \bar{\kappa}_{H}^{ik}$ are solely determined by $\mathcal{B}_{ij}$ and $\sigma_{0}^{ij}$, along with other thermodynamic quantities. The coupling to the massless modes $\delta \hat{c}^I$ is determined by one extra quantity, $\eta^I_i$. The reason is that the Josephson relation \eqref{eq:josephson} only involves the heat current $J^i_H$ and not the $U(1)$ current. The coupling to the $U(1)$ current has been considered from a hydrodynamic perspective in various contexts in \cite{Davison:2016hno,Delacretaz:2017zxd,Delacretaz:2019wzh,Armas:2020bmo}. It would be interesting to find a holographic model which realizes that.

In the following subsections we study the gap of hydrodynamic modes after turning on the pinning parameters $\delta\phi_{s}^{I}$ as well as the dispersion relations without pinning. Finally we compute the finite frequency retarded Green's functions which will help us extract the transport properties of the system.

\subsection{Pseudo-gapless modes}\label{sec:pinning}

In order to identify the gapped modes in our effective theory we need to study perturbations with the sources switched off and with wavevector $q^{i}=0$. They belong to the Goldstone mode sector, which leads us to consider the ansatz
\begin{align}
\delta\hat{T}=0,\quad \delta\hat{\mu}=0,\quad \delta \hat{c}^{I}=\delta c_{0}^{I}\,e^{-i\delta\omega_{g} t}\,.
\end{align}
This ansatz automatically solves the current conservation equations \eqref{eq:current_cons} while the Josephson relation \eqref{eq:josephson} gives the matrix equation for the vector of amplitudes $\delta c_{0}^{I}$
\begin{align}\label{eq:gap_veq}
\sum_{M}\left(i(\mathcal{M}^{-1})^{L}{}_{M} +\delta^{L}{}_{M}\delta\omega_{g}\right)\,\delta c^{M}_{0}=0\,,
\end{align}
where we have defined the matrix\footnote{We remind the reader that capital indices are \textit{not} being summed over.}
%\begin{align}
%\mathcal{M}^{L}{}_{M} &= \sum_{K}\left(\Omega^{-1}\right)^{L}{}_{K}\left(\delta^{K}{}_{M} -\sum_{I}\left( \theta^{-1}\right)^{K}{}_{I}\varrho^{I}_{i}\lambda^{i}_{M}\right)\nn
%&=\frac{\Phi^{(0)}_{L}s}{8\pi |\langle\mathcal{O}_{Z^{L}} \rangle| \delta\phi_{s}^{L}}\left( \delta^{L}_{M}-\left( \mathcal{B}^{-1}\right)^{ij}k^{L}_{i}k^{M}_{j}\Phi^{(0)}_{M}\right)\,.
%\end{align}
\begin{align}
\mathcal{M}^{L}{}_{M} &= \left[\langle \Omega^{L}\rangle\,\delta\phi^{L}_{s}\right]^{-1}\left[\theta^{L}\delta^{L}{}_{M} -\eta^{L}_{i}\lambda^{i}_{M}\right]= \frac{s}{4\pi \langle \Omega^{L}\rangle \delta\phi_{s}^{L}}\left( \Phi^{(0)}_{L}\delta^{L}_{M}-\left( \mathcal{B}^{-1}\right)^{ij}\eta^{L}_{i}\eta^{M}_{j}\right)\,.
\end{align}
In order for equation \eqref{eq:gap_veq} to have non-trivial solutions, the matrix multiplying the vector of amplitudes should not be invertible. Equating the determinant of this matrix to zero gives an algebraic equation which determines the $N_{Z}$ different values for the gaps $\delta\omega_{g}$ in terms of the eigenvalues of the matrix $\mathcal{M}^{-1}$.

In order to illustrate the effect of the magnetic field on the gaps of the theory we consider a simple case with $N_{Z}=N_{Y}=2$ and which apart from the $U(1)^{4}$ the model has a $\mathbb{Z}_{2}\times \mathbb{Z}_{2} $ symmetry which exchanges $Z^{1}\leftrightarrow Z^{2}$ and $Y^{1}\leftrightarrow Y^{2}$. This allows us to consider an isotropic background which can be achieved by choosing
\begin{align}\label{eq:isotropic_case}
k_{i}^{I}=k\,\delta^{I}_{i},\qquad k_{si}^{J}=k_{s}\,\delta^{J}_{i}\, \qquad \psi_{s}^{J}=\psi_{s},\qquad \delta\phi_{s}^{I}=\delta\phi_{s}\,.
\end{align}
The symmetries of the model along with the choice of boundary parameters leads to data of integration in which the internal indices can be suppressed, allowing us to write
\begin{align}
\langle \Omega^{I}\rangle&=\langle \Omega\rangle\,,\qquad \Phi^{I(0)}=\Phi^{(0)}\,, \qquad \Psi^{J(0)}=\Psi^{(0)}\,,\nn
g_{(0)ij}&=\delta_{ij}\,G_{(0)}\,,\qquad s=4\pi G_{(0)}\,,\qquad \sigma_{0}^{ij}=\tau^{(0)}\delta^{ij}\equiv\sigma_{0}\delta^{ij}\,.
\end{align}
This simplifies the quantities
\begin{align}
\mathcal{B}_{ij}&=4\pi\left[ \left(\frac{k_{s}^{2}\Psi^{(0)}}{4\pi}+\frac{k^{2}\Phi^{(0)}}{4\pi} +\frac{\sigma_{0}}{s}B^{2}\right)\,\delta_{ij}-\omega_{c}\,\varepsilon_{ij}\right]\,,
%&\equiv4\pi\left[ \left(\Gamma+\frac{k^{2}\Phi^{(0)}}{4\pi}+\frac{\sigma_{0}}{s}B^{2}\right) \delta_{ij}-\omega_{c}\,\varepsilon_{ij}\right]\,,
\end{align}
where we see that the magnetic field introduces an antisymmetric piece in the matrix $\mathcal{B}_{ij}$. In the above
\begin{align}
\omega_{c}=\rho B/s\,,
\end{align}
can be identified with the cyclotron mode frequency \cite{Hartnoll:2007ih,Hartnoll:2007ip}. As a consequence, the matrix $\mathcal{M}$ will now have complex eigenvalues leading to the gaps
\begin{align}\label{eq:gap_ex}
\delta\omega^{\pm}_{g}=-i\,\frac{4\pi\,\langle \Omega\rangle\,\delta\phi_{s}}{s\,\Phi^{(0)}} \left(1+ \frac{k^{2}\Phi^{(0)}/4\pi}{k_{s}^{2}\Psi^{(0)}/4\pi + \sigma_{0}B^{2}/s \pm i\omega_{c}}\right)\,.
\end{align}
for the pseudo-massless modes. It is easy to see that in zero magnetic field, these modes lie on the lower imaginary semi-axis and they agree with the expressions that were obtained in \cite{Donos:2019hpp}. Moreover, due to the isotropy of the model and background we are considering, they also lie at the same point. However, the characteristic frequency of our nearly gapless modes has a resonant frequency apart from a gap at finite magnetic field.

It is interesting to consider the extreme limits for the behaviour of the poles \eqref{eq:gap_ex} of our simple example. In the limit where $B$ is the smallest parameter\footnote{To be precise, the regime where \eqref{eq:gap_small_B} is valid is $\sigma_{0}B\ll\rho$ and $\omega_{c}\ll k^{2}\Phi^{(0)}$.} we have a perturbatively small correction to the results of \cite{Donos:2019hpp}
\begin{align}\label{eq:gap_small_B}
\delta\omega^{\pm}_{g}&=-i\,\frac{4\pi\,\langle \Omega\rangle\,\delta\phi_{s}}{s\,\Phi^{(0)}}\,\left(1+\frac{k^{2}\Phi^{(0)}}{k_{s}^{2}\Psi^{(0)}}\pm  4\pi i\frac{k^{2}\Phi^{(0)}}{k_{s}^{4}\Psi^{(0)\,2}} \omega_{c}+\mathcal{O}(B^{2})\right)\,.
%&=-i\,\frac{4\pi\,\langle \Omega\rangle\,\delta\phi_{s}}{s\,\Phi^{(0)}}\,\left(1+\frac{k^{2}\Phi^{(0)}/4\pi}{\Gamma}\pm i \frac{k^{2}\Phi^{(0)}/4\pi}{\Gamma^{2}} \omega_{c}+\mathcal{O}(B^{2})\right)\,.
\end{align}
We therefore see that for small magnetic fields the two modes split and they move horizontally in opposite directions in the complex plane. In the opposite limit, where $B$ is the largest scale in the system we have\footnote{In the regime $\sigma_{0}B\gg\rho$ and $B\gg k_{s}^{2}\Psi^{(0)},k^{2}\Phi^{(0)}$.}
\begin{align}\label{eq:gap_ex_largeB}
\delta\omega^{\pm}_{g}=-i\,\frac{4\pi\,\langle \Omega\rangle\,\delta\phi_{s}}{s\,\Phi^{(0)}}+\mathcal{O}(B^{-2})\,,
\end{align}
and the two frequencies become degenerate once again, at the value which is given by the $k\to0$ limit of \eqref{eq:gap_ex}. The same result is obtained if we keep the filling fraction $\rho/B$ finite while taking $B\to\infty$. It is interesting to note that the expression \eqref{eq:gap_ex_largeB} is the same with \cite{Donos:2019txg} but in a different thermal state. This situation is relevant to weak lattices and close to the phase transition at $T\sim T_{c}$.

%A third, distinct possibility which is relevant to weak lattices and magnetic fields is when $k_{s}^{2}\ll B\ll \rho$ giving,
%\begin{align}
%\delta\omega^{\pm}_{g}=\pm \frac{2k^{2} |\mathcal{O}_{Z} | \rho }{\tau^{(0) 2}\,B^{3}}\delta\phi_{s} +\cdots\,.
%\end{align}
%In this case we see that the resonant part dominates the poles related to the phase relaxation. 

A third, distinct possibility which is relevant to weak lattices and magnetic fields is when $\sigma_{0}B\ll\rho$ and $B\gg k_{s}^{2}\Psi^{(0)},B\ll k^{2}\Phi^{(0)}$, giving
\begin{align}\label{eq:pseudosp_gap}
\delta\omega^{\pm}_{g}=\pm \frac{k^{2} \langle \Omega\rangle}{s\, \omega_{c}}\delta\phi_{s} +\cdots\,.
\end{align}
In this case we see that the resonant part dominates the poles related to the phase relaxation. The expression \eqref{eq:pseudosp_gap} agrees with the prediction from hydrodynamics in the pseudo-spontaneous regime once we identify the pinning frequency $\omega_0^2\sim k^{2} \langle \Omega\rangle\delta\phi_{s}$ \cite{PhysRevB.18.6245,Gruner:1988zz,Delacretaz:2019wzh}.

Note that the full expression \eqref{eq:gap_ex} holds for finite magnetic field; in particular, we nowhere assumed that $B$ is perturbatively small. All the background quantities however, including the horizon values $\Psi^{(0)},\Phi^{(0)}$, depend implicitly on $B$ as well as $k_s$. Thus, the scaling of $\delta\omega_g$ with $B$ in the above expressions can be determined analytically (or, when not possible, numerically) from the properties of the ground state.\footnote{The only exception is \eqref{eq:gap_small_B}, since we expect the various background quantities to be continuously connected to the corresponding ones of the $B=0$ state.} In particular, it is not straightforward to compare our results to \cite{Baggioli:2020edn}, which found a $B^{1/2}$ scaling for large $B$ using a holographic massive gravity model.

\subsection{Finite frequency response}\label{sec:transport}

In this subsection we study linear response at zero wave number $q_i=0$. In order to achieve this we turn on all our sources with a finite frequency $\omega$ and look for a solution to the conservation law equations \eqref{eq:current_cons} and Josephson relation \eqref{eq:josephson}. A suitable ansatz for this purpose is
\begin{align}
\delta\hat{T}=\delta T_{[1]}\,e^{-i\omega t}\,,\quad \delta\hat{\mu}=\delta \mu_{[1]}\,e^{-i\omega t}\,,\quad \delta \hat{c}^{I}=\delta c_{[0]}^{I}\,e^{-i\omega t}\,.
\end{align}

After combining the transport heat current \eqref{eq:currents_dis_consti} and the Josephson relation \eqref{eq:josephson}, we obtain
\begin{align}
\label{eq:linresponse}
\delta c_{[0]}^{I}&=i\sum_{K,L}\left[\left(\omega+i \mathcal{M}^{-1} \right)^{-1}\right]^{I}{}_{L} \Lambda^{L}{}_{K} \left(\langle \Omega^{K}\rangle\,\xi_{S^{K}} -T \eta^{K}_{j}\,\bar{\alpha}^{jk}_{H}\,E_{k}-T \eta^{K}_{j}\,\bar{\kappa}^{jk}_{H}\,\zeta_{k}\right)\,,\nn
\left(\Lambda^{-1}\right)^{I}{}_{P}&\equiv\langle \Omega^{I}\rangle\,\delta\phi^{I}_{s}\, \mathcal{M}^{I}{}_{P} =\theta^{I}\delta^{I}{}_{P}-\eta^{I}_{i} \lambda^{i}_{P}\,.
%(\Lambda^{-1})^{I}{}_{P}&\equiv\sum_{K,L}\theta^{I}{}_{K} \Omega^{K}{}_{L} \mathcal{M}^{L}{}_{P} =\theta^{I}{}_{P}-\varrho^{I}_{i} \lambda^{i}_{P}\,.
\end{align}
Plugging this solution back in the constitutive relations \eqref{eq:current_cons} we obtain the VEVs for the scalar fields $S^{I}$ and the transport currents in terms of the sources $\zeta_{i}$, $E_{i}$ and $\xi_{S^{I}}$ according to
\begin{align}
\delta\langle S^{I}\rangle&=\langle \Omega^{I}\rangle\,\delta c_{[0]}^{I}\,,\nn
\delta J^{i}&=\sigma^{ij}_{H}E_{j}+T\alpha^{ij}_{H}\zeta_{j}+i\omega\,\sum_{I}\gamma^{i}_{I}\,\delta c_{[0]}^{I}\,,\nn
\delta J_{H}^{i}&=T\bar{\alpha}^{ij}_{H}E_{j}+T\bar{\kappa}^{ij}_{H}\zeta_{j}+i\omega\,\sum_{I}\lambda^{i}_{I}\,\delta c_{[0]}^{I}\,.
\end{align}
In order to extract the retarded Green's functions, we need to consider the derivative of the VEVs with respect to the time dependent sources. Since we are only considering linear response, we can write\footnote{Generally, the Green's functions of two operators $A$ and $B$ satisfy $G_{\dot AB}=-i\omega G_{AB} + i \langle\left[A,B\right]\rangle=-i\omega G_{AB} + i \left(\chi_{AB}-\chi_{BA}\right)$. However, the therodynamics of our model leads to vanishing susceptibilities when $A=S^I$ and $B$ is the electric or heat current.}
%\begin{align}
%\delta\langle S^{K}\rangle&=(i\omega)^{-1}\,G_{S^{K}Q^{k}}\,\zeta_{k}+(i\omega)^{-1}\,G_{S^{K}J^{k}}\,E_{k}+\sum_{I}G_{S^{K}S^{I}}\,\zeta_{S^{I}}\,,\nn
%\delta J^{i}&=(i\omega)^{-1}\,G_{J^{i}J^{k}}\,E_{k}+(i\omega)^{-1}\,G_{J^{i}Q^{k}}\,\zeta_{k}+\sum_{I}G_{J^{i}S^{I}}\,\zeta_{S^{I}}\,,\nn
%\delta J_{H}^{i}&=(i\omega)^{-1}\,G_{J^{i}_{H}J^{k}}\,E_{k}+(i\omega)^{-1}\,G_{J^{i}_{H}Q^{k}}\,\zeta_{k}+\sum_{I}G_{J^{i}_{H}S^{}}\,\zeta_{S^{I}}\,.
%\end{align}
\begin{align}
\delta\langle C \rangle&=(i\omega)^{-1}\,G_{CJ_{H}^{k}}\,\zeta_{k}+(i\omega)^{-1}\,G_{CJ^{k}}\,E_{k}+\sum_{I}G_{CS^{I}}\,\xi_{S^{I}}\,,
\end{align}
for any operator $C$ in our theory. After a little algebra we obtain the expressions
\begin{align}\label{eq:greensf}
\sigma^{ik}\equiv(i\omega)^{-1}\,G_{J^{i}J^{k}}&=\sigma_{H}^{ik}+\omega T\sum_{I,L,K}\gamma^{i}_{I}\left[\left(\omega+i \mathcal{M}^{-1} \right)^{-1}\right]^{I}{}_{L} \Lambda^{L}{}_{K} \eta^{K}_{j}\,\bar{\alpha}^{jk}_{H} \,,\nn
T\alpha^{ik}\equiv(i\omega)^{-1}\,G_{J^{i}J_{H}^{k}}&=T\alpha_{H}^{ik}+\omega T\sum_{I,L,K}\gamma^{i}_{I}\left[\left(\omega+i \mathcal{M}^{-1} \right)^{-1}\right]^{I}{}_{L} \Lambda^{L}{}_{K} \eta^{K}_{j}\,\bar{\kappa}^{jk}_{H}\,,\nn
T\bar{\alpha}^{ik}\equiv (i\omega)^{-1}\,G_{J_{H}^{i}J^{k}}&=T\bar{\alpha}_{H}^{ik}+\omega T\sum_{I,L,K}\lambda^{i}_{I}\left[\left(\omega+i \mathcal{M}^{-1} \right)^{-1}\right]^{I}{}_{L} \Lambda^{L}{}_{K}\eta^{K}_{j}\,\bar{\alpha}^{jk}_{H}\,,\nn
T\bar{\kappa}^{ik}\equiv (i\omega)^{-1}\,G_{J_{H}^{i}J_{H}^{k}}&=T\bar{\kappa}_{H}^{ik}+\omega T\sum_{I,L,K}\lambda^{i}_{I}\left[\left(\omega+i \mathcal{M}^{-1} \right)^{-1}\right]^{I}{}_{L} \Lambda^{L}{}_{K} \eta^{K}_{j}\,\bar{\kappa}^{jk}_{H}\,,\nn
G_{J^{i}S^{K}}&=-\langle \Omega^{K}\rangle\,\omega \,\sum_{I,L}\gamma^{i}_{I}\left[\left(\omega+i \mathcal{M}^{-1} \right)^{-1}\right]^{I}{}_{L} \Lambda^{L}{}_{K}\,,\nn
G_{S^{K}J^{i}}&=\langle \Omega^{K}\rangle\,\omega T\,\sum_{I,L}\left[\left(\omega+i \mathcal{M}^{-1} \right)^{-1}\right]^{K}{}_{I} \Lambda^{I}{}_{L}\eta^{L}_{j}\,\bar{\alpha}^{ji}_{H},\nn
G_{J_{H}^{i}S^{K}}&=-\langle \Omega^{K}\rangle\,\omega \,\sum_{I,L}\lambda^{i}_{I}\left[\left(\omega+i \mathcal{M}^{-1} \right)^{-1}\right]^{I}{}_{L} \Lambda^{L}{}_{K}\,,\nn
G_{S^{K}J_{H}^{i}}&=\langle \Omega^{K}\rangle\,\omega T\,\sum_{I,L}\left[\left(\omega+i \mathcal{M}^{-1} \right)^{-1}\right]^{K}{}_{I} \Lambda^{I}{}_{L}\eta^{L}_{j}\,\bar{\kappa}^{ji}_{H}\,,\nn
G_{S^{I}S^{K}}&=i\,\langle \Omega^{I}\rangle\langle \Omega^{K}\rangle\,\sum_{L}\left[\left(\omega+i \mathcal{M}^{-1} \right)^{-1}\right]^{I}{}_{L} \Lambda^{L}{}_{K}\,.
\end{align}
The non-trivial frequency dependence in the above thermoelectric conductivities includes only the horizon quantities $\bar{\alpha}^{ij}_{H},\bar{\kappa}^{ij}_{H}$, due to the fact that the Goldstone couples only to the heat current, \eqref{eq:josephson}. Moreover, from the above expressions we see that the gaps of the previous subsection determine the poles of the retarded Green's functions \eqref{eq:greensf}. From these expressions it is clear that the pseudo-gapless modes we discussed in section \ref{sec:pinning} couple to the conserved currents of our system. It should be emphasised that the Green’s functions in equation \eqref{eq:greensf} attain the most general form possible. In particular, $G_{S^I S^K}$ is simply controlled by the existence of a pole related to the pseudo-spontaneous symmetry breaking (subject to the presence of the gap), while $S^I$ couple to the currents only through their time derivatives, giving an additional factor of $\omega$ in the numerator.  Thus, we expect to see the same structure in more general theories of this type.

Given the time reversal symmetry of the theory, as a non-trivial check, we can see that the expressions above satisfy the Onsager relations $G_{CD}(\omega)\Bigm|_{B}=\varepsilon_{C}\varepsilon_{D}\,G_{DC}(\omega)\Bigm|_{-B}$, with $\varepsilon_{C,D}=\pm1$ depending on how the operators $C$ and $D$ transform under time reversal. In particular, we find that
\begin{align}\label{eq:onsager}
%&\bar{\alpha}^{ij}(\omega, 0)\Bigm|_B=\alpha^{ji}(\omega, 0)\Bigm|_{-B}\,,\nonumber\\
G_{J^{i}J_{H}^{j}}(\omega,0)\Bigm|_B&=G_{J_{H}^{j}J^{i}}(\omega,0)\Bigm|_{-B}\,,\qquad G_{SJ^{i}}(\omega,0)\Bigm|_B=-G_{J^{i}S}(\omega,0)\Bigm|_{-B}\,,\nonumber\\
G_{SJ^{i}_{H}}(\omega,0)\Bigm|_B&=-G_{J^{i}_{H}S}(\omega,0)\Bigm|_{-B}\,,\qquad G_{J^{i}_{H}J_{H}^{j}}(\omega,0)\Bigm|_B=G_{J_{H}^{j}J_{H}^{i}}(\omega,0)\Bigm|_{-B}\,,\nn
G_{J^{i}J^{j}}(\omega,0)\Bigm|_B&=G_{J^{j}J^{i}}(\omega,0)\Bigm|_{-B}\,,\qquad G_{S^{I}S^{K}}(\omega,0)\Bigm|_B=G_{S^{K}S^{I}}(\omega,0)\Bigm|_{-B}\,.
\end{align}
To show \eqref{eq:onsager}, it is enough to use the identities
\begin{align}\label{eq:onsager_coeffs}
&\mathcal{N}^{i}{}_{j}\Bigm|_B = \mathcal{N}_{j}{}^{i}\Bigm|_{-B}\,, \quad \mathcal{B}_{ij}\Bigm|_B=\mathcal{B}_{ji}\Bigm|_{-B}\,, \quad \gamma^i_I\Bigm|_B=T \eta^I_j\bar{\alpha}_H^{ji}\Bigm|_{-B}\,,\quad \lambda^i_I\Bigm|_B=T\eta^I_j \bar{\kappa}^{ji}_H\Bigm|_{-B}\,,\nn
&\sigma_H^{ij}\Bigm|_B=\sigma^{ji}_H\Bigm|_{-B}\,,\quad \bar{\alpha}_H^{ij}\Bigm|_B=\alpha^{ji}_H\Bigm|_{-B}\,,\quad \bar{\kappa}_H^{ij}\Bigm|_B=\bar{\kappa}^{ji}_H\Bigm|_{-B}\,,\quad \nn
&[\Lambda^{-1}\cdot(\omega+i\mathcal{M}^{-1})]^{I}{}_{M}\Bigm|_B=[\Lambda^{-1}\cdot(\omega+i\mathcal{M}^{-1})]^{M}{}_{I}\Bigm|_{-B}\,.
%&[\Lambda^{-1}\cdot(\omega+i\mathcal{M}^{-1})]^{I}{}_{M}\Bigm|_B=2i |\mathcal{O}_{Z^I}|\delta \phi_s^I\delta^{I}{}_{M}+\frac{\omega \,s}{4\pi}\left(\Phi^{(0)}_I\delta^{I}{}_{M}-(\mathcal{B}^{-1})^{ij}k_i^Ik_j^M\Phi^{(0)}_I\Phi^{(0)}_M\right)\Bigm|_B\nn
%&\qquad\qquad\qquad\qquad\qquad=[\Lambda^{-1}\cdot(\omega+i\mathcal{M}^{-1})]^{M}{}_{I}\Bigm|_{-B}\,.
\end{align}
which hold by construction.
%It is also useful to list the retarded Green's functions of the physically interesting operators $J^{i}$, $J_{H}^{i}$ and $S^{I}$. Using the definition \eqref{eq:pheat_current} we have
%\begin{align}
%T \hat{\alpha}^{ik}\equiv (i\omega)^{-1}\,G_{J^{i}J_{H}^{k}}&=T\alpha_{H}^{ik}+T\omega \sum_{I,L,J}\gamma^{i}_{I}\left[\left(\omega+i \mathcal{M}^{-1} \right)^{-1}\right]^{I}{}_{L} \Lambda^{L}{}_{J} \varrho^{J}_{j}\,\bar{\kappa}^{jk}_{H}\,,\nn
%T\hat{\bar{\alpha}}^{ik}\equiv (i\omega)^{-1}\,G_{J_{H}^{i}J^{k}}&=T\bar{\alpha}_{H}^{ik}+\omega T\sum_{I,L,J} \lambda^{i}_{I}\left[\left(\omega+i \mathcal{M}^{-1} \right)^{-1}\right]^{I}{}_{L} \Lambda^{L}{}_{J}\varrho^{J}_{j}\,\bar{\alpha}^{jk}_{H}\,,\nn
%T\hat{\bar{\kappa}}^{ik}\equiv(i\omega)^{-1}\,G_{J_{H}^{i}J_{H}^{k}}&=T\bar{\kappa}_{H}^{ik}+T\omega \sum_{I,L,J}\lambda^{i}_{I}\left[\left(\omega+i \mathcal{M}^{-1} \right)^{-1}\right]^{I}{}_{L} \Lambda^{L}{}_{J} \varrho^{J}_{j}\,\bar{\kappa}^{jk}_{H}\,,\nn
%G_{S^{J}J_{H}^{k}}&=2|\langle \mathcal{O}_{Z^{J}}\rangle | T\,\omega \,\sum_{I,L}\left[\left(\omega+i \mathcal{M}^{-1} \right)^{-1}\right]^{J}{}_{I} \Lambda^{I}{}_{L}\varrho^{L}_{j}\,\bar{\kappa}^{jk}_{H}\,.
%\end{align}
%The retarded Green's functions with the order interchanged is guaranteed to satisfy the Onsager relations due to \eqref{eq:onsager_coeffs} and \eqref{eq:forOnsager}. Notice that in the end the coefficients $w_{I}^{i}$ all drop out and the Green's functions are all fixed by horizon quantities and VEVs of operators in the thermal state.

The explicit expressions \eqref{eq:greensf} show that the limits $\omega\to0$ and $\mathcal{M}\to0$ do not commute, as also discussed in \cite{Donos:2019tmo,Donos:2019hpp}. Specifically, taking the DC limit $\omega\to0$ while the gap remains finite, reduces the conductivities to the corresponding horizon conductivities, first obtained in \cite{Donos:2015bxe}, as well as the expected Goldstone mode susceptibility $\chi_{IK}$\footnote{In purely spontaneous phases the susceptibilities diverge \cite{Forster}, but here this divergence is regulated by the perturbative explicit source $\delta\phi_{s}^{I}$.}
\begin{align}\label{eq:greensfDC}
\sigma^{ik}_{DC}&=\sigma_{H}^{ik}\,,\qquad T\alpha^{ik}_{DC}=T\alpha_{H}^{ik}\,,\qquad T\bar{\alpha}^{ik}_{DC}=T\bar{\alpha}_{H}^{ik}\,,\nn T\bar{\kappa}^{ik}_{DC}&=T\bar{\kappa}_{H}^{ik}\,,\qquad \chi_{IK}\equiv G_{S^{I}S^{K}}=\frac{\langle\Omega^{I}\rangle}{\delta\phi^{I}_{s}} \delta^{I}{}_{K}\,,
\end{align}
with the rest of the correlators vanishing. However, when we first take $\delta\phi^{I}_{s}\to0$, transport at zero-frequency gets modified by the Goldstone modes
\begin{align}\label{eq:greensfGold}
&\sigma^{ik}_{DC}=\sigma_{H}^{ik}+T\sum_{I,K}\gamma^{i}_{I}\Lambda^{I}{}_{K} \eta^{K}_{j}\,\bar{\alpha}^{jk}_{H} \,,\qquad \qquad &&T\alpha^{ik}_{DC}=T\alpha_{H}^{ik}+ T\sum_{I,K}\gamma^{i}_{I}\Lambda^{I}{}_{K} \eta^{K}_{j}\,\bar{\kappa}^{jk}_{H}\,,\nn
&T\bar{\alpha}^{ik}_{DC}=T\bar{\alpha}_{H}^{ik}+ T\sum_{I,K}\lambda^{i}_{I}\Lambda^{I}{}_{K}\eta^{K}_{j}\,\bar{\alpha}^{jk}_{H}\,, &&T\bar{\kappa}^{ik}\left(\omega=0\right)=T\bar{\kappa}_{H}^{ik}+ T\sum_{I,K}\lambda^{i}_{I}\Lambda^{I}{}_{K} \eta^{K}_{j}\,\bar{\kappa}^{jk}_{H}\,,\nn
&G_{J^{i}S^{K}}\left(\omega=0\right)=-\langle \Omega^{K}\rangle\, \sum_{I}\gamma^{i}_{I}\Lambda^{I}{}_{K}\,, &&G_{S^{K}J^{i}}\left(\omega=0\right)=\langle \Omega^{K}\rangle\, T\,\sum_{I}\Lambda^{K}{}_{I}\eta^{I}_{j}\,\bar{\alpha}^{ji}_{H},\nn
&G_{J_{H}^{i}S^{K}}\left(\omega=0\right)=-\langle \Omega^{K}\rangle \,\sum_{I}\lambda^{i}_{I}\Lambda^{I}{}_{K}\,, &&G_{S^{K}J_{H}^{i}}\left(\omega=0\right)=\langle \Omega^{K}\rangle\, T\,\sum_{I}\Lambda^{K}{}_{I}\eta^{I}_{j}\,\bar{\kappa}^{ji}_{H}\,,
\end{align}
while $G_{S^{I}S^{K}}$ diverges as
\begin{align}\label{eq:greensfGoldSS}
&G_{S^{I}S^{K}}\sim i\,\langle \Omega^{I}\rangle\langle \Omega^{K}\rangle\, \frac{\Lambda^{I}{}_{K}}{\omega}\,.
\end{align}

\subsection{Decoupling the Goldstone modes}\label{subsec:decoupling}

Given the results of the previous subsection we can proceed to obtain Kubo formulas, which can be taken as the fundamental definition of the corresponding transport coefficients in a generic theory with the symmetry breaking pattern we are considering. We first extract the transport coefficient $\Lambda^{I}{}_{K}$ as
\begin{align}\label{eq:KuboLambda}
\Lambda^{I}{}_{K}&=\frac{1}{\langle \Omega^{I}\rangle\langle \Omega^{K}\rangle} \lim_{\omega\to0} \lim_{\delta\phi^{I}_{s}\to0} \left(-i\omega\, G_{S^{I}S^{K}} \right)\,,	
\end{align}
which is a finite quantity, given by the combination of transport coefficients shown in \eqref{eq:linresponse} in our specific holographic model. We can then express $\gamma^{i}_{I}$ and $\lambda^{i}_{I}$ as
\begin{align}\label{eq:Kubo}
\gamma^{i}_{I}&=-\sum_{K}\frac{1}{\langle \Omega^{K}\rangle} \lim_{\omega\to0} \lim_{\delta\phi^{K}_{s}\to0}G_{J^{i}S^{K}} \left(\Lambda^{-1}\right)^{K}{}_{I}\,,\nn
\lambda^{i}_{I}&=-\sum_{K}\frac{1}{\langle \Omega^{K}\rangle} \lim_{\omega\to0} \lim_{\delta\phi^{K}_{s}\to0}G_{J_H^{i}S^{K}} \left(\Lambda^{-1}\right)^{K}{}_{I}\,.
\end{align}
The order of the limits is important, as was also noticed in \cite{Delacretaz:2019wzh}. We first need take the gap to zero in order to include the effects of the Goldstone modes in the low frequency regime, and then take $\omega\to0$.

Similarly we can define
\begin{align}\label{eq:Kubo2}
\tilde{\gamma}^{i}_{I}&=\sum_{K}\frac{1}{\langle \Omega^{K}\rangle} \lim_{\omega\to0} \lim_{\delta\phi^{K}_{s}\to0} \left(\Lambda^{-1}\right)^{K}{}_{I}\, G_{S^{K}J^{i}}\,,\nn
\tilde{\lambda}^{i}_{I}&=\sum_{K}\frac{1}{\langle \Omega^{K}\rangle} \lim_{\omega\to0} \lim_{\delta\phi^{K}_{s}\to0} \left(\Lambda^{-1}\right)^{K}{}_{I}\, G_{S^{K}J_{H}^{i}} \,,
\end{align}
which are the time reversed versions of $\gamma^{i}_{I}$ and $\lambda^{i}_{I}$ and satisfy
\begin{align}\label{eq:gl_ons}
	\gamma^{i}_{I}\Bigm|_B= \tilde{\gamma}^{i}_{I}\Bigm|_{-B}\,,\qquad\qquad \lambda^{i}_{I}\Bigm|_B= \tilde{\lambda}^{i}_{I}\Bigm|_{-B}\,,	
\end{align}
In our specific holographic model, we have explicitly computed $\gamma^{i}_{I},\lambda^{i}_{I},\tilde\gamma^{i}_{I},\tilde\lambda^{i}_{I}$, see \eqref{eq:transc_defs} and \eqref{eq:onsager_coeffs}, or \eqref{eq:greensf}. They are related by
\begin{align}\label{eq:gl_hol_relations}
	\gamma^{i}_{I}= \frac{\rho}{T\,s} \mathcal{N}^{i}{}_{j}\lambda^{j}_{I}\,,\qquad\qquad	\tilde\gamma^{i}_{I}= \frac{\rho}{T\,s} \tilde\lambda^{j}_{I}\mathcal{N}_{j}{}^{i}\,,
\end{align}
since, as explained in section \ref{sec:eft}, only the heat current enters the Josephson relation. However, more generally, equations \eqref{eq:Kubo},\eqref{eq:Kubo2} will hold in a generic theory in which we do not have explicit expressions for the low energy Green's functions, as long as the expressions \eqref{eq:Kubo}, \eqref{eq:Kubo2} remain finite as $\delta\phi_{s}^I\to0$.

%\vzb{Is it ok that $\delta\phi^{I}_{s}$ appears in Kubo formulas? Should we write it in terms of \eqref{eq:greensfDC}? Or write them as:}
%\begin{align}\label{eq:Kubo2}
%\gamma^{i}_{I}&=- \sum_{K}\lim_{\omega\to0} \lim_{\delta\phi^{K}_{s}\to0} \frac{G_{J^{i}S^{K}}}{\langle\Omega_{K}\rangle} \left(\theta^{I}\delta^{I}{}_{K}-\eta^{I}_{i} \lambda^{i}_{K}\right)^{-1} = i \sum_{K}\lim_{\omega\to0} \lim_{\delta\phi^{K}_{s}\to0} \frac{G_{J^{i}S^{K}}\,G_{S^{I}S^{K}}}{\langle\Omega_{K}\rangle} \,,
%\end{align}
It would be interesting to also define modified electric and heat current operators $\mathcal{J}^i,\tilde{\mathcal{J}}^i$ and $\mathcal{J}^i_H,\tilde{\mathcal{J}}^i_H$  which decouple from the Goldstone modes. This is satisfied as long as we demand that the Green's functions $G_{\mathcal{J}^i S^I}$, $G_{\mathcal{J}^i_H S^I}$, $G_{S^I \tilde{\mathcal{J}}^i}$, $G_{S^I \tilde{\mathcal{J}}^i_H}$ vanish as $\omega\to0,\omega_{g}\to0$, irrespective of the order of limits. Within the hydrodynamic regime, \eqref{eq:Kubo},\eqref{eq:Kubo2} imply that
\begin{align}\label{eq:mathcalJ_def}
\mathcal{J}^i& =J^i +\sum_{I}\frac{\gamma^{i}_{I}}{\langle\Omega^{I}\rangle} \partial_{t}S^I\,,\qquad\qquad &&\mathcal{J}^i_H =J^i_H +\sum_{I}\frac{\lambda^{i}_{I}}{\langle\Omega^{I}\rangle} \partial_{t}S^I\,,\nn
\tilde{\mathcal{J}}^i &=J^i +\sum_{I}\frac{\tilde\gamma^{i}_{I}}{\langle\Omega^{I}\rangle} \partial_{t}S^I\,, &&\tilde{\mathcal{J}}^i_H =J^i_H +\sum_{I}\frac{\tilde\lambda^{i}_{I}}{\langle\Omega^{I}\rangle} \partial_{t}S^I\,,
\end{align}
indeed satisfy
\begin{align}\label{eq:mathcalJ_G_def}
G_{\mathcal{J}^i S^I}=0\,,\qquad G_{\mathcal{J}^i_H S^I}=0\,,\qquad G_{S^I \tilde{\mathcal{J}}^i}=0\,,\qquad G_{S^I \tilde{\mathcal{J}}^i_H}=0\,.
\end{align}

Note that these currents are related by
\begin{align}\label{eq:mathcalJ_ons}
\mathcal{J}^i\Bigm|_{B}& = \tilde{\mathcal{J}}^i \Bigm|_{-B}\,,\qquad\qquad \mathcal{J}^i_H\Bigm|_{B}= \tilde{\mathcal{J}}^i_H \Bigm|_{-B}\,,
\end{align}
which implies that $\mathcal{J}^i,\mathcal{J}^i_H$ are not vector operators on backgrounds with $B\neq0$, since they do not have definite transformation properties under time reversal. In other words, equation \eqref{eq:mathcalJ_G_def} shows that the Goldstone modes $S^I$ do not source $\mathcal{J}^i,\mathcal{J}^i_H$, but $\tilde{\mathcal{J}}^i,\tilde{\mathcal{J}}^i_H$ are the operators which do not source $S^I$. Note however that the sums $\mathcal{J}^i+\tilde{\mathcal{J}}^i,\,\mathcal{J}^i_H+\tilde{\mathcal{J}}^i_H$ define good vector operators.

In a hydrodynamic theory with the constitutive relations \eqref{eq:currents_dis_consti}, we can immediately see that the combinations \eqref{eq:mathcalJ_def} simply remove the contributions of the Goldstone modes $\delta\hat{c}^I$. It is then straightforward to check that the corresponding Green's functions satisfy
\begin{align}\label{eq:greensf_cal}
(i\omega)^{-1}\,G_{\mathcal{J}^{i}\mathcal{J}^{k}}&=\sigma_{H}^{ik}\,, \qquad\qquad &&(i\omega)^{-1}\,G_{\mathcal{J}^{i}\mathcal{J}_{H}^{k}}=T\alpha_{H}^{ik}\,,\nn
(i\omega)^{-1}\,G_{\mathcal{J}_{H}^{i}\mathcal{J}^{k}}&=T\bar{\alpha}_{H}^{ik}\,, &&(i\omega)^{-1}\,G_{\mathcal{J}_{H}^{i}\mathcal{J}_{H}^{k}}=T\bar{\kappa}_{H}^{ik}\,.
\end{align}
Thus, from a holographic perspective, the combinations \eqref{eq:mathcalJ_def} isolate the horizon contribution to the electric and heat currents. As expected, the finite-frequency poles related to the pseudo-gapless modes cancel out in the above Green's functions, which turn out to be frequency-independent for low frequencies up to $\omega_{g}$.

The Green's functions for the time-reversed currents $\tilde{\mathcal{J}}^i,\tilde{\mathcal{J}}^i_H$ are simply the time-reversed versions of \eqref{eq:greensf_cal} and so they also satisfy Onsager relations similar to \eqref{eq:onsager}. This can be seen by combining \eqref{eq:mathcalJ_ons} and \eqref{eq:onsager_coeffs}.

We can proceed further by recalling the relation \eqref{eq:gl_hol_relations} between the transport coefficients entering in \eqref{eq:mathcalJ_def}. We then see that the combinations
\begin{align}\label{eq:Jinc_def}
J^{i}_{dec}&\equiv T s \mathcal{J}^i-\rho\, \mathcal{N}^{i}{}_{k}\mathcal{J}^k_{H} = T s J^{i}-\rho\, \mathcal{N}^{i}{}_{k}J^{k}_{H}\,,\nn
\tilde{J}^{i}_{dec}&\equiv T s \tilde{\mathcal{J}}^i-\rho\, \tilde{\mathcal{J}}^k_{H} \mathcal{N}_{k}{}^{i}= T s J^{i}-\rho\, J^{k}_{H}\mathcal{N}_{k}{}^{i}\,,
\end{align}
do not include contributions from the Goldstone modes, and can be solely expressed in terms of the original currents $J^i, J^i_H$. As before, $J^{i}_{dec}+\tilde{J}^{i}_{dec}$ is a well-defined vector operator.

For the retarded Green's function we find
\begin{align}\label{eq:GJinc}
(i\omega)^{-1}\,G_{J^{i}_{dec}J_{dec}^{j}}
&=(Ts)^{2}\,\sigma^{ij}-Ts\rho\,\left(T\mathcal{N}^{j}{}_{k}\alpha^{ik}+T\mathcal{N}^{i}{}_{k}\bar{\alpha}^{kj}\right)+\rho^{2}\mathcal{N}^{i}{}_{k}\mathcal{N}^{j}{}_{l}T\bar{\kappa}^{kl}\nn
&=(Ts)^{2}\,\sigma^{ij}_{H}-Ts\rho\,\left(T\mathcal{N}^{j}{}_{k}\alpha^{ik}_{H}+T\mathcal{N}^{i}{}_{k}\bar{\alpha}^{kj}_{H}\right)+\rho^{2}\mathcal{N}^{i}{}_{k}\mathcal{N}^{j}{}_{l}T\bar{\kappa}^{kl}_{H}\nn
&=(Ts)^{2}\,\sigma^{ij}_{0}\,,
\end{align}
which turns out to be given by the horizon quantity $\sigma^{ij}_{0}$ defined in \eqref{eq:matrix_defs}. Similarly
%\begin{align}\label{eq:GJinctilde}
%(i\omega)^{-1}\,G_{\tilde{J}^{i}_{dec}\tilde{J}_{dec}^{j}}(\omega)&=(Ts)^{2}\,\sigma^{ij}(\omega)-Ts\rho\,\left(T\alpha^{ik}(\omega)\mathcal{N}_{k}{}^{j}+T\bar{\alpha}^{kj}(\omega)\mathcal{N}_{k}{}^{i}\right)+\rho^{2}T\bar{\kappa}^{kl}(\omega) \mathcal{N}_{k}{}^{i} \mathcal{N}_{l}{}^{j}\nn
%&=(Ts)^{2}\,\sigma^{ij}_{H}-Ts\rho\,\left(T\alpha^{ik}_{H}\mathcal{N}_{k}{}^{j}+T\bar{\alpha}^{kj}_{H}\mathcal{N}_{k}{}^{i}\right)+\rho^{2}T\bar{\kappa}^{kl}_{H} \mathcal{N}_{k}{}^{i} \mathcal{N}_{l}{}^{j}\nn
%&=(Ts)^{2}\,\sigma^{ij}_{0}\,,
%\end{align}
\begin{align}\label{eq:GJinctilde}
(i\omega)^{-1}\,G_{\tilde{J}^{i}_{dec}\tilde{J}_{dec}^{j}}(\omega)&=(Ts)^{2}\,\sigma^{ij}_{0}\,.
\end{align}
We thus observe that the horizon quantity $\sigma^{ij}_{0}$ defined in \eqref{eq:matrix_defs} corresponds to the conductivity of the part of the $U(1)$ current which decouples from the heat current $J_H$ and the Goldstone modes. In the special case of time-reversal invariant backgrounds with $B=0$, we have that $\mathcal{N}^{i}{}_{k}=\delta^{i}{}_{k}=\mathcal{N}_{k}{}^{i}$, and thus $\mathcal{J}^i = \tilde{\mathcal{J}}^i ,\,\mathcal{J}^i_H= \tilde{\mathcal{J}}^i_H $. Then both decoupled combinations \eqref{eq:Jinc_def} reduce to the current considered in \cite{Davison:2015bea,Davison:2015taa}. Furthermore, in the absense of a background lattice, the latter can be identified with the incoherent current which decouples from the conserved momentum operator \cite{Donos:2018kkm}.

Finally, note that all of the above results hold in the strong holographic lattice limit that we are considering in our paper, where the low frequency transport properties are determined by the Goldstone modes and the momentum non-conservation poles are outside our hydrodynamic regime. However, the explicit sources $\delta\phi_{s}^{I}$ also relax momentum apart from the massless modes with a relaxation rate $\sim(\delta\phi_{s}^{I})^2$. So, had we not included such a background lattice, the momentum poles would dominate over the Goldstone mode poles in the hydrodynamic regime.

%Further results, on phases with pseudo-sponanteous symmetry breaking from the field theory prespective, and the construction of decoupled or incoherent currents, will be discussed elsewhere \cite{}.

\subsection{Hydrodynamic modes}\label{sec:diffusion}
In this subsection we wish to extract the dispersion relations of the hydrodynamic modes in our system at zero pinning. Similarly to the previous section, we switch off all the sources and we look for solutions of the form
\begin{align}
\delta\hat{T}=\delta T_{0}\,e^{-i\omega t+i q_{i}x^{i}},\quad \delta\hat{\mu}=\delta\mu_{0}\,e^{-i\omega t+i q_{i}x^{i}},\quad \delta \hat{c}^{I}=\delta c_{0}^{I}\,e^{-i\omega t+i q_{i}x^{i}}\,,
\end{align}
which solve the conservation law equations \eqref{eq:current_cons} and Josephson relation \eqref{eq:josephson}. Similarly to the previous subsection, the resulting system of equations reduces a linear system of equations for the vector of amplitudes
\begin{align}\label{eq:matrixeqv2}
\hat{\mathbb{S}}(\omega,q_{i},B)
\left( \begin{array}{c}
\delta T_{0}/T\\\delta\mu_{0}\\-i\omega\,\delta c^{L}_{0}
\end{array}\right)=0\,,
\end{align}
where we have defined the matrix
\begin{align}\label{eq:sdef}
\hat{\mathbb{S}}(\omega,q_{i},B)&=\left(
\begin{array}{ccc}
T(-\omega c_{\mu}-iq_{i}q_{j}\bar{\kappa}_{H}^{ij}) & T(-\omega\xi-iq_{i}q_{j}\bar{\alpha}_{H}^{ij}) &  q_{i}(T\nu_{L}^{i}-\lambda_{L}^{i})\\
T(-\omega\xi-iq_{i}q_{j}\alpha^{ij}_{H}) & -\omega\chi_{q}-iq_{i}q_{j}\sigma^{ij}_{H}&  q_{i}\left( \beta^{i}_{L}-\gamma^{i}_{L}\right)\\
q_{i}T(\nu_{I}^{i}-\eta_{j}^{I}\bar{\kappa}_{H}^{ji} ) & q_{i}( \beta_{I}^{i}-\eta^{I}_{j}T\bar{\alpha}_{H}^{ji}) & -i (\Lambda^{-1})^{I}{}_{L}+\omega^{-1}q_{i}q_{j}w^{ij}_{IL}
\end{array}
\right)\nn
&=\left(\begin{array}{cc}
(-\omega\mathbf{X}(B)-i\,\mathbf{\Sigma}(B))_{\alpha\beta}& \mathbf{m}(B)_{\alpha L}\\
\mathbf{m}^{\prime}(B)_{L\alpha} & (-i\mathbf{\Theta}(B)+\omega^{-1}\,\mathbf{W}(B))_{IL}
\end{array}
\right)\,,
\end{align}
where we used the notation of \eqref{eq:linresponse}. The indices of the matrices we have defined above take the values $\alpha,\beta=1,2$ and $I,L=1,\ldots,N_{Z}$ . The dispersion relations of the hydrodynamic $2+N_{Z}$ modes $\omega=\omega(q^{i})$ are then determined by demanding that $\det\hat{\mathbb{S}}=0$. The expressions for the dispersion relations are going to be rather complicated in general. However, we can obtain some of their interesting characteristics by closely examining the expression \eqref{eq:sdef} for the matrix $\hat{\mathbb{S}}$.

Equations \eqref{eq:onsager_coeffs} also imply the following relations between the various submatrices in \eqref{eq:sdef}
\begin{align}\label{eq:transposem}
&(\mathbf{m}(B))^{T}=\mathbf{m}^{\prime}(-B)\,,\qquad (\mathbf{\Theta}(B))^{T}=\mathbf{\Theta}(-B)\,,\qquad (\mathbf{\Sigma}(B))^{T}=\mathbf{\Sigma}(-B)\,,\nn
&(\mathbf{X}(B))^{T}=\mathbf{X}(-B)=\mathbf{X}(B)\,\qquad (\mathbf{W}(B))^{T}=\mathbf{W}(-B)=\mathbf{W}(B).
\end{align}
The above show that
\begin{align}\label{eq:transposes}
\hat{\mathbb{S}}(\omega, q_{i}, -B)=(\hat{\mathbb{S}}(\omega, q_{i}, B))^{T}\,,
\end{align}
which further implies that the dispersion relations are independent of the sign of $B$, since the determinant is invariant under transposition. However, the Kernel of $\hat{S}$ which solves equation \eqref{eq:matrixeqv2} will depend on it and thus the actual modes will change as we change the sign of $B$. 

By directly exploiting these properties, we show below that the frequencies $\omega(q_{i})$ of our hydrodynamic modes are pure imaginary. Finally, we notice that the transformation $q_{i}\to \lambda\,q_{i}$, $\omega\to \lambda^{2}\,\omega$ and $\delta c^{I}\to \lambda^{-1}\,\delta c^{I}$ is a symmetry of equation \eqref{eq:matrixeqv2}. This shows that all our modes are diffusion-like with $\omega(\lambda\,q_{i})=\lambda^{2}\omega(q_{i})$.

In contrast, the effective field theory of \cite{Delacretaz:2019wzh}, as well as the holographic model of \cite{Baggioli:2020edn}, both include a real quadratic part in the dispersion relation for the magnetophonon mode in the purely spontaneous or pseudo-spontaneous symmetry breaking regime. In our case, the modes are purely diffusive as a result of being in the strong translational symmetry breaking regime.

Let us now prove that the modes are pure imaginary. We find convenient to split the mode vector which solves \eqref{eq:matrixeqv2} according to
\begin{align}
	\left|v_{1} \right>=\left( \begin{array}{c}\delta T_{0}/T\\\delta\mu_{0} \end{array}\right),\qquad \left|v_{2} \right>=-i\omega(q_{i})\,\left(\begin{array}{c} \vdots\\ \delta c_{0}^{L}\\ \vdots \end{array}\right)\,.
\end{align}
Then for the background with $B\to -B$ there is a different mode with $\left| \tilde{v}_{1}\right>$ and $\left| \tilde{v}_{2}\right>$ but with the same dispersion relation $\omega(q_{i})$. This can be justified by using the transformation property \eqref{eq:transposes} and the comment below it.

We can write
\begin{align}\label{eq:linearsys1}
	-\left(\omega\mathbf{X}(B)+i\,\mathbf{\Sigma}(B)\right)\left|v_{1}\right>+\mathbf{m}(B)\left|v_{2}\right>=&0\nn
	\mathbf{m}^{\prime}(B)\left|v_{1}\right>+\left( -i\mathbf{\Theta}(B)+\omega^{-1}\,\mathbf{W}(B)\right)\left|v_{2}\right>=&0\,,
\end{align}
while for the time reversed configuration with $B\to -B$ we have
\begin{align}\label{eq:linearsys2}
	-\left(\omega\mathbf{X}(-B)+i\,\mathbf{\Sigma}(-B)\right)\left|\tilde{v}_{1}\right>+\mathbf{m}(-B)\left|\tilde{v}_{2}\right>=&0\nn
	\mathbf{m}^{\prime}(-B)\left|\tilde{v}_{1}\right>+\left( -i\mathbf{\Theta}(-B)+\omega^{-1}\,\mathbf{W}(-B)\right)\left|\tilde{v}_{2}\right>=&0\,.
\end{align}
From the above systems and after using \eqref{eq:transposem} we obtain the relation
\begin{align}
	i\omega=\frac{\omega \bar{\omega} \left(\left<\tilde{v}_{1}\right| \mathbf{X} \left|v_{1}\right>+\left<v_{1}\right|\mathbf{X}^{T} \left|\tilde{v}_{1}\right>\right)+\left<\tilde{v}_{2}\right|\mathbf{W} \left|v_{2}\right>+\left<v_{2}\right|\mathbf{W} ^{T}\left|\tilde{v}_{2}\right>}{\left<\tilde{v}_{1}\right| \mathbf{\Sigma} \left|v_{1}\right>+\left<v_{1}\right|\mathbf{\Sigma}^{T} \left|\tilde{v}_{1}\right>+\left<\tilde{v}_{2}\right|\mathbf{\Theta} \left|v_{2}\right>+\left<v_{2}\right|\mathbf{\Theta} ^{T}\left|\tilde{v}_{2}\right>}\,,
\end{align}
showing that $i\omega$ has to be a real number.

For $B=0$, the matrices $\mathbf{X}$, $\mathbf{W}$, $\mathbf{\Sigma}$ and $\mathbf{\Theta}$ are symmetric. We also know that the vectors $\left|v_1 \right>$ and $\left|v_2 \right>$ coincide with the vectors $\left|\tilde{v}_1 \right>$ and $\left|\tilde{v}_2 \right>$. This observation shows that if the matrices $\mathbf{X}$ and $\mathbf{W}$ are positive definite, then $i\omega>0$. In other words, at zero magnetic field, thermodynamic stability implies dynamical stability in the hydrodynamic sector that we focussed on.

%%%%%%%%%%%%%%%%
\section{Numerical checks}\label{sec:numerics}
In this section we numerical confirm the results presented in section \ref{sec:lin_hydro}, and in particular the formula for the dispersion relations of the hydrodynamic modes coming from equation \eqref{eq:sdef}, the gap \eqref{eq:gap_ex} and the optical conductivities \eqref{eq:greensf}. This is achieved by focusing on the model of \cite{Donos:2019tmo}, which is a truncation of  the general bulk action \eqref{eq:bulk_action} down to  the four-dimensional Einstein-Maxwell theory coupled to six real scalars, $\phi$, $\psi$, $\chi_i$ and $\sigma_i$ with $i=1,2$
\begin{align}\label{eq:bulk_action2}
S&=\int d^4 x \sqrt{-g}\,\Bigl(R-V(\phi)-\frac{3}{2}\left(\partial\phi \right)^2 -\frac{1}{2}\left(\partial\psi \right)^2 -\frac{1}{2}\theta(\phi)\left[\left(\partial\chi_{1} \right)^2 +\left(\partial\chi_{2} \right)^2 \right]\nn
&\qquad\qquad\qquad  -\frac{1}{2}\theta_{1}(\psi)\left[\left(\partial\sigma_{1} \right)^2 +\left(\partial\sigma_{2} \right)^2 \right] -\frac{\tau(\phi, \psi)}{4}\,F^{2} \Bigr) \,,
\end{align}
where
\begin{align}\label{model}
V(\phi, \psi) &=-6 \cosh \phi\,,\qquad\qquad &&\theta(\phi)=12 \sinh^2(\delta\, \phi)\,,\nn
\tau(\phi, \psi)& =\cosh(\gamma\, \phi)\,, &&\theta_1(\psi) =\psi^2\,.
\end{align}
The variation of the above action gives rise to the following field equations of motion
\begin{align}\label{eq:eom}
&R_{\mu\nu}-\frac{\tau}{2} (F_{\mu\rho}F_{\nu}{}^{\rho}-\frac{1}{4}g_{\mu\nu}F^2)-\frac{1}{2}g_{\mu\nu} V-\frac{3}{2}\partial_\mu\phi\partial_\nu\phi-\frac{1}{2}\partial_\mu\psi\partial_\nu\psi\notag\\
&\quad-\sum_{i}(\frac{\theta}{2}\partial_\mu\chi_i\partial_\nu\chi_i+\frac{\theta_1}{2}\partial_\mu\sigma_i\partial_\nu\sigma_i)=0\,,\nonumber\\
& \frac{3}{\sqrt{-g}}\partial_\mu\left(\sqrt{-g}\,\partial^\mu\phi\right)-\partial_{\phi}V-\frac{1}{4}\partial_{\phi}\tau\, F^{2}-\frac{1}{2}\theta^{\prime}\,\sum_{i}(\partial \chi_i)^{2}=0\,,\nonumber\\
& \frac{1}{\sqrt{-g}}\partial_\mu\left(\sqrt{-g}\,\partial^\mu\psi\right)-\partial_{\psi}V-\frac{1}{4}\partial_{\psi}\tau\, F^{2}
-\frac{1}{2}\theta'_{1}\,\sum_{i}(\partial \sigma_i)^{2}=0\,,\nonumber\\
& \frac{1}{\sqrt{-g}}\partial_\mu\left(\theta_1\sqrt{-g}\,\partial^\mu\sigma_i\right)=0\,,\quad \frac{1}{\sqrt{-g}}\partial_\mu\left(\theta\sqrt{-g}\,\partial^\mu\chi_i\right)=0\,,\nonumber\\
& \partial_\mu(\sqrt{-g}\, \tau F^{\mu \nu})=0\,.
\end{align}
The simplest solution to the above equations is the unit radius vacuum $AdS_4$,  which is dual to a $d=3$ CFT with a conserved $U(1)$ charge. In this work we choose to place the CFT at finite temperature and deform it by a chemical potential, an external magnetic field and a background lattice. Within this theory, we are interested in thermal states that correspond to density waves.
Putting all the ingredients together, the solutions we are after are  captured by the ansatz \eqref{eq:ansatz}, which we rewrite here for convenience
\begin{align} \label{eq:DC_ansatz}
ds^{2}&=-U(r)\,dt^{2}+\frac{1}{U(r)}\,dr^{2}+e^{2V_{1}(r)}\,dx^1 dx^1 +e^{2V_{2}(r)}\,dx^2 dx^2\,,\nn
A&=a(r)\,dt\,+B \,x^1 dx^2,\nn
\phi&=\phi(r)\,,\qquad\qquad \chi^{I}=k^I_{i} x^{i}\,,\nn
\psi&=\psi(r)\,,\qquad\qquad \sigma^{I}=k^I_{si} x^{i}\,,
\end{align}
where $I=1,2$, $i=1,2$. For simplicity we choose $k^I_{i}=k_{i} \delta ^{Ii}, k^I_{si}=k_{si} \delta ^{Ii} $.

Let us now move on to discuss the boundary conditions.  In the IR, we demand the presence of a regular Killing horizon at $r=0$ by imposing the following expansion
\begin{align}\label{nhexpbh_num}
&U\left(r\right)=4\pi\,T\,r+\dots\,,\qquad &&V_{i}=V_{i}^{(0)}+\dots\,, \qquad a=a^{(0)}\,r+\dots\,,\nn
&\phi=\phi^{(0)}(x)+\dots\,, &&\psi=\psi^{(0)}(x)+\dots\,,  
\end{align}
which is specified in terms of 6 constants. In the UV, we demand the conformal boundary expansion
\begin{align}\label{asymptsol_num}
U&\to (r+R)^2+\dots+W\,(r+R)^{-1}+\dots\,, &&V_{1}\to \log(r+R)+\dots+W_p (r+R)^{-3}+\dots,\nn V_{2}&\to \log(r+R)+\dots\, &&a\to\mu+Q\,(r+R)^{-1}+\dots\,,\nn
\phi&\to \phi_{s}\,(r+R)^{-1}+ \phi_{v}\,(r+R)^{-2}+\dots\,, &&\psi\to  \psi_{s}\,+\dots+\psi_{v}\,(r+R)^{-3}+\dots\,.
\end{align}
Just like in \cite{Donos:2019tmo}, the scalar fields $(\psi,\sigma)$ are taken to constitute the anisotropic Q-lattice in which both translational invariance and the two $U(1)_\psi$ symmetries are explicitly broken, while the density wave phase is supported by $(\phi,\chi)$ and breaks the the two $U(1)_\phi$ symmetries spontaneously. As such, the thermal states of interest correspond to taking $\psi_s \neq 0$ and  $\phi_s = 0$. Thus, this expansion is parametrised by 8 constants. Overall we have 14 constants appearing in the expansions, in comparison to the 11 integration constants of the problem. Thus, for fixed $\gamma, \delta,B, k_{i},k_{si}$ and temperatures below a critical one $T<T_c$, we expect to find a 3 parameter family of solutions, labelled by $\psi_{s},\mu,T$. 

\begin{figure}[h!]
	\centering
	\includegraphics[width=0.48\linewidth]{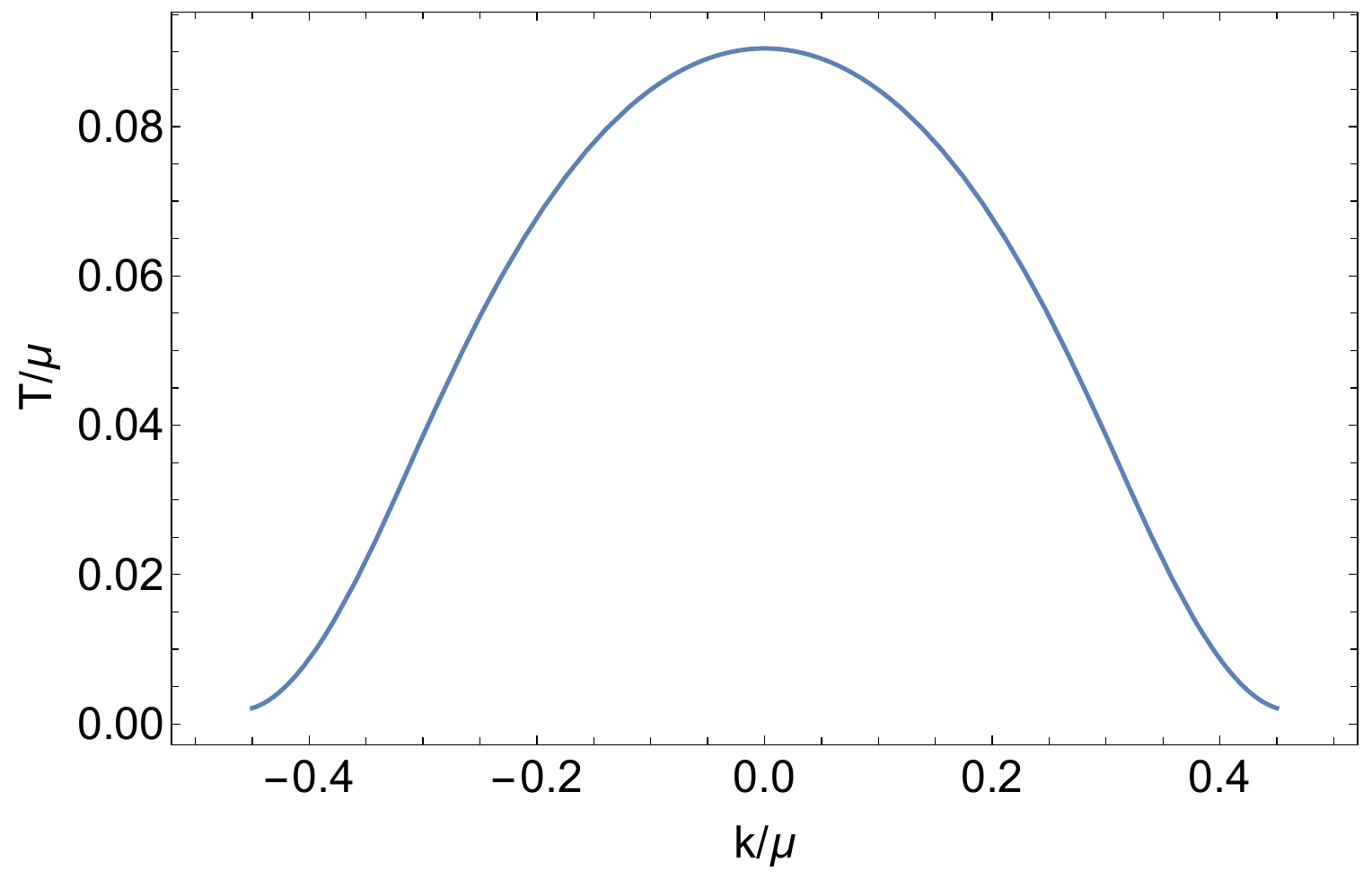}
	\caption{Plot of the critical temperature at which the background Q-lattice becomes unstable as a function of $k$ for $(k_{s1},k_{s2},\psi_s,\gamma, \delta, \mu,B)=(\tfrac{3}{10},\tfrac{3}{10},4,3,1,1,\tfrac{1}{10})$. We see that the most unstable mode corresponds to $k=0$.}
	\label{fig:BellCurve}
\end{figure}

In figure \ref{fig:BellCurve} we plot the critical temperature, $T_c$, as a function of $k=k_1=k_2$ for a particular choice of parameters. This is obtained by considering linearised fluctuations around the normal phase of the system ($\phi=0,\chi=0$) and exhibits the usual ``Bell Curve''  shape.

%%%%%%%%%%%%

\subsection{Quasinormal modes}
We now move on to compute quasinormal modes for the backgrounds constructed above. For simplicity, we focus only on isotropic backgrounds characterised by $k_1=k_2 \equiv k, k_{s1}=k_{s2} \equiv k_s, V_1=V_2$.  We consider perturbations of the form
\begin{align}\label{eq:deltag}
&\delta ds^2= - U \delta h_{tt} dt^2+2U \delta h_{t\,x_i}dt dx_i+e^{2 V_1} \left( h_{11} dx_1^2+h_{22} dx_2^2+2 h_{12}dx_1 dx_2\right)\,,
\end{align}
together with $(\delta a_t, \delta a_{1},\delta a_{2}, \delta\phi,\delta\psi,\delta\chi_1, \delta\chi_2,\delta\sigma_1, \delta\sigma_2)$, where the variations are taken to have the form
\begin{equation}
\delta f(t,r,x_1)=e^{-i \omega v(t,r)+i q x_1} \delta f(r)\,,
\end{equation}
with $v_{EF}$ the Eddington-Finkelstein coordinate defined as
\begin{align}\label{eq:EF_num}
v_{EF}(t,r,x_1)=t+\int_{\infty}^{r}\frac{dy}{U(y)}\,.
\end{align}
Compared to the analytic setup of the problem in section \ref{sec:lin_hydro}, we have chosen $S$ in  \eqref{eq:vef} such that $S'=U^{-1}$, as well as a radial gauge in which all perturbations with an $r$ index vanish. Such a gauge is not compatible with the way we constructed the modes in section \ref{sec:lin_hydro}, but the physical information of the quasinormal modes in the end should of course be the same. Note also that our choice for the momentum $q^i$ to point in the direction $x_1$ is without loss of generality, because the background is isotropic. Plugging this ansatz in the equations of motion, we obtain 5 first order ODEs and 10 second order giving rise to 25 integration constants.

Let us now discuss the boundary conditions that we need to impose on these fields. In the IR, we impose infalling boundary conditions at the horizon, which without oss of generality is set at $r = 0$
\begin{align}
&\delta h_{tt}=c_1 \,r+\dots\,,\nonumber\\
&\delta h_{t \,x_1}=c_2 +\dots\,,&\delta h_{t \,x_2}=c_3 +\dots\,,\nonumber\\
&\delta h_{x_1x_1}=c_4+\dots\,,&\delta h_{x_2 x_2}=-c_4+\dots\,,\nonumber\\
&\delta h_{x_1 \,x_2}=c_5 +\dots\,,&\delta a_t =c_6 \,r+\dots\,,\nonumber\\
&\delta a_{x_1}=c_7 +\dots\,,&\delta a_{x_2} =c_8 +\dots\,,\nonumber\\
&\delta\phi=c_9+\dots\,,&\delta\psi=c_{10}+\dots\,,\nonumber\\
&\delta\chi_1=c_{11}+\dots\,,&\delta\sigma_1=c_{12}+\dots\,,\nonumber\\
&\delta\chi_2=c_{13}+\dots\,,&\delta\sigma_2=c_{14}+\dots\,,
\end{align}
where the constants $c_1,c_2,c_3$ and $c_6$ are not free but are fixed in terms of the others. Thus, for fixed value of $q$, the expansion is fixed in terms of 11 constants, $\omega,c_4,c_5,c_7,c_8,c_9,c_{10},c_{11},c_{12},c_{13},c_{14}$. 

On the other hand, in the UV, the most general expansion with  $\phi_s=0$  is given by
\begin{align}\label{eq:pertUV}
&\delta h_{tt}=\delta h_{tt}^{(s)}+\dots\,,\nonumber\\
&\delta h_{t x_1}=\delta h_{t\,x_1}^{(s)}+\dots\,,\qquad &&\delta h_{t x_2}=\delta h_{t\,x_2}^{(s)}+\dots,,\nonumber\\
&\delta h_{x_1x_1}=\delta h_{x_1\,x_1}^{(s)}+\dots\,, &&\delta h_{x_2 x_2}=\delta h_{x_2\,x_2}^{(s)}+\dots+\frac{\delta h_{x_2\,x_2}^{(v)}}{(r+R)^3}+\dots\,,\nonumber\\
&\delta h_{x_1\, x_2}=\delta h_{x_1\,x_2}^{(s)}+\dots+\frac{\delta h_{x_1\,x2}^{(v)}}{(r+R)^3}+\dots\,, &&\delta a_t =a_{t}^{(s)}+\dots\,,\nonumber\\
&\delta a_{x_1}=a_{x_1}^{(s)}+\frac{a_{x_1}^{(v)}}{(r+R)}+\dots\,, &&\delta a_{x_2}=a_{x_2}^{(s)}+\frac{a_{x_2}^{(v)}}{(r+R)}+\dots\,,\nonumber\\
&\delta\phi=\frac{\delta\phi^{(s)}}{(r+R)}+\frac{\delta\phi^{(v)}}{(r+R)^2}+\dots\,, &&\delta\psi=\delta\psi^{(s)}+\dots+\frac{\delta\psi^{(v)}}{(r+R)^3}+\dots\,,\nonumber\\
&\delta\chi_1=\delta\chi_1^{(v)}+\dots\,, &&\delta\sigma_1=\delta\sigma_1^{(s)}+\dots+\frac{\delta\sigma_1^{(v)}}{(r+R)^3}+\dots\,,\nonumber\\
&\delta\chi_2=\delta\chi_2^{(v)}+\dots\,, &&\delta\sigma_2=\delta\sigma_2^{(s)}+\dots+\frac{\delta\sigma_2^{(v)}}{(r+R)^3}+\dots\,.
\end{align}
For the computation of quasinormal modes, we need to ensure that we remove all the sources from the UV expansion up to a combination of coordinate reparametrisations and gauge transformations
\begin{align}
[\delta g_{\mu\nu}+\mathcal{L}_{\tilde{\zeta}} g_{\mu\nu}]\to 0\,,\notag\\
[\delta A+\mathcal{L}_{\tilde{\zeta}} A+d\Lambda]\to 0\,,
\end{align} 
where the gauge transformations are of the form
\begin{align}\label{eq:coord_transf}
x^\mu\to x^\mu+\tilde{\zeta}^\mu\,,\qquad\tilde{\zeta}&=e^{-i\omega t+i q x_1}\,\zeta^\mu \,\partial_\mu\,,\notag\\
A_\mu\to A_\mu+\partial_\mu \Lambda\,,\qquad\Lambda&=e^{-i\omega t+i q x_1}\,(\lambda+\lambda_2 x_2)\,,
\end{align}
for $\zeta^\mu$, $\lambda$ constants. This requirement demands that the sources appearing in \eqref{eq:pertUV} take the form
\begin{align}
\delta h_{t t}^{(s)}&=2 i \omega\, \zeta_1-2 \zeta_2 \,,\notag\\
\delta h_{t x_{1}}^{(s)}&=iq\, \zeta_1+i \omega \,\zeta_3\,, \qquad \qquad	&&\delta h_{t x_{2}}^{(s)}=i \omega \,\zeta_4\,,\notag\\
\delta h_{x_1\, x_{1}}^{(s)}&=-2 \zeta_2-2 i q \,\zeta_3\,,  &&\delta h_{x_2\, x_{2}}^{(s)}=-2 \zeta_2\,,\notag\\
\delta h_{x_1\, x_{2}}^{(s)}&=-i q \zeta_4\,,  &&\delta a_{t}^{(s)}=i \mu\,\omega\, \zeta_1+i \omega \lambda\,,\notag\\
\delta a_{x_1}^{(s)}&=-i \mu\,q\, \zeta_1-i q\,\lambda+B\, \zeta_4\,,  &&\delta a_{x_2}^{(s)}=-B\, \zeta_3\,,\notag\\
\delta\phi^{(s)}&=0\,, &&\delta\psi^{(s)}=0\,,\notag\\ 
\delta \sigma_1^{(s)}&=-k_{s}\,\zeta_3\,, &&\delta \sigma_2^{(s)}=-k_{s}\,\zeta_4\,,
\end{align}
where $\lambda_2=B\zeta_3$. Therefore, the UV expansion is fixed in terms of 15 constants: $\zeta_1,\zeta_2,\zeta_3, \zeta_4,\lambda$ and $\delta h_{x_2\,x_2}^{(v)},\delta h_{x_1\,x_2}^{(v)},a_{x_1}^{(v)},a_{x_2}^{(v)},\delta\phi^{(v)},\delta\psi^{(v)},\delta\sigma_1^{(v)},\delta\sigma_2^{(v)},\delta\chi_1^{(v)},\delta\chi_2^{(v)}$. 

\begin{figure}[h!]
	\centering
	\includegraphics[width=0.45\linewidth]{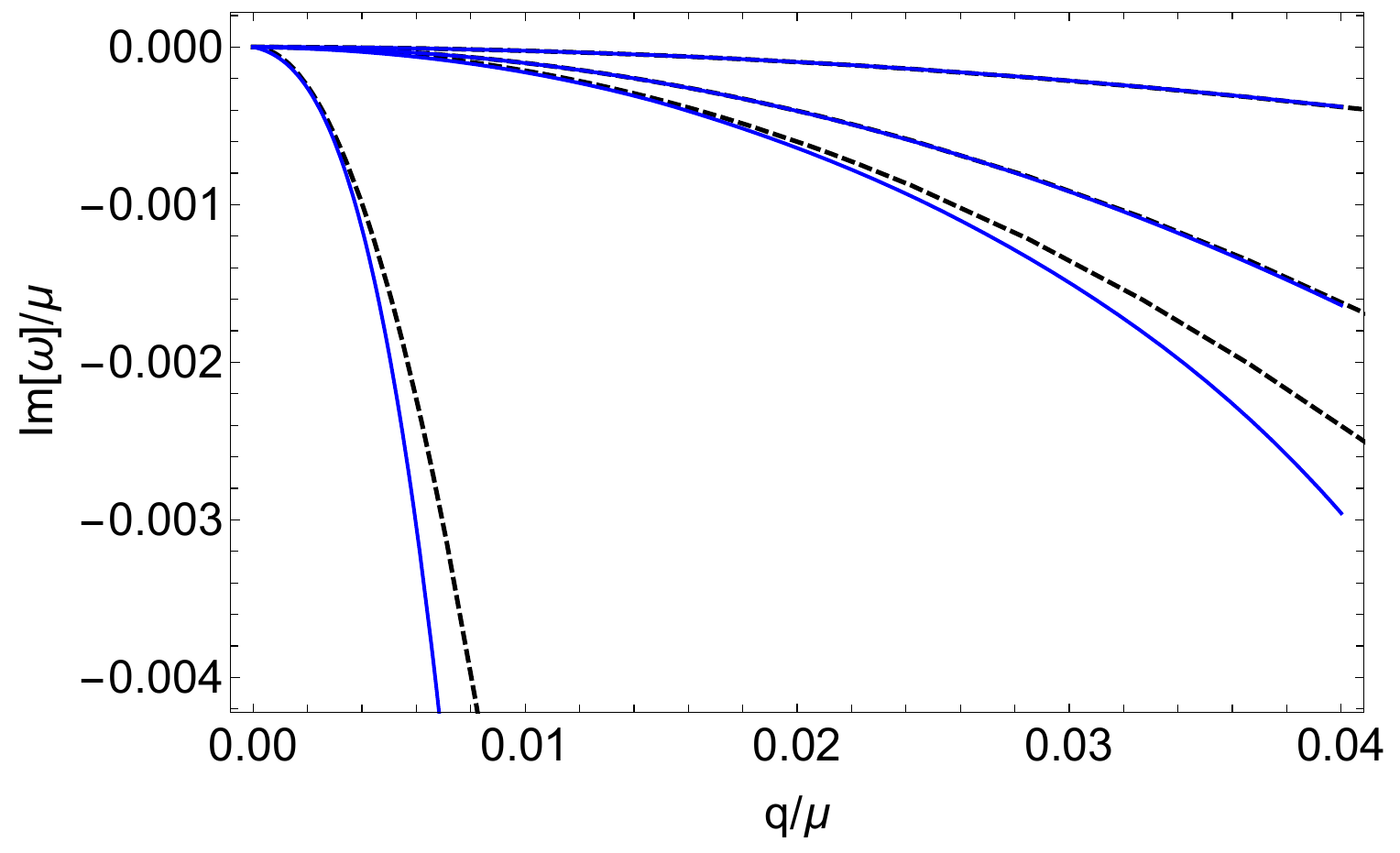}\quad\includegraphics[width=0.45\linewidth]{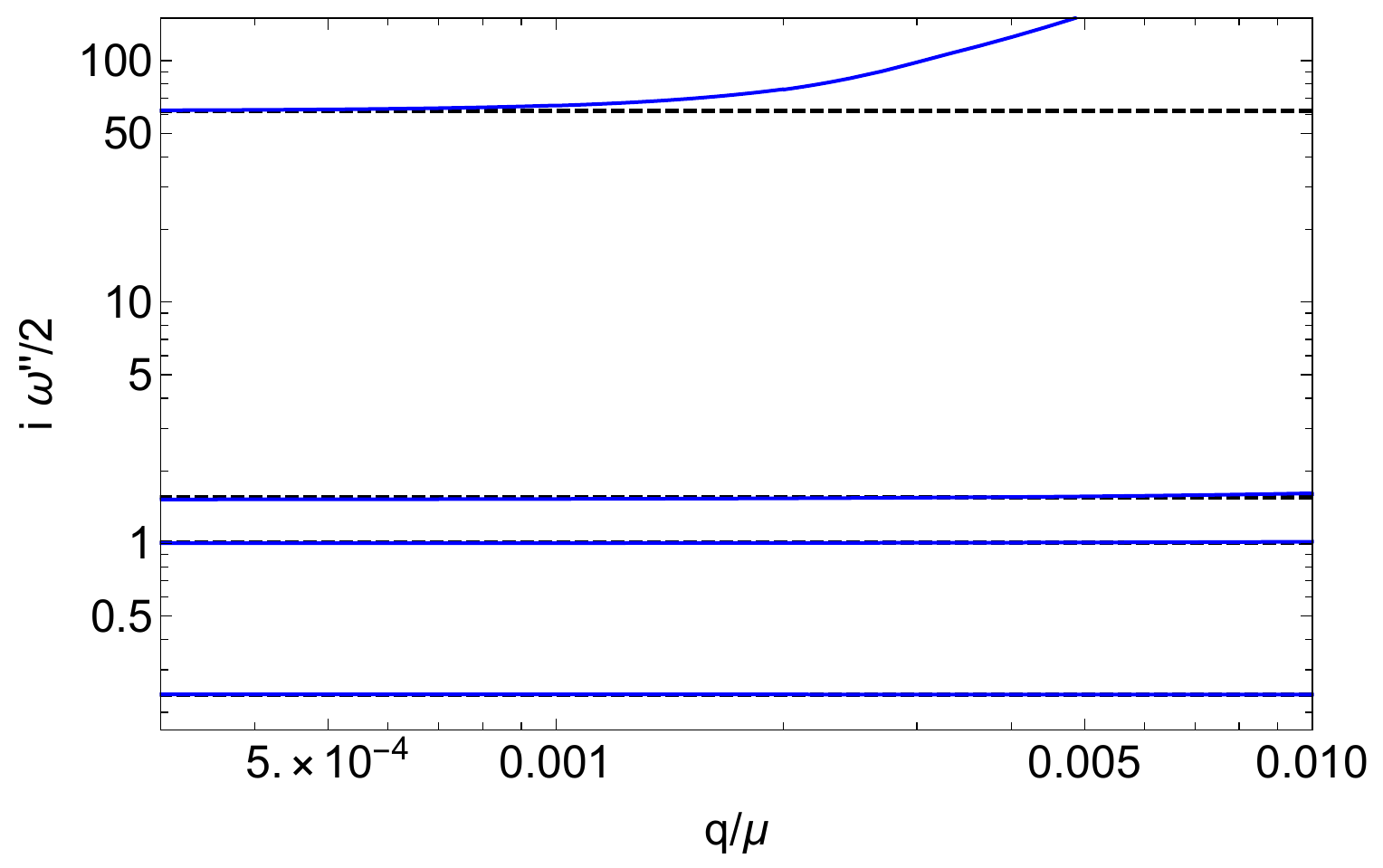}
	\caption{In the left panel we show the dispersion relations for the four diffusive modes in our system. The dashed lines represent the dispersion relations obtained from the linear system in \eqref{eq:matrixeqv2}. In the right panel we plot $i\omega(q)''/2$ as a function of $q$ for each one of the four diffusive modes that we have found numerically. In the limit of small momenta the value of this function converges to the diffusion constants we find from equation \eqref{eq:matrixeqv2}, shown here with dashed lines. In this figure we used $(\phi_s,\psi_s,T,\mu,k,k_s,\gamma,\delta,B)=(0,4,\tfrac{1}{100},1,\tfrac{3}{20},\tfrac{3}{10},3,\tfrac{1}{2},\tfrac{1}{100})$. }
	\label{fig:qnms}
\end{figure}

Overall, for fixed $q,B$, we have 26 undetermined constants, of which one can be set to unity because of the linearity of the equations. This matches precisely the 25 integration constants of the problem and thus we expect our solutions to be labelled by $q$ and $B$. We proceed to solve numerically this system of equations subject to the above boundary conditions using a double-sided shooting method. Figure \ref{fig:qnms} (left) shows the dispersion relations for the four hydrodynamic quasinormal modes in our system for a particular choice of the background configuration. We also illustrate with dashed lines the dispersion relations\footnote{Note that, in order to evaluate the quasinormal modes using the analytic formula \eqref{eq:sdef}, we need to compute the derivatives  $w^i_I$ and $w^{ij}_{IJ}$.  In order to compute these correctly one needs to consider backgrounds with general  $k^I_i$, i.e. $k^1_1\ne k^1_2\ne k^2_1\ne k^2_2$ .} fixed by the linear system \refeq{eq:matrixeqv2}. Figure \ref{fig:qnms} (right) shows the analytically predicted diffusion constants from \refeq{eq:matrixeqv2}, and the ones computed numerically from the $q\to0$ limit of the function $i\omega(q)''/2$.\footnote{Note that, as explained below \eqref{eq:EF_num}, we have chosen the momentum $q$ to point in $x_1$, and thus, in this setting, the diffusion matrices $D_{ij}$ become diffusion constants $D$ for each mode.}

Let us now make some remarks on figure \ref{fig:qnms}. First of all, we note that all the modes we find are diffusive and purely imaginary, as expected from the analysis in section \ref{sec:diffusion}. We also see a good quantitative agreement of the numerical solution and the analytical expressions in the regime of validity of hydrodynamics, where $q$ is parametrically smaller than all the dimensionful scales in the system. Actually, we expect the radius of convergence of hydrodynamics to be set by the collision points of the hydrodynamic modes with the first non-hydrodynamic mode \cite{Withers:2018srf,Grozdanov:2019uhi,Jansen:2020hfd}. In figure \ref{fig:qnms} we chose the parameters such that the lattice is weak; in this case, the lowest lying non-hydrodynamic mode is the momentum relaxation/cyclotron mode. One of the thermoelectric modes, the steepest curve in figure \ref{fig:qnms}, interacts with this non-hydrodynamic mode as $q$ is increased, leading to the quickest deviation from the analytic quadratic dispersion relations. The top curve describes the incoherent thermoelectric mode which decouples from momentum, and agrees very well with the analytic expression even for very large $q$.\footnote{The above characterization of the modes as thermoelectric versus Goldstone is done by examining the system as $k\to0$; in general all the modes are coupled.} This is similar to the case without magnetic field, see \cite{Donos:2019hpp} for further details.

\subsection{Pseudo-gapless modes and two-point fucntions}
In this subsection we outline the numerical computation of the pseudo-gapless modes as well as certain two-point functions involving the currents $J,Q$ in the presence of pinning, $\phi_s\ne0$. We perform a calculation similar to the one for quasinormal modes, but we now consider fluctuations with $q=0$ around a background configuration that has a small but finite source $\phi_s\ne0$. Looking at the ansatz \eqref{eq:deltag}, it is consistent to set $\delta h_{tt},\delta h_{x_1\,x_1},\delta h_{x_1\,x_2},\delta h_{x_2\,x_2},\delta a_{t}, \delta \phi, \delta \psi=0$. We are thus left we 6 second order and 2 first order equations for the remaining fluctuations, giving rise to 14 integration constants.

The IR expansion close to the horizon ($r=0$) takes a similar form as above, namely
\begin{align}\label{eq:expIRgap}
\delta h_{t \,x_1}&=c_2 +\dots\,, \qquad\qquad &&\delta h_{t \,x_2}=c_3+\dots\,,\nn
\delta a_{x_1}&=c_7 +\dots\,, &&\delta a_{x_2}=c_8 +\dots\,,\nn
\delta\chi_1&=c_{11}+\dots\,,&&\delta\sigma_1=c_{12}+\dots\,,\nn
\delta\chi_2&=c_{13}+\dots\,,&&\delta\sigma_2=c_{14}+\dots\,,
\end{align}
where the constants $c_2,c_3$ are fixed in terms of the others. We see that the expansion is fixed in terms of 7 constants, $\omega, c_7, c_8, c_{11}, c_{12}, c_{13}, c_{14}$. 

On the other hand, the UV expansion changes slightly in comparison to \eqref{eq:pertUV} because $\phi_s\ne0$. In particular, it is given by
\begin{align}\label{eq:pertUVgap}
&\delta h_{t x_1}=\delta h_{t\,x_1}^{(s)}+\dots\,,\qquad\qquad &&\delta h_{t x_2}=\delta h_{t\,x_2}^{(s)}+\dots\,,\nn
&\delta a_{x_1}=a_{x_1}^{(s)}+\frac{a_{x_1}^{(v)}}{(r+R)}+\dots\,, &&\delta a_{x_2}=a_{x_2}^{(s)}+\frac{a_{x_2}^{(v)}}{(r+R)}+\dots\,,\nn
&\delta\chi_1=\delta\chi_1^{(s)}+\frac{\delta\chi_1^{(v)}}{(r+R)}+\dots\,, &&\delta\sigma_1=\delta\sigma_1^{(s)}+\dots+\frac{\delta\sigma_1^{(v)}}{(r+R)^3}+\dots\,,\nn
&\delta\chi_2=\delta\chi_2^{(s)}+\frac{\delta\chi_2^{(v)}}{(r+R)}+\dots\,, &&\delta\sigma_2=\delta\sigma_2^{(s)}+\dots+\frac{\delta\sigma_2^{(v)}}{(r+R)^3}+\dots\,.
\end{align}
Once again, we remove all the sources from the UV expansion apart from an external electric field $E$ and a temperature gradient $\zeta$ in the $x_1$ direction, up to a combination of coordinate reparametrisations and gauge transformations. This is done by imposing the following constraints on the sources in \eqref{eq:pertUVgap}
%\begin{align}
%	\delta h_{t x_{1}}^{(s)}&=i \omega \,\zeta_3\,,\quad &&\delta h_{t x_{2}}^{(s)}=i \omega \,\zeta_4\,,\quad &&&\delta a_{x_1}^{(s)}=B\, \zeta_4\,,\quad &&&&\delta a_{x_2}^{(s)}=-B\, \zeta_3\,,\notag\\
%	\delta \sigma_1^{(s)}&=-k_{s}\,\zeta_3\,, &&\delta \sigma_2^{(s)}=-k_{s}\,\zeta_4\,, &&&\delta \chi_1^{(s)}=-k\,\zeta_3\,, &&&&\delta \chi_2^{(s)}=-k\,\zeta_4\,.
%\end{align}
\begin{align}\label{eq:UV_exp_q0}
\delta h_{t x_{1}}^{(s)}&=i \omega \,\zeta_3+\frac{\zeta}{i\,\omega}\,,\, &&\delta h_{t x_{2}}^{(s)}=i \omega \,\zeta_4\,,&&&\delta a_{x_1}^{(s)}=B\, \zeta_4+\frac{(E-\mu\,\zeta)}{i\,\omega}\,,\, &&&&\delta a_{x_2}^{(s)}=-B\, \zeta_3\,,\notag\\
\delta \sigma_1^{(s)}&=-k_{s}\,\zeta_3\,, &&\delta \sigma_2^{(s)}=-k_{s}\,\zeta_4\,, &&&\delta \chi_1^{(s)}=-k\,\zeta_3\,,\quad &&&&\delta \chi_2^{(s)}=-k\,\zeta_4\,.
\end{align}

\begin{figure}[h!]
	\centering
	\includegraphics[width=0.45\linewidth]{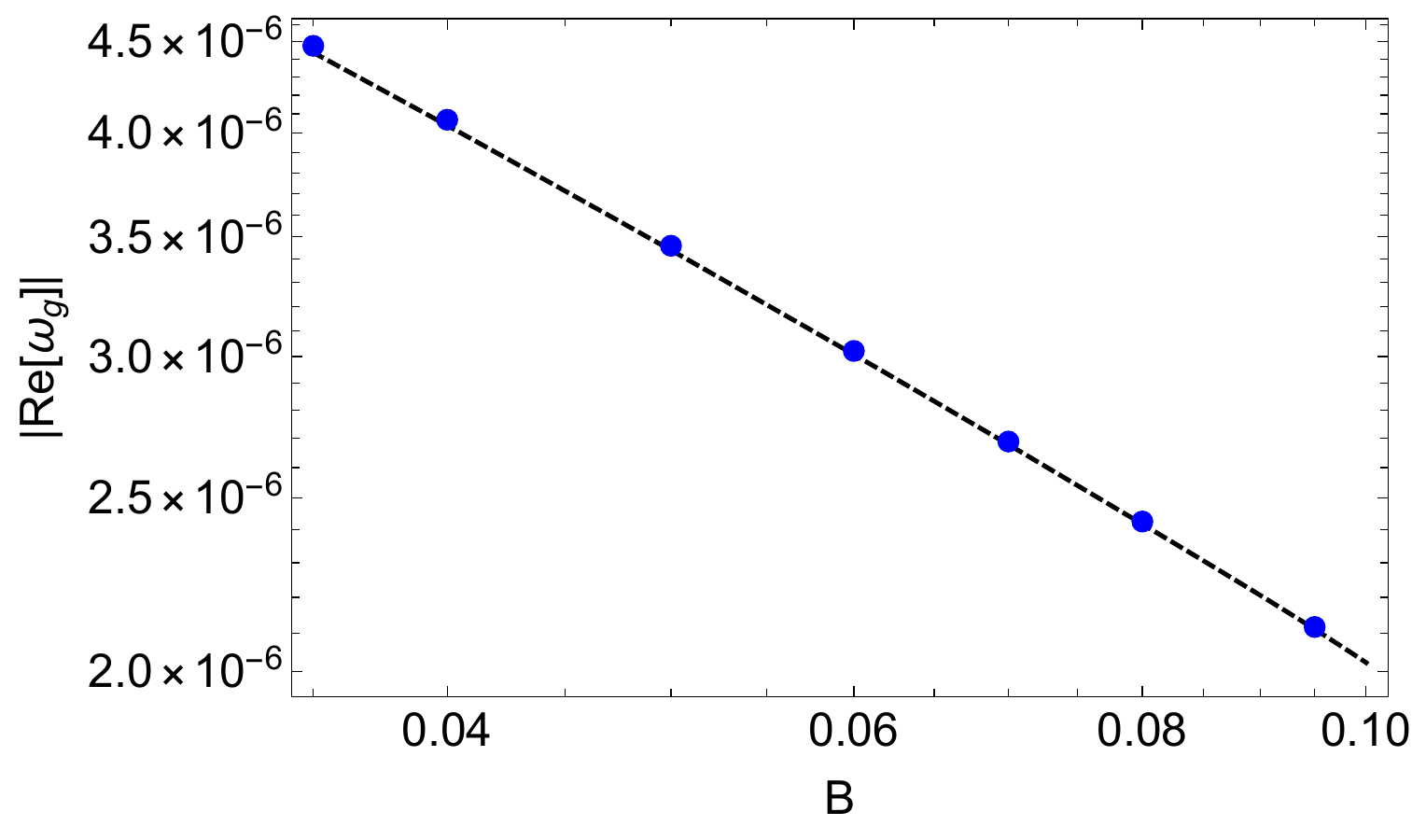}\quad\includegraphics[width=0.45\linewidth]{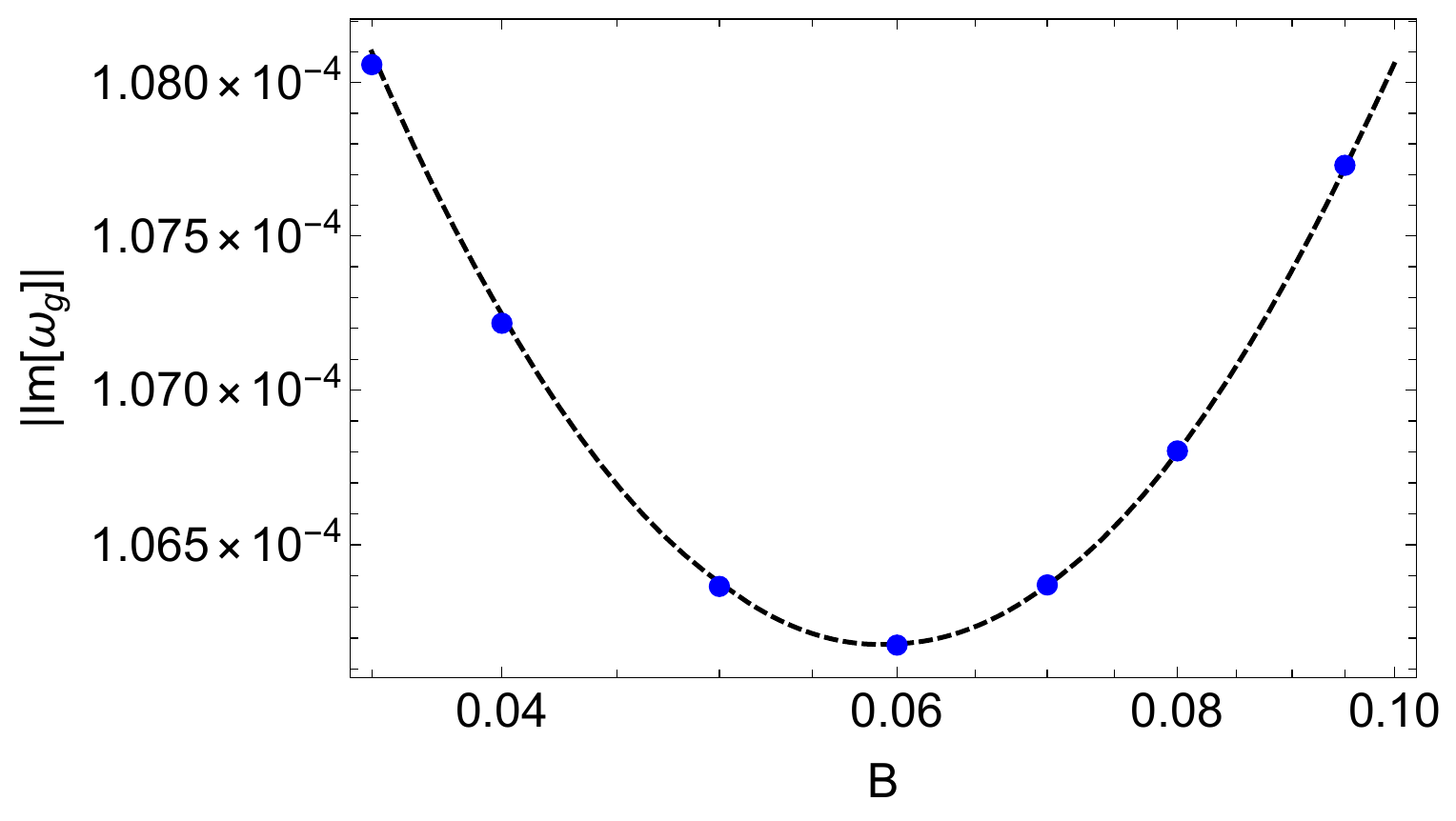}\\
	\includegraphics[width=0.45\linewidth]{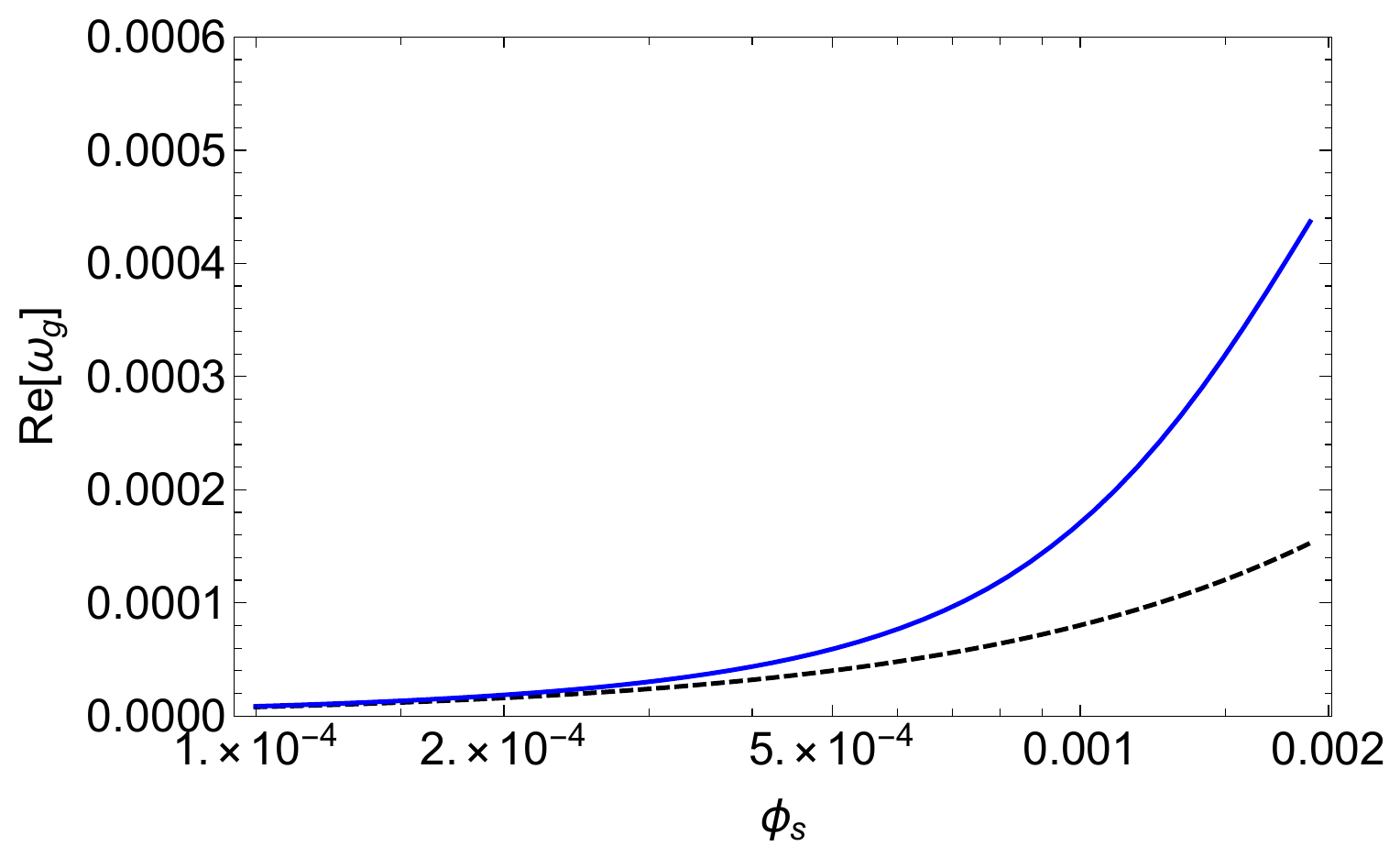}\quad\includegraphics[width=0.45\linewidth]{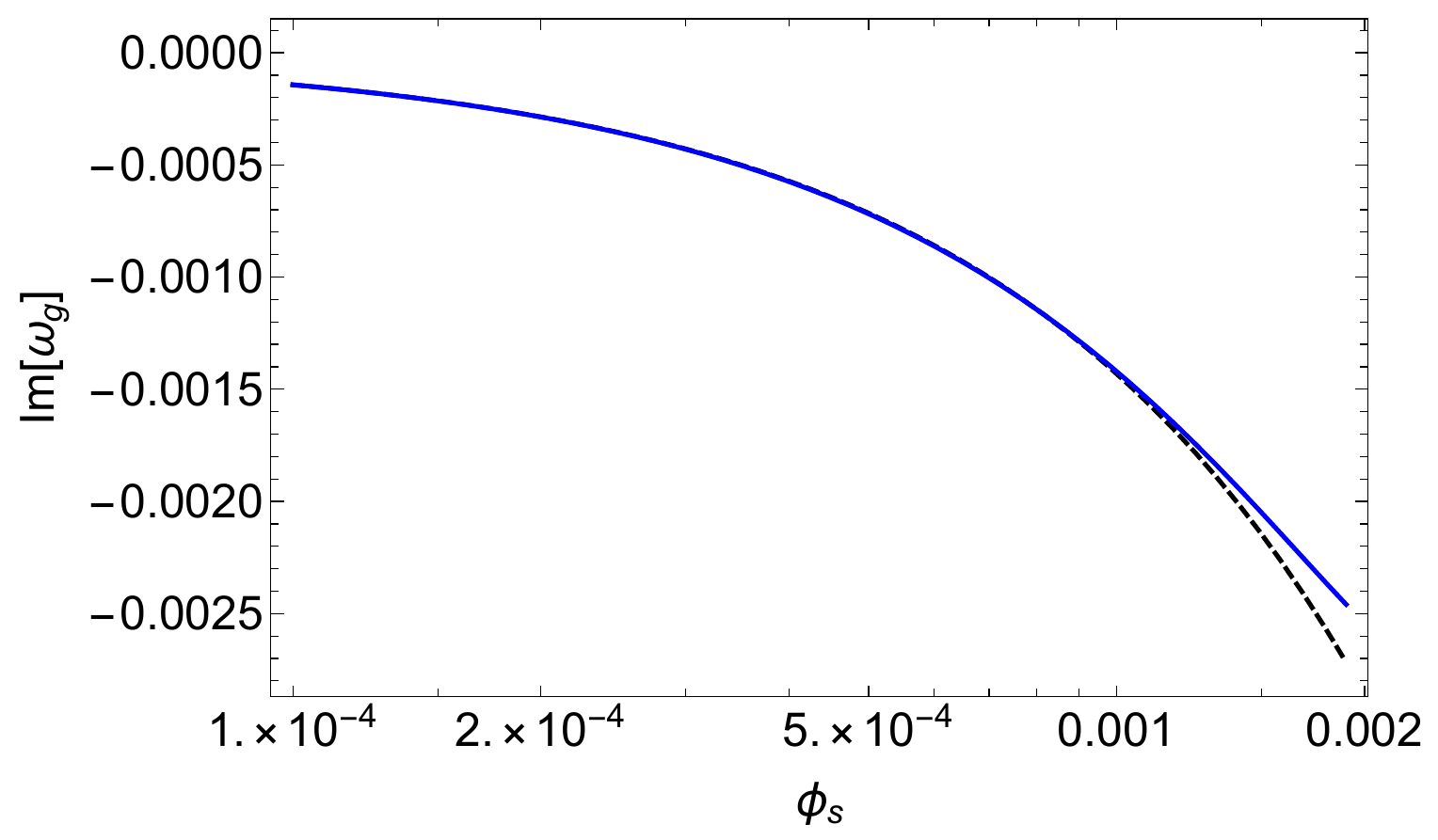}\\
	\caption{In the top row we plot the real and imaginary part of the gap, $\omega_g$, as functions of the magnetic field $B$ for $\phi_s=10^{-4}$. In the bottom row we plot the real and imaginary part of $\omega_g^+$ as functions of the magnetic field $\phi_s$ for $B=1/200$. The dashed lines represent the analytic formula of the previous section, which is exact in $B$ and perturbative in $\phi_s$. Here $(\psi_s,T,\mu,k,k_s,\gamma,\delta)=(4,\tfrac{1}{100},1,\tfrac{3}{20},\tfrac{3}{10},3,\tfrac{1}{2})$. }
	\label{fig:gap}
\end{figure}

Let us first consider the case of the pseudo-gapless modes by setting $(E,\zeta)=(0,0)$. We see that the UV expansion is fixed in terms of 8 constants: $\zeta_3,\zeta_4, a_{x_1}^{(v)}, a_{x_2}^{(v)},$ $ \delta\sigma_1^{(v)},\delta\chi_1^{(v)}, \delta\sigma_2^{(v)}, \delta\chi_2^{(v)}$. Overall, we have 15 undetermined constants, one of which can be set to unity because of the linearity of the equations. This matches precisely the 14 integration constants of the problem and thus we expect to find a discrete set of solutions, labelled by $ B$. We proceed to solve numerically this system of equations subject to the above boundary conditions using a double-sided shooting method aiming to identify the two pseudo-gapless modes of equations \eqref{eq:gap_ex}. Note that the two modes have equal imaginary parts and opposite real parts. In figure \ref{fig:gap} we plot the real and imaginary part of these modes as a function of the pinning parameter, $\phi_s$, and the external magnetic field, $B$, and we compare with the analytic formulas which are depicted with dashed lines. We see that the numerical and analytic calculations are in good agreement. The reader is reminded that the analytic computation is perturbative in $\phi_s$, but exact in $B$.

%\subsection{Thermoelectric conductivities}
%In this subsection we outline the computation of the conductivities in the presence of pinning, $\phi_s\ne0$. Just like in the case of the gap, we consider $q=0$, and we are thus allowed to consistently set $\delta h_{tt},\delta h_{x_1\,x_1},\delta h_{x_1\,x_2},\delta h_{x_2\,x_2},\delta a_{t}, \delta \phi, \delta \psi=0$; this leaves us with  6 second order and 2 first order equations for the remaining \vzg{}fluctuations. The difference is that now we need to solve these equations subject to different boundary conditions in the UV compatible with the presence of an external electric field \vzg{$E$} and a temperature gradient \vzg{$\zeta$} in the $x_1$ direction.

%The \vzg{expansions} in the IR and the UV remain the same as in \eqref{eq:expIRgap} and \eqref{eq:pertUVgap} respectively. The sources satisfy the following constraints \vzg{}
%\begin{align}
%	\delta h_{t x_{1}}^{(s)}&=i \omega \,\zeta_3+\frac{\zeta}{i\,\omega}\,,\, &&\delta h_{t x_{2}}^{(s)}=i \omega \,\zeta_4\,,\, &&&\delta a_{x_1}^{(s)}=B\, \zeta_4+\frac{(E-\mu\,\zeta)}{i\,\omega}\,,\, &&&&\delta a_{x_2}^{(s)}=-B\, \zeta_3\,,\notag\\
%	\delta \sigma_1^{(s)}&=-k_{s}\,\zeta_3\,, &&\delta \sigma_2^{(s)}=-k_{s}\,\zeta_4\,, &&&\delta \chi_1^{(s)}=-k\,\zeta_3\,,\quad &&&&\delta \chi_2^{(s)}=-k\,\zeta_4\,.
%\end{align}

We finally consider the computation of the conductivities. From \eqref{eq:UV_exp_q0} we see that, for fixed $(E,\zeta)$, we have 7 constants in the IR and 8 in the UV. Comparing with the 14 integration constants in the problem, we expect to find a $1$-parameter family of solutions labelled by $\omega$. Using the linearity of the equations we set $(E,\zeta)=(1,0)$ or $(E,\zeta)=(0,1)$ depending on which source we want to keep. The diffusion currents are then given by 
\begin{align}
\delta \hat J^{x_1}&=\,E-\left(\mu+\frac{i Q}{\omega}\right)\,\zeta+a_1^{(v)}+i B \omega \zeta_4\,,\nonumber\\
\delta \hat J^{x_2}&=a_2^{(v)}-i B \omega \zeta_3-M^{12}\,\zeta\,,\nonumber\\
\delta \hat Q^{x_1}&=-\mu a_1^{(v)}+3 i \frac{k_s \psi_s^2}{\omega}\delta \sigma_1^{(v)}+12 i \delta^2 \frac{k \phi_s^2}{\omega}\delta\chi_1^{(v)}-E\,\left(\mu+\frac{i Q}{\omega}\right)\nonumber\\
&-i\frac{\zeta}{\omega}\left(3 W+3 S_V\phi_s+\frac{3}{2}i k_s^2\psi_s^2\omega-2 \mu Q+i \mu^2\omega\right)\nonumber\\
&+\frac{\zeta_3}{4}\,\left(48 k^2 \delta ^2 \phi_s^2+3 k_s^2\phi_s^2\psi_s^2+2 k_s^4\psi_s^4-2 k_s^2\psi_s^2\omega^2+4B^2\right)+\frac{i B}{\omega}a_2^{(v)}-i B\omega\mu\zeta_4\,,\nonumber\\
\delta \hat Q^{x_2}&=-\mu a_2^{(v)}+3 i \frac{k_s \psi_s^2}{\omega}\delta \sigma_2^{(v)}+12 i \delta^2 \frac{k \phi_s^2}{\omega}\delta\chi_2^{(v)}- \frac{i B}{\omega} E+\frac{i\,\zeta}{\omega}\,B\left(\mu+\frac{iQ}{\omega}\right)\nonumber\\
&+\frac{\zeta_4}{4}\,\left(48 k^2 \delta ^2 \phi_s^2+3 k_s^2\phi_s^2\psi_s^2+2 k_s^4\psi_s^4-2 k_s^2\psi_s^2\omega^2+4 B^2\right)-\frac{i B}{\omega}a_1^{(v)}+i B\omega\mu\zeta_3\nonumber\\
&-M^{12}\,E-2M^{12}_T \,\zeta\,.
\end{align}

\begin{figure}[h!]
	\centering
	\includegraphics[width=0.45\linewidth]{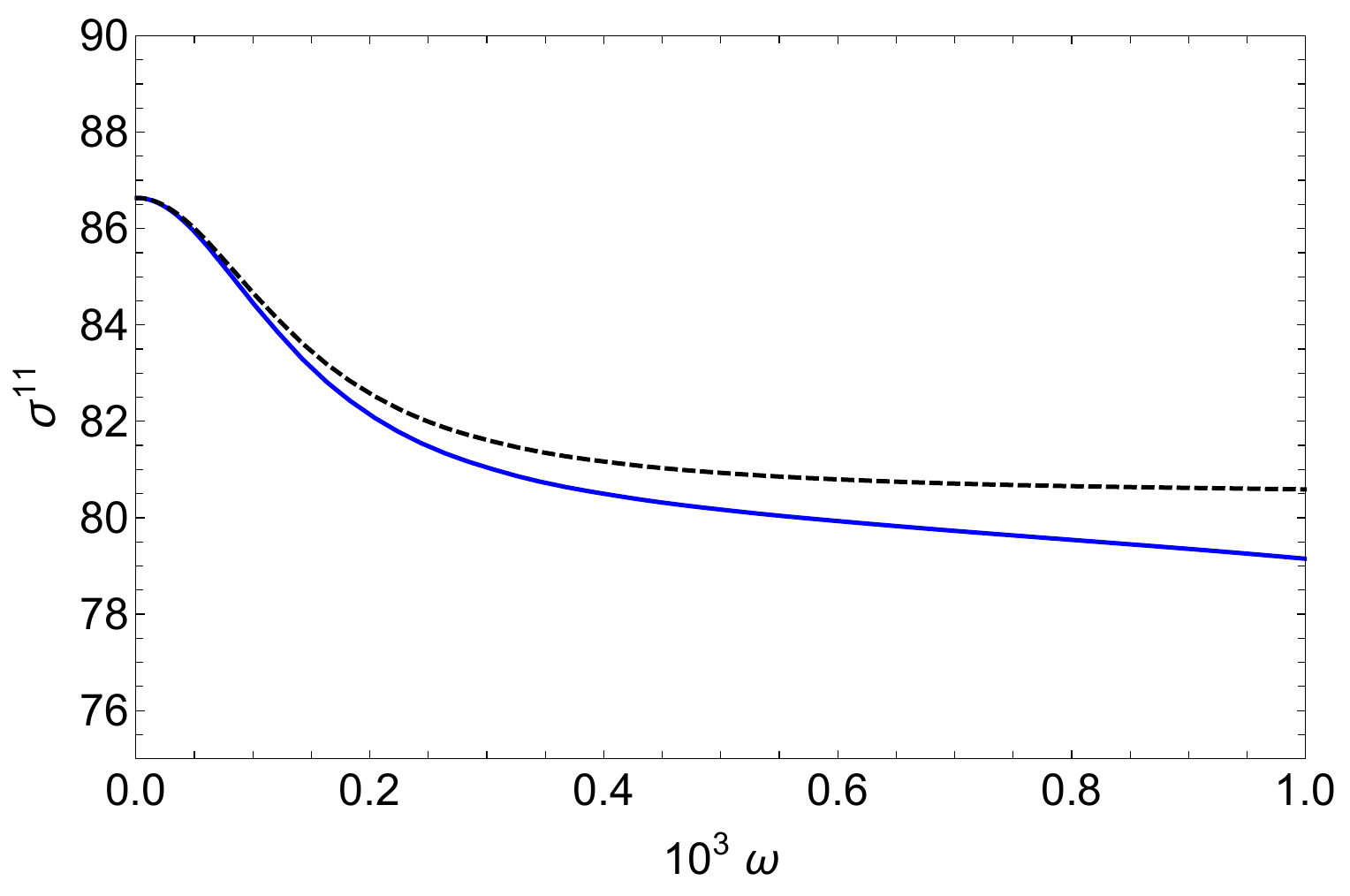}\quad\includegraphics[width=0.45\linewidth]{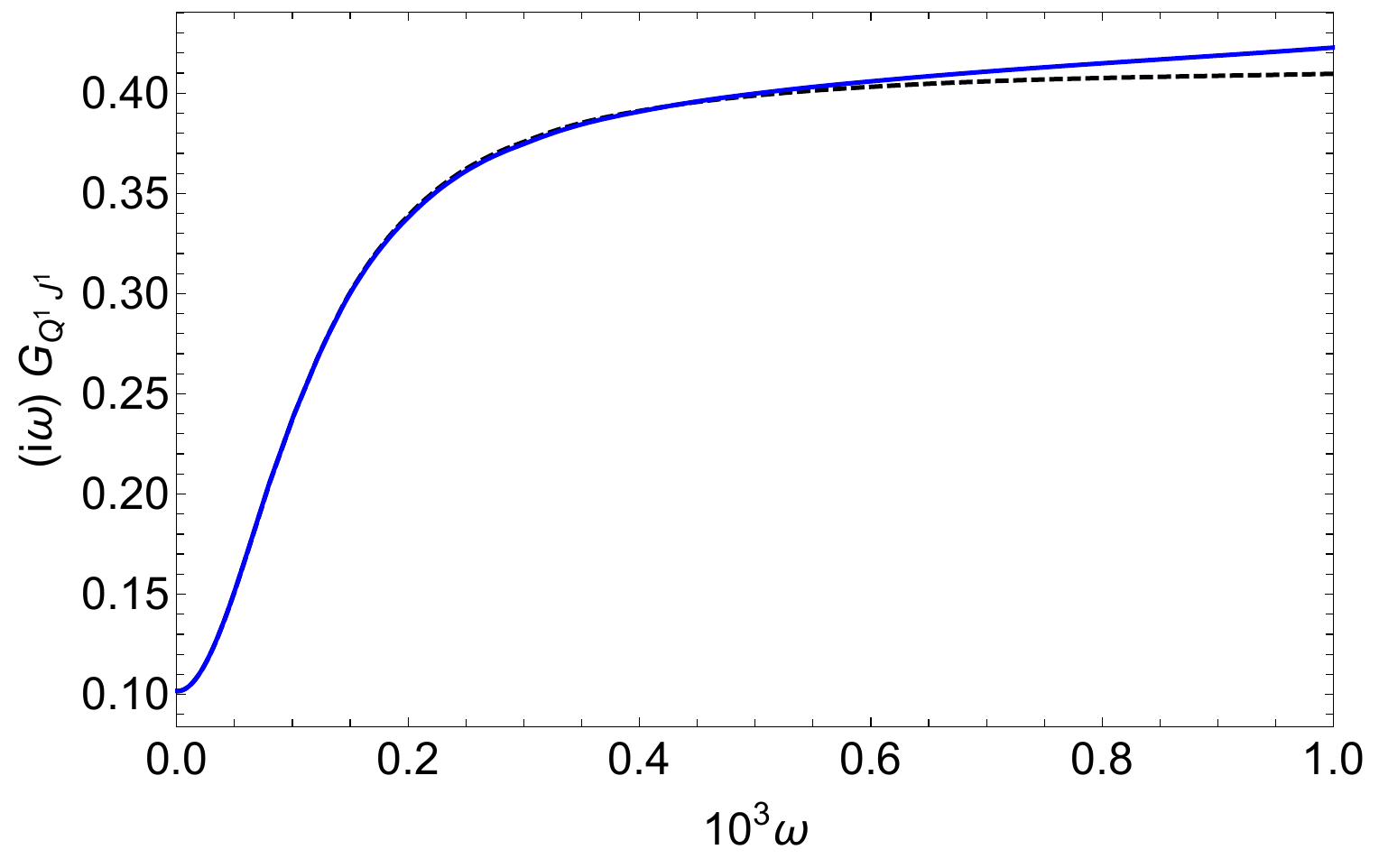}\\
	\includegraphics[width=0.45\linewidth]{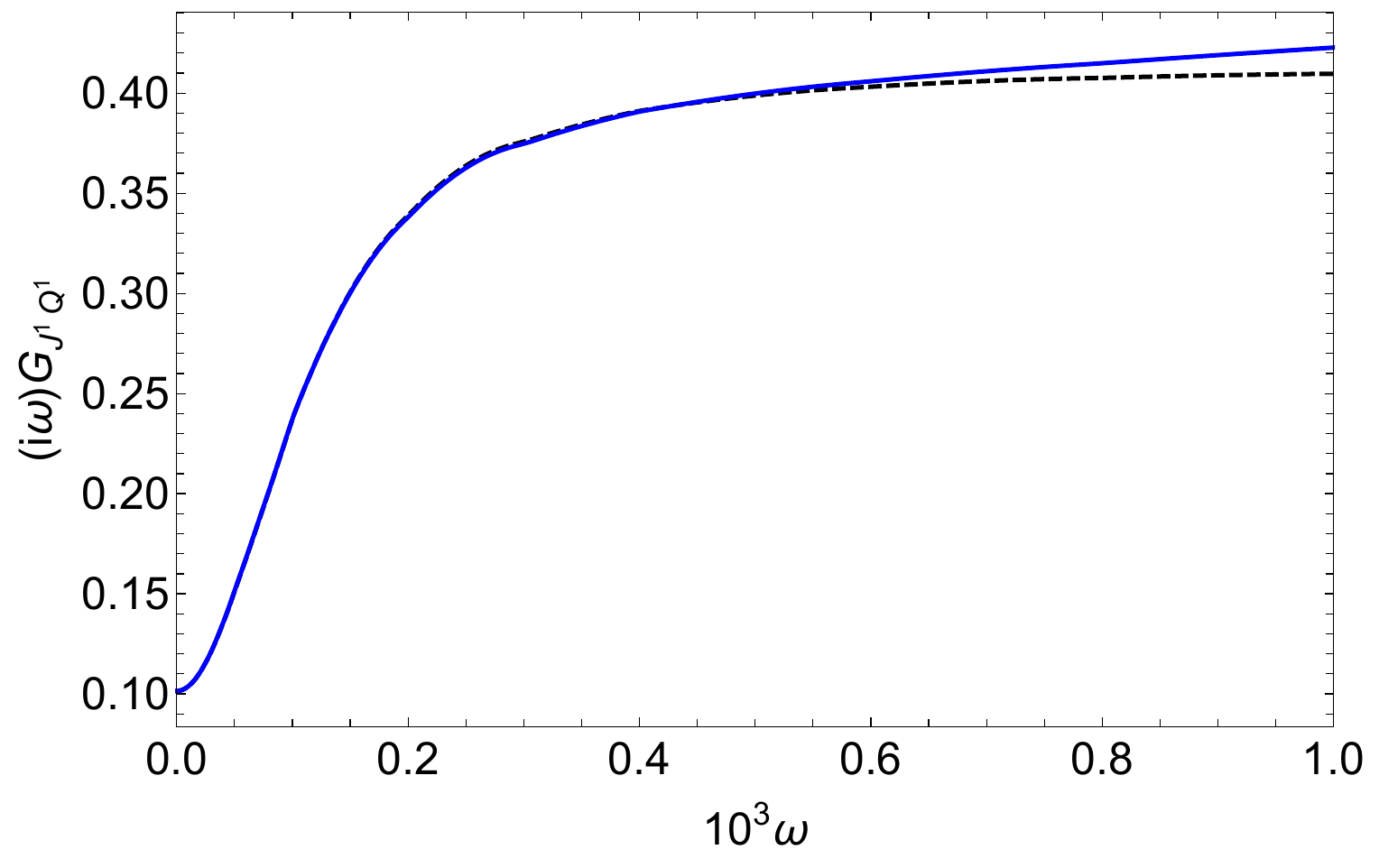}\quad\includegraphics[width=0.45\linewidth]{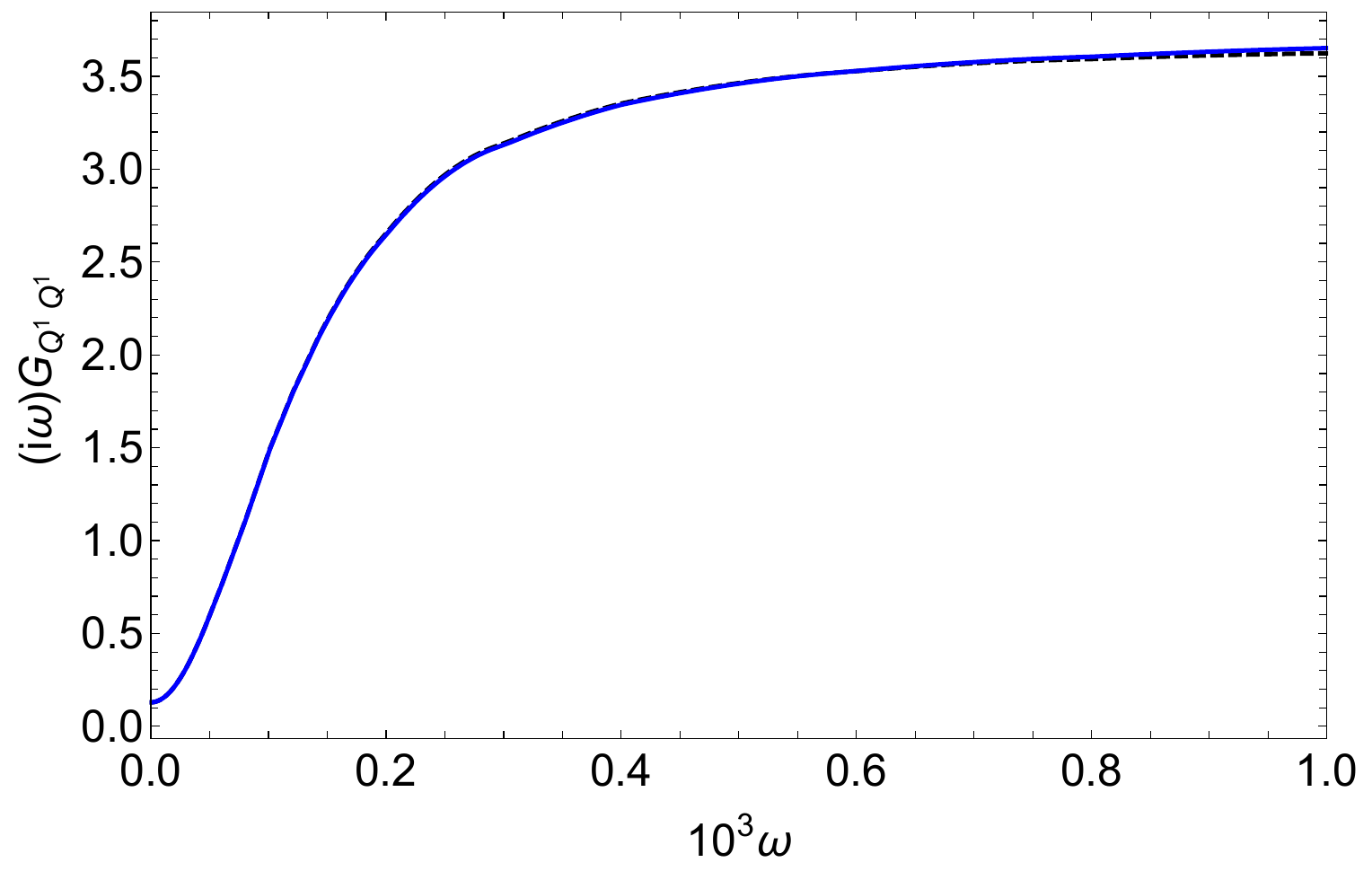}\\
	\includegraphics[width=0.45\linewidth]{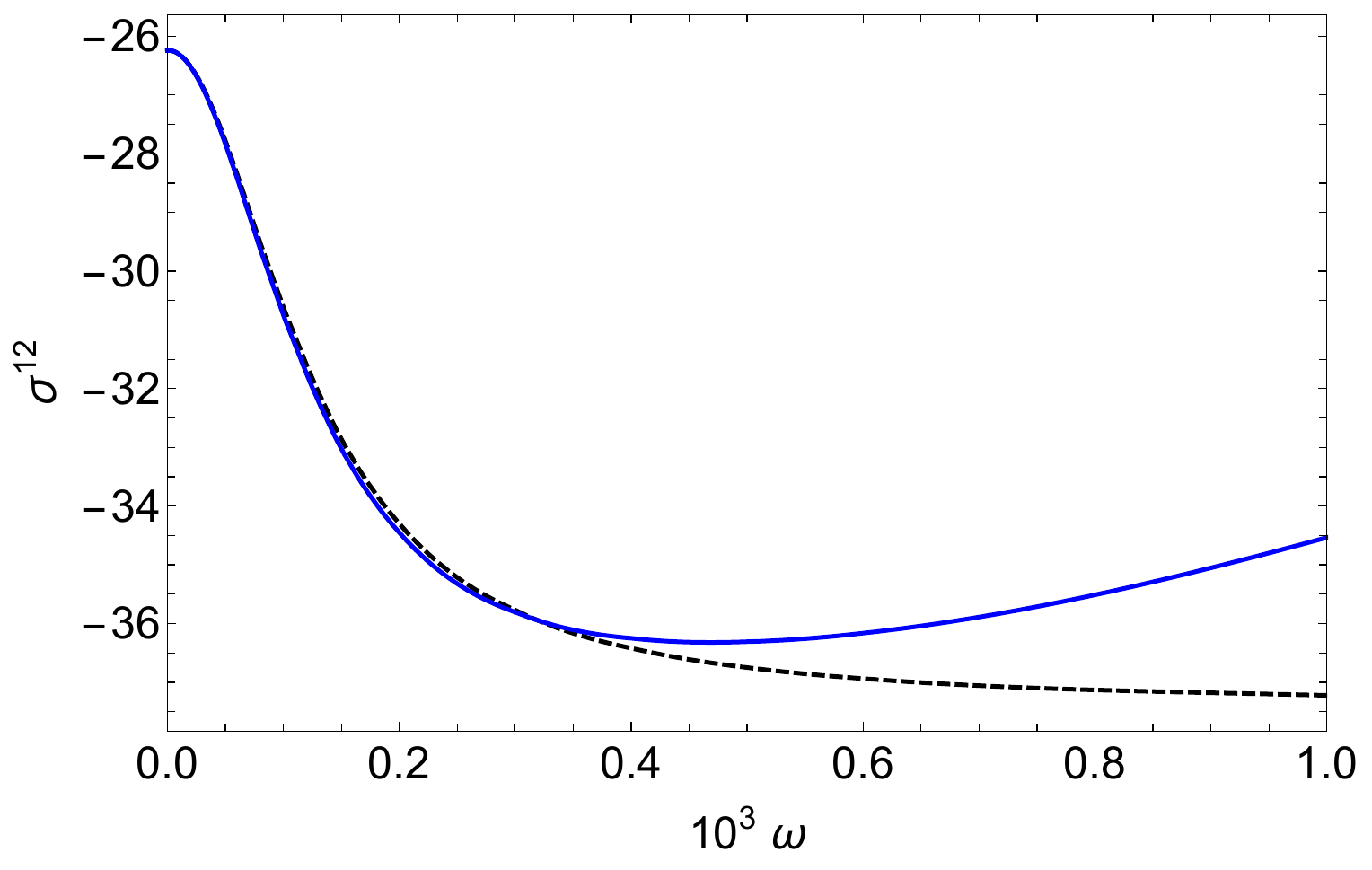}\quad\includegraphics[width=0.45\linewidth]{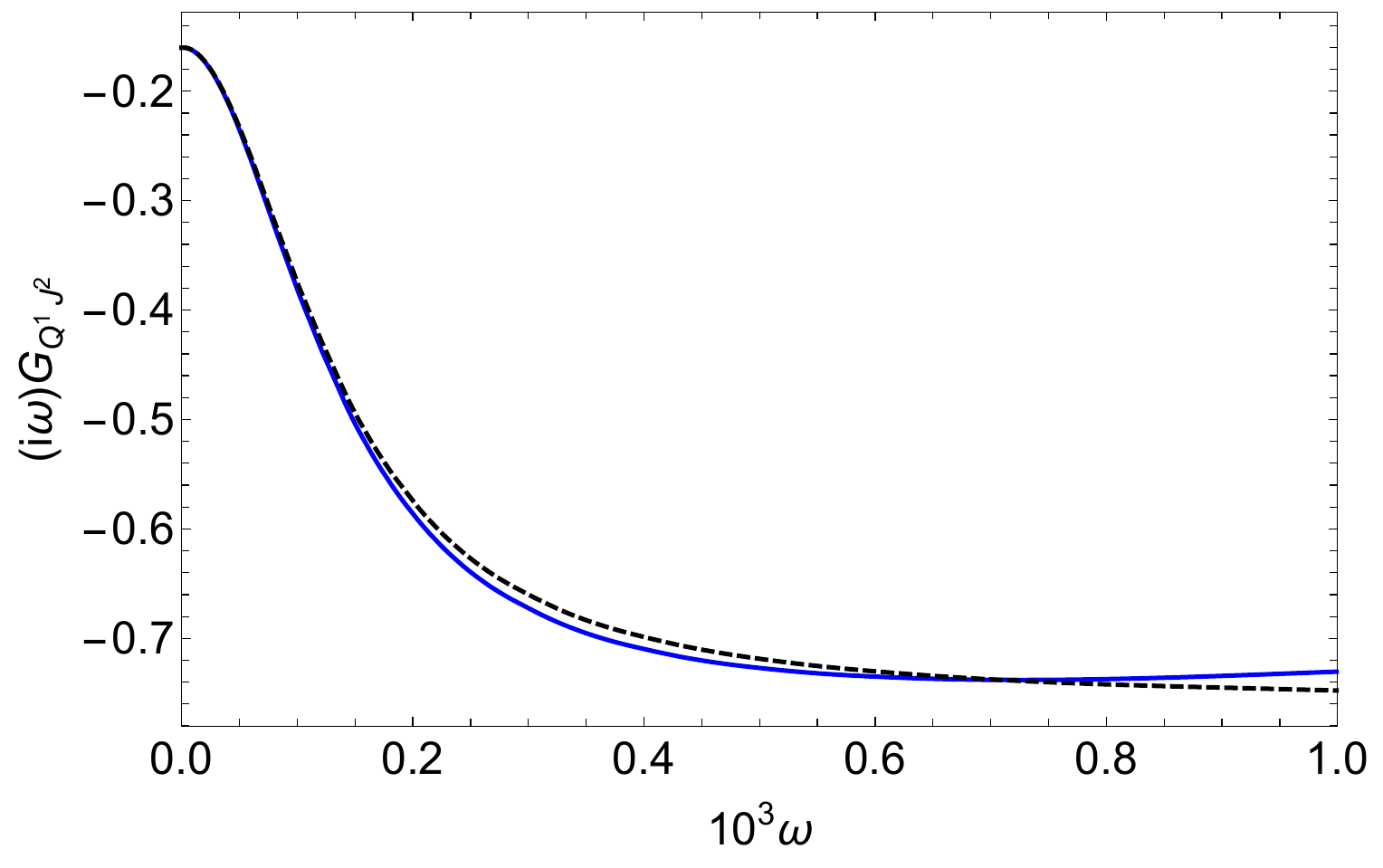}\\
	\includegraphics[width=0.45\linewidth]{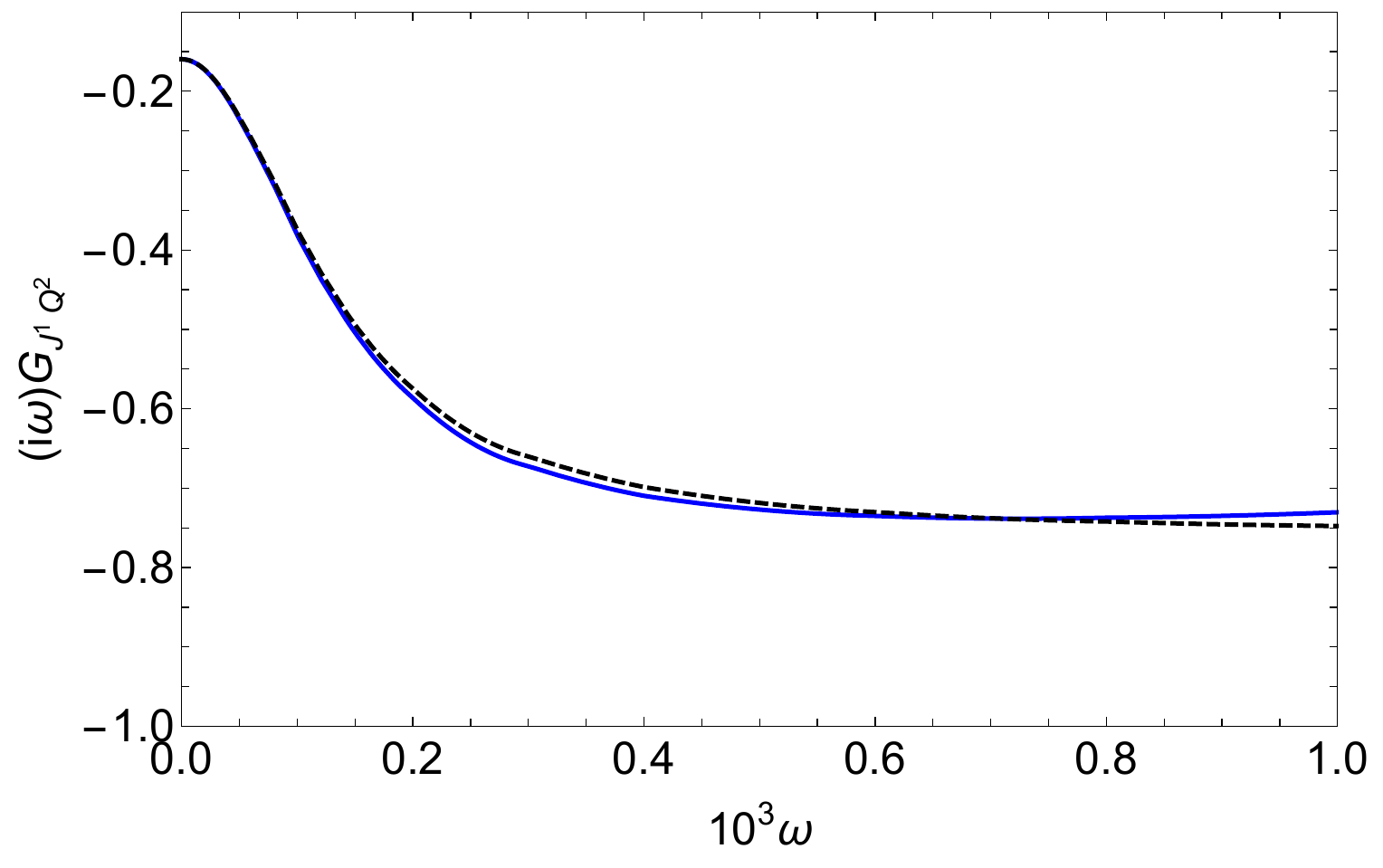}\quad\includegraphics[width=0.45\linewidth]{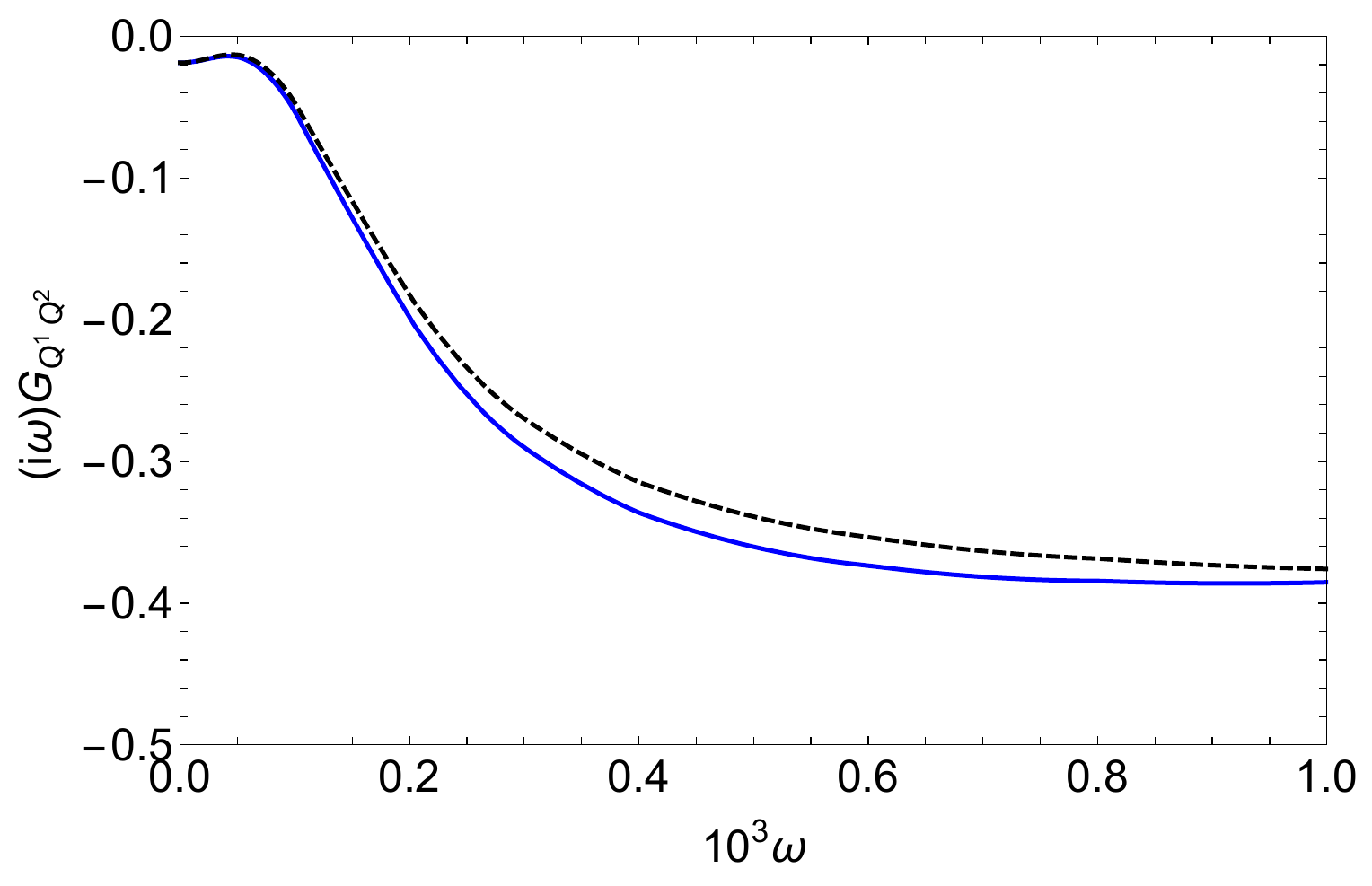}\\
	\caption{Plots of  the components $(1,1)$ and $(1,2)$ of the thermoelectric conductivities as functions of the frequency. The dashed lines correspond to the analytic formulas \eqref{eq:greensf}. Here $(\phi_s,\psi_s,T,\mu,k,k_s,\gamma,\delta,B)=(10^{-4},4,\tfrac{1}{100},1,\tfrac{3}{20},\tfrac{3}{10},3,\tfrac{1}{2},\tfrac{1}{100})$. }
	\label{fig:conductivities}
\end{figure}

Carrying out the numerical shooting computation, we calculate the $(1,1)$ and $(1,2)$ components of the two-point functions $ (i\omega)^{-1}G_{J^iJ^j}, (i\omega)^{-1}G_{J^iQ^j},\, (i\omega)^{-1}G_{Q^jJ^i}$, $(i\omega)^{-1} G_{Q^iQ^j}$. For fixed $B$ and $\phi_s$,  these  quantities are plotted %as functions of the frequency 
in figure \ref{fig:conductivities} with solid lines. In order to compare our numerics with the analytic results of section \ref{sec:lin_hydro}, we use the definition \eqref{eq:pheat_current} to write
\begin{align}
&(i\omega)^{-1}G_{J^iJ^j}=\sigma^{ij}\,,\nonumber\\
& (i\omega)^{-1}G_{J^iQ^j}=T\alpha^{ij}-\sum_{I}\frac{w^j_I}{\langle \Omega^I\rangle} G_{J^iS^I},\,\nonumber\\
& (i\omega)^{-1}G_{Q^jJ^i}=T\bar\alpha^{ij}+\sum_{I}\frac{w^i_I}{\langle \Omega^I\rangle} G_{S^IJ^j}\,,\nonumber\\
&(i\omega)^{-1} G_{Q^iQ^j}=T \bar \kappa^{ij}-\sum_{I}\frac{w^j_I}{\langle \Omega^I\rangle}G_{J_H^iS^I}+\sum_{I}\frac{w^i_I}{\langle \Omega^I\rangle}\left(G_{S^IJ_H^j}-i\omega\sum_{K}\frac{w^i_K}{\langle \Omega^K\rangle }G_{S^I S^K}\right)\,,
\end{align}
and thus obtain analytic expressions using \eqref{eq:greensf}, which are depicted in figure \ref{fig:conductivities} with dashed lines. We see that the two are in good quantitative agreement at small frequencies. The reader is reminded that in this calculation we have set $\zeta_S^I=0$ and we only included sources in the $x_1$ direction; thus we can not compute the $(2,1)$ and $(2,2)$ components of the the two-point functions.

\section{Discussion}\label{sec:discussion}
In this paper we constructed the effective theory of hydrodynamics which captures holographic phases in which translations are broken explicitly and spontaneously. We have significantly extended the construction of \cite{Donos:2019hpp} to include an arbitrary number $N_{Z}$ of gapless degrees of freedom emerging from spontaneous density waves and we also included a background magnetic field.

%Effective theories describing two Goldstone modes arising from spontaneous breaking of translations in magnetic fields have been considered before in \cite{}. According to the effective model of \cite{} and ignoring the coupling to the heat current, the dispersion relations of these two modes are real and quadratic at leading order in the derivative expansion. A holographic model which incorporates the two gapless modes along with the coupling to the heat current was studied in \cite{\cite{Baggioli:2020edn}} and a complex quadratic dispersion relation was found.

A holographic model which incorporates the two Goldstone modes arising from spontaneous breaking of translations in magnetic fields, along with the coupling to the heat current was studied in \cite{Baggioli:2020edn} and a complex quadratic dispersion relation was found. In our setup, the strength of the explicit breaking is large compared to the wavelength of the hydrodynamic fluctuations. In section \ref{sec:diffusion} we analytically derived an equation whose roots yield the dispersion relations of the hydrodynamic modes governing our system. Despite not being able to write down the dispersion relations of all of our $2+N_{Z}$ hydrodynamic modes in closed form, we prove that they are purely imaginary and diffusive, unlike \cite{Baggioli:2020edn}.

In our construction we have also included the corresponding $N_{Z}$ perturbative deformation parameters which pin down the density waves and introduce $N_{Z}$ gaps in our theory. Interestingly, we have shown that apart from the gap, the magnetic field field causes the corresponding poles to move off the imaginary axis due to resonance effects. In section \ref{sec:transport} we computed the retarded Green's functions of the operators relevant to the hydrodynamic description of the system. As one might expect, the poles due to pinning have a direct effect on the transport properties of our system as can be seen from the explicit form of the Green's functions in equation \eqref{eq:greensf}.

Finally, an important byproduct in our work is the identification of the correct current in \eqref{eq:pheat_current} which describes the transfer of entropy as can be seen by the conservation equation in the last line of \eqref{eq:current_cons}. Given this definition, the variation of the free energy $w^{i}_{I}$ with respect to the wavenumbers $k^{I}_{i}$ drops out of the corresponding Green's functions \eqref{eq:greensf}. Moreover, the gaps and the resonance frequencies which can be found by solving the eigenvalue problem equation \eqref{eq:gap_veq} are also independent of $w^{i}_{I}$.

There are various open questions which one could further explore. It would be interesting to consider second order hydrodynamic perturbation theory and examine what the second law implies for the transport coefficients in phases with spontaneous and explicit symmetry breaking. Additionally, it is important to examine how transport in such phases is constrained from purely field theoric considerations, such as the Ward identities, and also investigate the possible experimental significance of the decoupled/incoherent currents we defined in this paper. Finally, it would be enlightning to move away from homogeneity and explore what kind of novel effects inhomogeneous models with similar symmetry breaking patterns might exhibit.

\section*{Acknowledgements}
We would like to thank Blaise Goutéraux for useful discussions. AD is supported by STFC grant ST/T000708/1. CP is supported by the European Union’s Horizon 2020 research and innovation programme under the Marie Skłodowska-Curie grant agreement HoloLif No 838644. The work of VZ was supported by the China Postdoctoral Science Foundation (International Postdoctoral Fellowship Program 2018), the National Natural Science Foundation of China (NSFC) (Grant number 11874259), and is supported by the European Research Council (ERC) under the European Union’s Horizon 2020 research and innovation programme (grant agreement No.758759).

\appendix

\section{Perturbations in the bulk}
In this appendix we derive the constitutive relations \eqref{eq:currents_dis_consti} for the transport part of the heat and electric currents along with the Josephson relations \eqref{eq:josephson} for the density wave degrees of freedom. In order to do this, we solve for the perturbation $\delta\chi^{I}$ through its equation of motion \eqref{eq:chi_eom}. We only need to do this up to second order in our $\varepsilon$ expansion \eqref{eq:X_epsilon_exp}, which we carry out in Appendix \ref{app:chi}. Then, in Appendix \ref{app:js} we derive the constitutive relations for the currents by relating the field theory and horizon currents densities of our holographic model.

\subsection{Perturbations for $\chi^{I}$}\label{app:chi}
After perturbatively expanding the equation of motion \eqref{eq:chi_eom}, we have
\begin{align}\label{eq:chi_eom_pert}
&-\partial_{t}\left( \sqrt{-g}\,\Phi_{I}g^{tt}g^{ij}k_{i}^{I}\,\delta g_{tj}\right)-\partial_{r}\left( \sqrt{-g}\,\Phi_{I}g^{rr}g^{ij}k_{i}^{I}\,\delta g_{rj}\right)\nn
&+\partial_{j}\left(\delta\left( \sqrt{-g}\,\Phi_{I}g^{ij}\right) k_{i}^{I}\right)+\partial_{\mu}\left( \sqrt{-g}\,\Phi_{I}g^{\mu\nu} \partial_{\nu}\delta \chi^{I}\right)=0\,.
\end{align}
From the form of the solution close to the conformal boundary at $r\to \infty$, we can infer a relation between the sources $\zeta_{S_{I}}$ of the operators $S_{I}$ and their vevs $\delta\langle S_{I}\rangle$. This will essentially give a Josephson type of equation for the variable $\delta \hat{c}^{I}$ through equation \eqref{eq:currents_consti}. In the next subsections we will solve equation \eqref{eq:chi_eom_pert} in an $\varepsilon$ expansion.

\subsubsection{Field theory interpretation at order $\mathcal{O}(\varepsilon)$}

At order $\mathcal{O}(\varepsilon)$ we obtain the equation
\begin{align}
-\sqrt{g}\,\Phi_{I}k_{i}^{I}g^{ij}\zeta_{j[1]}-\partial_{r}\left( \sqrt{g}\,\Phi_{I}Ug^{ij}k_{i}^{I}\delta g_{rj[1]}\right)+iq_{j}\,\delta\left( \sqrt{g}\,\Phi_{I}g^{ij}\right)_{[0]}k_{i}^{I}\nn
+\partial_{r}\left( \sqrt{g}\,\Phi_{I}U \partial_{r}\delta\chi_{[1]}^{I}\right)-i\omega_{[1]}\partial_{r}\left( \sqrt{g}\, \Phi_{I}U\partial_{r}S \,\delta c^{I}_{[0]}\right)=0\,.
\end{align}
We integrate this equation for $\delta\chi_{[1]}^{I}$ while insisting on the near horizon behaviour \eqref{eq:gen_exp}. After doing so, we obtain the asymptotic behavior
\begin{align}
\delta\chi_{[1]}^{I}=\frac{r^{2\Delta_{I}-3}}{(2\Delta_{I}-3)\phi^{I}_{v}{}^{2}}\Bigl[&w_{I}^{j}\zeta_{j[1]}+\sqrt{g_{(0)}}\Phi_{I}^{(0)}\left(g^{ij}_{(0)}k_{i}^{I}v_{j[1]}-i\omega_{[1]}\,\delta c^{I}_{[0]}\right)\\
&+iq_{i}\left(\nu^{i}_{I}\delta T_{[0]}+\beta^{i}_{I}\delta\mu_{[0]}\right)\Bigr] +\cdots\,.
\end{align}

Demanding that the operator $S_I$ is not sourced at this order, we must have
\begin{align}\label{eq:deltachi1_source}
\sqrt{g_{(0)}}\Phi_{I}^{(0)}\left(i\omega_{[1]}\,\delta c^{I}_{[0]} -g^{ij}_{(0)}k_{i}^{I}v_{j[1]}\right) =w_{I}^{j}\zeta_{j[1]}+iq_{i}\left(\nu^{i}_{I}\delta T_{[0]}+\beta^{i}_{I}\delta\mu_{[0]}\right)\,.
\end{align}

\subsubsection{Field theory interpretation at order $\mathcal{O}(\varepsilon^{2})$}

At order $\mathcal{O}(\varepsilon^{2})$ we obtain the equation
\begin{align}
-\sqrt{g}\,\Phi_{I}k_{i}^{I}g^{ij}\zeta_{j[2]}-\partial_{r}\left( \sqrt{g}\,\Phi_{I}Ug^{ij}k_{i}^{I}\delta g_{rj[2]}\right)+iq_{j}\,\delta\left( \sqrt{g}\,\Phi_{I}g^{ij}\right)_{[1]}k_{i}^{I}\nn
-q_{i}q_{j}\sqrt{g}\Phi_{I}g^{ij}\,\delta c^{I}_{[0]}+\partial_{r}\left( \sqrt{g}\,\Phi_{I}U \partial_{r}\delta\chi_{[2]}^{I}\right)-i\omega_{[2]}\partial_{r}\left( \sqrt{g}\, \Phi_{I}U\partial_{r}S \,\delta c^{I}_{[0]}\right)=0\,,
\end{align}
where we have used that $\delta T_{[0]}=\delta \mu_{[0]}=\omega_{[1]}=0$ shown in subsection \ref{subsec:horizon_constraint} below. Following similar steps as above, we obtain the asymptotic expansion
\begin{align}
\delta\chi_{[2]}^{I}=\frac{r^{2\Delta_{I}-3}}{(2\Delta_{I}-3)\phi^{I}_{v}{}^{2}}\Bigl[&w_{I}^{j}\zeta_{j[2]}+\sqrt{g_{(0)}}\Phi_{I}^{(0)}\left(g^{ij}_{(0)}k_{i}^{I}v_{j[2]}-i\omega_{[2]}\,\delta c^{I}_{[0]}\right)\nn
&+iq_{i}\left(\nu^{i}_{I}\delta T_{[1]}+\beta^{i}_{I}\delta\mu_{[1]}-i\sum_{J}w_{IJ}^{ij}\delta c^{J}_{[0]}q_{j}\right)\Bigr]+\cdots\,.
\end{align}
This result, along with equation \eqref{eq:currents_consti}, allows us to write the Josephson relation \eqref{eq:josephson}.

\subsection{Constitutive relations for the thermoelectric currents}\label{app:js}
In this subsection we will relate the horizon current densities \eqref{eq:J_hor2} to the boundary quantities $\delta J^{i}$ and $\delta Q^{i}$ that appear in the current conservation equation \eqref{eq:conservation_laws}. 

The bulk electric current is defined as
\begin{align}\label{eq:el_def_app}
\delta	J^i_{bulk} = \sqrt{-g}\, \tau\, \delta F^{ir}\,.
\end{align}
The equations of motion \eqref{eq:eom1} imply
\begin{align}\label{eq:drel_bulk}
\partial_{r} \delta J^{i}_{bulk} = \partial_{t} \left(\sqrt{-g}\, \tau\, \delta F^{ti}\right) + \partial_{j} \delta \left(\sqrt{-g}\, \tau\, F^{ji}\right)\,.
\end{align}

Following \cite{Banks:2015wha}, for any vector $\varLambda^{\mu}$ in the bulk we can define the bulk two-form
\begin{align}\label{eq:g2form}
G^{\mu\nu}= -2\nabla^{[\mu} \varLambda^{\nu]} - \tau\, \varLambda^{[\mu} F^{\nu]\rho} A_\rho -\frac{1}{2}\left(\varLambda^\rho A_\rho-f\right) \tau F^{\mu\nu}\,,
\end{align}
where $\varLambda^\mu F_{\mu\nu} = \partial_\nu f +\beta_\nu$, with $\beta$ a $1$-form and $f$ a globally defined function. After using the equations of motion \eqref{eq:eom1}, its divergence can be brought to the form
\begin{align}\label{eq:nablag2form}
\nabla_\mu G^{\mu\nu}&=V\, \varLambda^{\nu} +2\nabla^{\nu}\nabla_{\rho}\varLambda^{\rho} -2\nabla_{\mu}\nabla^{(\mu}\varLambda^{\nu)} +\frac{1}{2}\tau F^{\nu\rho}\beta_\rho -\frac{1}{2}A_\rho \mathcal{L}_\varLambda \left(\tau\,F^{\nu\rho}\right)\nn
&-\frac{\tau}{2}\,F^{\nu\rho}A_{\rho}\,\nabla_{\mu}\varLambda^{\mu} +\left(\sum_{I} G_{I}\,\partial^\nu\phi^{I}\partial_\rho\phi^{I} +\sum_{J} W_{J}\,\partial^\nu\psi^{J}\partial_\rho\psi^{J}\right) \varLambda^\rho\nn
&+\left(\sum_{I}\Phi_{I}\,\partial^\nu\chi^{I}\partial_\rho\chi^{I} +\sum_{J}\Psi_{J}\,\partial^\nu\sigma^{J}\partial_\rho\sigma^{J}\right) \varLambda^\rho \,.
\end{align}

We now consider $\varLambda^\mu=\partial_t$, and a general perturbation around the background ansatz \eqref{eq:ansatz} (not necessarily of the form \eqref{eq:sep_var}). The bulk heat current is defined as
\begin{align}\label{eq:heat_def_app}
\delta Q^i_{bulk} = \sqrt{-g}\,G^{ir} &=U^2\sqrt{g}\,g^{ij}\left(\partial_r\left(\frac{\delta g_{jt}}{U}\right)-\partial_j\left(\frac{\delta g_{rt}}{U}\right) \right) -a_t\, \delta J_{bulk}^i\nn
&=U^{1/2} \sqrt{g} \left[2 K^i{}_t+U^{1/2} g^{ij}\,\partial_t \delta g_{rj}\right] -a_t \,\delta J_{bulk}^i\,,
\end{align}
where we have used the result of Appendix B of \cite{Donos:2017ihe} for the extrinsic curvature component
\begin{align}\label{eq:extr_curv}
K^i{}_{\,t}&=\frac{1}{2} U^{3/2} g^{ij} \left[ \partial_r \left(\frac{\delta g_{jt}}{U}\right) - \partial_j \left(\frac{\delta g_{rt}}{U}\right)-\frac{\partial_t \delta g_{rj}}{U}\right]\,.
\end{align}
Writing $\tilde{t}^\mu{}_\nu=-2 K^\mu{}_\nu+X \delta^\mu{}_\nu+Y^\mu{}_\nu$, where $X=2K+\cdots$ and $Y$ are additonal terms that come from the counterterms, we recognize $\tilde{t}^\mu{}_\nu$ as the field theory stress tensor, when evaluated on the boundary. 
%Thus
%\begin{align}
%\delta Q^i_{bulk} =U^{1/2} \sqrt{g}\left [Y^i{}_t- \tilde{t}^i{}_t+U^{1/2} g^{ij} \partial_t \delta g_{rj}\right]-a_t \,\delta J_{bulk}^i\,.
%\end{align}
Evaluating \eqref{eq:heat_def_app} at the boundary, this gives
\begin{align}
\delta Q_{bulk}^i\Big|_\infty =-  \left(r^{-2} \,t^i{}_t+\mu\, \delta J_{bulk}^i\Big|_\infty\right)\,,
\end{align}
where $t^\mu{}_\nu=r^{5} \tilde{t}^\mu{}_\nu$. Note that the contribution from $Y^i{}_t$, as coming from \eqref{eq:bdy_action}, and contribution from the term involving a time derivative are subleading even in the precense of sources. This result matches the expression for the boundary heat current obtained from the variation of the action in the presence of the sources as in \eqref{eq:sep_var}
\begin{align}
\delta S&=\int d^{3}x  \,\sqrt{-h}\,\left[ \frac{1}{2}\,r^{-5}\,t^{\mu\nu}\,\delta g_{\mu\nu}+r^{-3} J^\mu\delta A_\mu\right]\,,
\end{align}
where $h_{\mu\nu}=g_{\mu\nu}-n_\mu\,n_\nu$ and $n$ is the unit norm normal vector. 
%Thus, we can identify $\delta Q_{bulk}^i\Big|_\infty$ with the field theory heat current.
Furthermore, equation \eqref{eq:nablag2form} implies the radial dependence
\begin{align}\label{eq:drheat_bulk}
\partial_r \delta Q^i_{bulk} & = \partial_j \left(\sqrt{-g}G^{ji}\right) + \partial_t \left(\sqrt{-g}G^{ti}\right) -2\sqrt{-g}\,\partial^i\partial_t\log\sqrt{-g} +2\sqrt{-g} g^{i \rho} \nabla^{\mu}\partial_{t} g_{\mu\rho} \nn
& -\sqrt{-g}\,\tau F^{i\rho}\partial_t A_\rho +\frac{1}{2}\partial_t \left(\sqrt{-g}\,A_\rho \tau\,F^{i\rho}\right)\nn
& -\sqrt{-g} \left(\sum_{I} G_{I}\,\partial^i \phi^{I}\partial_t \phi^{I} +\sum_{J} W_{J}\,\partial^i \psi^{J}\partial_t \psi^{J}\right)\nn
&-\sqrt{-g}\left(\sum_{I}\Phi_{I}\,\partial^i\chi^{I}\partial_t \chi^{I} +\sum_{J}\Psi_{J}\,\partial^i \sigma^{J}\partial_t\sigma^{J}\right) \,.
\end{align}
%
%The first term in \eqref{eq:drheat_bulk} can be identified with the local thermal magnetisation density
%\begin{align}\label{eq:thermal_mag_app}
%	M^{ij}_{T}(r) &= \sqrt{-g}G^{ij} =-2\sqrt{-g}\,\nabla^{[i} k^{j]} -\sqrt{-g}\,A_t\, \tau F^{ij}\,,\nn
%	&=-\sqrt{g_d}\,a_t\, \tau F^{(B)ij} +\sqrt{g_d}\,g^{im}_d g^{jn}_d\left(\partial_n \delta g_{mt} -\partial_m \delta g_{nt}\right) -\delta\left(\sqrt{-g}\,A_t\, \tau F^{ij}\right)\,,
%\end{align}
%
%We also compute 
%\begin{align}\label{eq:gti_app}
%	\sqrt{-g}G^{ti} =& -2\sqrt{-g}\,\nabla^{[i} k^{j]}-\frac{1}{D-2}\sqrt{-g}\,\tau\,A_{\rho} F^{i\rho} -\sqrt{-g}\,A_t\, \tau F^{ti}\,,\nn
%	=& -\frac{1}{D-2}\sqrt{g_d}\,\tau\,F_{jk}^{(B)} F^{(B)ik} x^j +\sqrt{g_d}\,g^{ij}_d \left( \delta g_{jr} \partial_r U -U\partial_t \delta g_{jt} +U\partial_j \delta g_{tt}\right) \nn
%	&-\frac{1}{D-2}\delta\left(\sqrt{-g}\,\tau\,A_{\rho} F^{i\rho}\right) -\delta\left(\sqrt{-g}\,A_t\, \tau F^{ti}\right)\,.
%\end{align}
%
%The inhomogeneous piece may look alarming, but it will cancel with the last term of the second line in \eqref{eq:drheat_bulk}.
%
%
%%One can check that for our choice of $k^\mu$, only the last term in \eqref{eq:drheat_bulk} contributes to order $\mathcal{O}\left(\varepsilon^2\right)$, which leads to the radial evolution for $\delta Q^1_{bulk}$ presented in \eqref{eq:drheat_bulk_1}, and the relation \eqref{eq:bdy_hor_heat} between the boundary and horizon heat currents.

\subsubsection{The boundary currents at order $\mathcal{O}(\varepsilon)$}
Expanding the radial evolution for the electric current \eqref{eq:drel_bulk} to order $\mathcal{O}(\varepsilon)$ we obtain
\begin{align}
\partial_{r}\delta J^i_{bulk[1]}=-\tau\sqrt{g}g^{lk}g^{ij}\varepsilon_{kj}\,B\,\zeta_{l[1]}+iq_{j}\,\delta\left( \tau\sqrt{g}g^{jk}g^{il}\right)_{[0]}\,\varepsilon_{kl}B\,,
\end{align}
which can integrate from the horizon up to the conformal boundary at infinity to find
\begin{align}
\delta J^i_{\infty[1]}&=\delta J^i_{(0)[1]}-M^{ij}\,\zeta_{j[1]}+iq_{j}\left(\partial_{T}M^{ij}\,\delta T_{[0]}+\partial_{\mu}M^{ij}\delta\mu_{[0]} \right)\nn
&=\sqrt{g_{(0)}}\tau^{(0)}g^{ij}_{(0)} \left( -i q_{j} \delta\mu_{[0]} +E_{j[1]}+v_{j[1]}a_{t}^{(0)}+B\varepsilon_{jl}g^{lk}_{(0)}v_{k[1]}\right)-M^{ij}\,\zeta_{j[1]}\nn
&+iq_{j}\left(\partial_{T}M^{ij}\,\delta T_{[0]}+\partial_{\mu}M^{ij}\delta\mu_{[0]} \right)\,.
\end{align}

For the radial evolution of the heat current, after expanding equation \eqref{eq:drheat_bulk} we obtain
\begin{align}
\partial_{r}\delta Q_{bulk[1]}^{i}&=\sqrt{g}\tau \varepsilon^{ij}\,B\,(E_{j[1]}-2a_{t}\,\zeta_{j[1]})-iq_{j}\,\delta\left( \sqrt{-g}\,\tau a_{t} F^{ji}\right)_{[0]}\nn
&+i\omega_{[1]}\,\sqrt{g}\,g^{ij} \sum_{I}\Phi_{I} k_{j}^{I}\delta c_{[0]}^{I}\,.
\end{align}
Integrating from the horizon to infinity we obtain
\begin{align}
\delta Q_{\infty[1]}^{i}&=\delta Q_{(0)[1]}^{i}+i\omega_{[1]}\sum_{I}w^{i}_{I}\,\delta c_{[0]}^{I}-M^{ij}\,E_{j[1]}-2\,M_{T}^{ij}\,\zeta_{j[1]}\nn
&+i q_{j}\,\left( \partial_{T}M_{T}^{ij}\,\delta T_{[0]}+\partial_{\mu}M_{T}^{ij}\,\delta \mu_{[0]}\right)\nn
&=4\pi T\sqrt{g_{(0)}}g^{ij}_{(0)}\,v_{j[1]}+i\omega_{[1]}\sum_{I}w^{i}_{I}\,\delta c_{[0]}^{I}\nn
&-M^{ij}\,E_{j[1]}-2\,M_{T}^{ij}\,\zeta_{j[1]}+i q_{j}\,\left( \partial_{T}M_{T}^{ij}\,\delta T_{[0]}+\partial_{\mu}M_{T}^{ij}\,\delta \mu_{[0]}\right)\,.
\end{align}

\subsubsection{The boundary currents at order $\mathcal{O}(\varepsilon^{2})$}
Using the fact that $\delta T_{[0]}=\delta \mu_{[0]}=\omega_{[1]}=0$ (shown in subsection \ref{subsec:horizon_constraint} below), we proceed to compute the currents at next order in $\mathcal{O}(\varepsilon)$. 

The radial evolution equation \eqref{eq:drel_bulk} for the electric current gives
\begin{align}
\partial_{r}\delta J^i_{bulk[2]}=-\tau\sqrt{g}g^{lk}g^{ij}\varepsilon_{kj}\,B\,\zeta_{l[2]}+iq_{j}\,\delta\left( \tau\sqrt{-g}g^{jk}g^{il}\right)_{[1]}\,\varepsilon_{kl}B\,,
\end{align}
which can be integrated to give the expression
\begin{align}\label{eq:J2v2}
\delta J^i_{\infty[2]}&=\delta J^i_{(0)[2]}-M^{ij}\,\zeta_{j[2]}+iq_{j}\left(\partial_{T}M^{ij}\,\delta T_{[1]}+\partial_{\mu}M^{ij}\delta\mu_{[1]}+iq_{l}\sum_{I} \partial_{k^{I}_{l}}M^{ij}\delta c^{I}_{[0]}\right)\nn
&=\sqrt{g_{(0)}}\tau^{(0)}g^{ij}_{(0)} \left( -i q_{j} \delta\mu_{[1]} +E_{j[2]}+v_{j[2]}a_{t}^{(0)}+B\varepsilon_{jl}g^{lk}_{(0)}v_{k[2]}\right)-M^{ij}\,\zeta_{j[2]}\nn
&\qquad +iq_{j}\left(\partial_{T}M^{ij}\,\delta T_{[1]}+\partial_{\mu}M^{ij}\delta\mu_{[1]}+iq_{l}\sum_{I} \partial_{k^{I}_{l}}M^{ij}\delta c^{I}_{[0]} \right)\,.
\end{align}

For the heat current we have
\begin{align}
\partial_{r}\delta Q_{bulk[2]}^{i}&=\sqrt{g}\tau \varepsilon^{ij}\,B\,(E_{j[2]}-2a_{t}\,\zeta_{j[2]})-iq_{j}\,\delta\left( \sqrt{-g}\,\tau a_{t} F^{ji}\right)_{[1]}\nn
&+i\omega_{[2]}\,\sqrt{g}\,g^{ij} \sum_{I}\Phi_{I} k_{j}^{I}\delta c_{[0]}^{I}\,.
\end{align}
Integrating from the horizon to infinity we obtain
\begin{align}\label{eq:Q2v2}
\delta Q_{\infty[2]}^{i}&=\delta Q_{(0)[2]}^{i}+i\omega_{[2]}\sum_{I}w^{i}_{I}\,\delta c_{[0]}^{I} -M^{ij}\,E_{j[2]}-2\,M_{T}^{ij}\,\zeta_{j[2]}\nn 
&+i q_{j}\,\left( \partial_{T}M_{T}^{ij}\,\delta T_{[1]}+\partial_{\mu}M_{T}^{ij}\,\delta \mu_{[1]}+iq_{l}\sum_{I}\partial_{k^{I}_{l}}M_{T}^{ij}\delta c^{I}_{(0)}\right)\nn
&=Ts\,\,v^{i}_{[2]}+i\omega_{[2]}\sum_{I}w^{i}_{I}\,\delta c_{[0]}^{I} -M^{ij}\,E_{j[2]}-2\,M_{T}^{ij}\,\zeta_{j[2]}\nn
&+i q_{j}\,\left( \partial_{T}M_{T}^{ij}\,\delta T_{[1]}+\partial_{\mu}M_{T}^{ij}\,\delta \mu_{[1]}+iq_{l}\sum_{I}\partial_{k^{I}_{l}}M_{T}^{ij}\delta c^{I}_{(0)}\right)\,.
\end{align}

\subsection{Horizon vector constraint}\label{subsec:horizon_constraint}

In this subsection, following \cite{Donos:2019hpp}, we use the vector constraint \eqref{imomconstraint} in order to show that $\delta T_{[0]}=\delta \mu_{[0]}=\omega_{[1]}=0$, as well as solve for the horizon fluid velocity $v^{i}_{[2]}$ in terms of the zero modes and the sources
.
At $\mathcal{O}(\varepsilon)$, the vector constraint \eqref{imomconstraint} gives
\begin{align}\label{eq:vec1o1}
\mathcal{B}_{ij} \, v_{[1]}^{j} + \frac{iq_{i}}{\sqrt{g_{(0)}}} \left(\rho\,\delta\mu_{[0]} +s\,\delta T_{[0]} \right) +i q_{k} \, B\varepsilon_{ij} g_{(0)}^{jk} \tau^{(0)} \delta\mu_{[0]} -i\omega_{[1]}\sum_{I}\Phi_{I}^{(0)}k_{i}^{I}\,\delta c_{[0]}^{I} &=0\,,
\end{align}
where we have defined
\begin{align}\label{eq:curlyB_def}
\mathcal{B}_{ij}=\sum_{J}\Psi_{J}^{(0)}k^{J}_{si}k^{J}_{sj}+\sum_{I}\Phi_{I}^{(0)}k_{i}^{I}k_{j}^{I}+\tau^{(0)}B^{2}\,\varepsilon_{ik}\varepsilon_{jl}g^{kl}_{(0)}-\frac{4\pi\rho}{s}B\varepsilon_{ij}\,.
\end{align}
%At $\mathcal{O}(\varepsilon)$, the scalar constraints at the horizon \eqref{hamconstraint}-\eqref{gaugeconstraint} read
%\begin{align}\label{eq:first_conflicting_relation}
%0&=i\omega_{[1]} \sqrt{g_{(0)}} g^{ij}_{(0)} \left(\frac{\partial g^{(0)}_{ij}}{\partial T}\,\delta T_{[0]}+\frac{\partial g^{(0)}_{ij}}{\partial \mu}\,\delta \mu_{[0]} \right)\,,\nn
%0& =i\omega_{[1]} \,\left[\frac{1}{2}\tau^{(0)}a^{(0)} g^{ij}_{(0)}\frac{\partial g^{(0)}_{ij}}{\partial T} +\partial_{\phi^I} \tau^{(0)}a^{(0)}\, \frac{\partial \phi^{I(0)}}{\partial T}+\partial_{\psi^J} \tau^{(0)}a^{(0)}\, \frac{\partial \psi^{J(0)}}{\partial T}+\tau^{(0)} \frac{\partial a^{(0)}}{\partial T}\right]\delta T_{[0]}\nn
%&+i\omega_{[1]} \,\left[\frac{1}{2}\tau^{(0)}a^{(0)} g^{ij}_{(0)}\frac{\partial g^{(0)}_{ij}}{\partial \mu} +\partial_{\phi^I} \tau^{(0)}a^{(0)}\, \frac{\partial \phi^{I(0)}}{\partial \mu}+\partial_{\psi^J} \tau^{(0)}a^{(0)}\, \frac{\partial \psi^{J(0)}}{\partial \mu}+\tau^{(0)} \frac{\partial a^{(0)}}{\partial T}\right]\delta \mu_{[0]} \,.
%\end{align}
We now note that the boundary currents $\delta J_{\infty}^{i}$ and $\delta Q_{\infty}^{i}$ are of order $\mathcal{O}(\varepsilon)$, and so the Ward identities \eqref{eq:current_cons} give\footnote{As explained in \cite{Donos:2019hpp}, we can alternatively get this system by considering the horizon scalar constraints \eqref{hamconstraint}-\eqref{gaugeconstraint} at order $\mathcal{O}(\varepsilon^2)$.}
\begin{align}\label{dcmat}
	i\omega_{[1]}\left( \begin{array}{cc}
		T^{-1}c_{\mu} &\xi\\
		\xi &\chi_q
	\end{array} \right)
	\left( \begin{array}{c}
		\delta T_{[0]}\\
		\delta\mu_{[0]}
	\end{array} \right)=0\,.
\end{align}

We now consider the two possibilities for $\omega_{[1]}$:

\textbf{$\omega_{[1]} \neq 0$:} Provided the matrix of susceptibilities in \eqref{dcmat} is invertible, as is generically the case, we deduce from \eqref{dcmat} that $\delta T_{[0]} = \delta\mu_{[0]} = 0$. However, we can then combine \eqref{eq:deltachi1_source} and \eqref{eq:vec1o1}, leading to
\begin{align}\label{eq:orderone_firstcase}
&\sqrt{g_{(0)}}\left(\sum_{J}\Psi_{J}^{(0)}k^{J}_{si}k^{J}_{sj}+\tau^{(0)}B^{2}\,\varepsilon_{ik}\varepsilon_{jl}g^{kl}_{(0)}-\frac{4\pi\rho}{s}B\varepsilon_{ij}\right) \, v_{[1]}^{j} -\sum_{I}k_{i}^{I}w_{I}^{j}\zeta_{j[1]}=0\,.
\end{align}
In order to find quasinormal modes we set the sources to zero $\zeta_{i[1]}=0$, which then leads to $v_{[1]}^{j}=0$.\footnote{Assuming that the matrix multiplying $v_{[1]}^{j}$ in \eqref{eq:orderone_firstcase} is invertible in generic backgrounds.} This in turn leads to the trivial perturbation with $\delta c_{[0]}^{I} = 0$ as well.

\textbf{$\omega_{[1]} = 0$:} In this case, \eqref{dcmat} contributes nothing new. However, at next order in $\varepsilon$, the continuity equations will lead to another version of \eqref{dcmat}, but with $\omega_{[1]} \rightarrow \omega_{[2]}$. The only way this relation can avoid conflicting with the combination of \eqref{eq:deltachi1_source} and \eqref{eq:vec1o1} is if $\delta T_{[0]} = \delta\mu_{[0]} = 0$.
%
%\vz{Maybe this comment is enough} Thus, in the presence of momentum relaxation the hydrodynamic modes generated by our thermodynamic perturbations are diffusive, seeded by a zero mode with $\delta c_{[0]}^{I} \neq 0$, $\delta T_{[0]}=\delta\mu_{[0]}=0$.

We now solve the horizon vector constraint at order $\mathcal{O}(\varepsilon^{2})$ to write
%\begin{align}\label{eq:v2sol}
%v_{[2]}^{j}=i\omega_{[2]}\,\sum_{I}C^{j}_{I}\,\delta c^{I}_{[0]}+S^{ji}\,\left(\zeta_{i[2]}-iq_{i}T^{-1}\,\delta T_{[1]} \right)+K^{jk}\,\left( E_{k[2]}-iq_{k}\delta \mu_{[1]}\right)\,,
%\end{align}
%where we have defined
%\begin{align}\label{eq:matrix_defs_app}
%&C^{j}_{I}=\left(\mathcal{B}^{-1}\right)^{ji}k^{I}_{i}\Phi_{I}^{(0)},\qquad S^{ji}=4\pi T \left(\mathcal{B}^{-1}\right)^{ji},\qquad K^{jk}=\frac{4\pi\rho}{s}\,\left(\mathcal{B}^{-1}\right)^{ji}\,\mathcal{N}_{i}{}^{k},\nn
%&\mathcal{N}_{ij}=g_{(0)ij}+\frac{s\tau^{(0)}}{4\pi\rho}B\,\varepsilon_{ij},\qquad \mathcal{N}_{i}{}^{k}\equiv \mathcal{N}_{ij}g^{jk}_{(0)}  ,\qquad \mathcal{N}^{k}{}_{i}\equiv g^{kj}_{(0)}\mathcal{N}_{ji} \,,
%\end{align}
\begin{align}\label{eq:v2sol}
Ts\,v_{[2]}^{j}=i\omega_{[2]}\,\sum_{I}\lambda^{j}_{I}\,\delta c^{I}_{[0]}+T\bar{\kappa}_{H}^{ji}\,\left(\zeta_{i[2]}-iq_{i}T^{-1}\,\delta T_{[1]} \right)+T\bar{\alpha}_{H}^{jk}\,\left( E_{k[2]}-iq_{k}\delta \mu_{[1]}\right)\,,
\end{align}
where we have defined
\begin{align}\label{eq:matrix_defs_app}
&\sigma_{0}^{ij}=\frac{\tau^{(0)}s}{4\pi}g^{ij}_{(0)}\,,\qquad \mathcal{N}_{i}{}^{k}=\delta_{i}{}^{k}+\frac{B}{\rho}\varepsilon_{ij}\sigma_{0}^{jk}\,, \qquad \eta^{I}_{i}=\frac{1}{4\pi T}\Phi_{I}^{(0)}k^{I}_{i}\,,\nn
&\bar{\alpha}_{H}^{ik}=4\pi\rho\,\left(\mathcal{B}^{-1}\right)^{ij}\mathcal{N}_{j}{}^{k}\,,\qquad \bar{\kappa}_{H}^{ik}=4\pi T s\,\left( \mathcal{B}^{-1}\right)^{ik}\,,\qquad \lambda^{j}_{I}=T\bar{\kappa}_{H}^{ji}\eta^{I}_{i}\,,
\end{align}
where $\mathcal{B}_{ij}$ is given in \eqref{eq:curlyB_def} and indices in $\mathcal{N}$ are raised and lowered with the horizon metric $g_{(0)ij}$. The expressions \eqref{eq:J2v2} and \eqref{eq:Q2v2} for the currents contain the horizon fluid velocity $v^{i}_{[2]}$. Substituting \eqref{eq:v2sol}, leads to the constitutive relations \eqref{eq:currents_dis_consti} presented in the main text. Essentially the explicit lattice has allowed us to integrate out the fluid velocity, and this was conveniently done by solving the constraints at the black hole horizon.

\bibliographystyle{utphys}
\bibliography{refs}{}

\end{document}